\documentclass{aa} 
\usepackage{graphicx}   

\usepackage{txfonts}

\usepackage[colorlinks=true, citecolor=blue]{hyperref}

\newcommand{\simgt}{\,\rlap{\lower 3.5 pt \hbox{$\mathchar \sim$}} \raise 1pt \hbox {$>$}\,}
\newcommand{\simlt}{\,\rlap{\lower 3.5 pt \hbox{$\mathchar \sim$}} \raise 1pt \hbox {$<$}\,}

\newcommand{\BE}{\begin{equation}}
\newcommand{\EE}{\end{equation}}
\newcommand{\BEA}{\begin{eqnarray}}
\newcommand{\EEA}{\end{eqnarray}}

\newcommand{\DV}{\ifmmode{\Delta v_{\mathrm{Ly}\alpha}}\else $\Delta v_{\mathrm{Ly}\alpha}$\xspace\fi}

\newcommand{\HI}{\ifmmode{\textsc{hi}}\else H\textsc{i}\fi\xspace}
\newcommand{\HII}{\ifmmode{\textsc{hii}}\else H\textsc{ii}\fi\xspace}

\newcommand{\OIII}{[O\textsc{iii}]\xspace}

\newcommand{\Msun}{\ifmmode{M_\odot}\else $M_\odot$\xspace\fi}
\newcommand{\MUV}{\ifmmode{M_\textsc{uv}}\else M$_\textsc{uv}$\xspace\fi}
\newcommand{\fesc}{\ifmmode{f_\textrm{esc}^{\textrm{Ly}\alpha}}\else $f_\textrm{esc}^{\textrm{Ly}\alpha}$\xspace\fi}
\newcommand{\lya}{\ifmmode{\mathrm{Ly}\alpha}\else Ly$\mathrm{\alpha}$\xspace\fi}
\newcommand{\ewlya}{\ifmmode{\mathrm{EW_{Ly\alpha}}}\else EW$_{\mathrm{Ly\alpha}}$\xspace\fi}
\newcommand{\fwhmlya}{\ifmmode{\mathrm{FWHM_{Ly\alpha}}}\else FWHM$_{\mathrm{Ly\alpha}}$\xspace\fi}
\newcommand{\fwhmha}{\ifmmode{\mathrm{FWHM_{H\alpha}}}\else FWHM$_{\mathrm{H\alpha}}$\xspace\fi}
\newcommand{\Ha}{\ifmmode{\mathrm{H}\alpha}\else H$\alpha$\xspace\fi}

\newcommand{\nh}[1][]{\ifmmode{\overline{n}_\textsc{h}^{#1}}\else $\overline{n}_\textsc{h}$\xspace\fi}

\newcommand{\xHI}{\ifmmode{x_\HI}\else $x_\HI$\xspace\fi}
\newcommand{\xHImean}{\ifmmode{\overline{x}_\HI}\else $\overline{x}_\HI$\xspace\fi}
\newcommand{\xHIImean}{\ifmmode{\overline{x}_\HII}\else $\overline{x}_\HII$\xspace\fi}
\newcommand{\trec}{\ifmmode{t_\textrm{rec}}\else $t_\textrm{rec}$\xspace\fi}
\newcommand{\clump}[1][]{\ifmmode{C_\HII^{#1}}\else $C_\HII$\xspace\fi}
\newcommand{\xiion}{\ifmmode{\xi_\mathrm{ion}}\else $\xi_\mathrm{ion}$\xspace\fi}

\newcommand{\Nion}{\ifmmode{\dot{N}_{\mathrm{ion}}}\else $\dot{N}_\mathrm{ion}$\xspace\fi}
\newcommand{\Rion}[1][]{\ifmmode{R_\mathrm{ion}^{#1}} \else $R_\mathrm{ion}$\xspace\fi}

\newcommand{\Tb}{\ifmmode{T_{21}}\else $T_{21}$\xspace\fi}
\newcommand{\aesc}{\ifmmode{\alpha_\mathrm{esc}}\else $\alpha_\mathrm{esc}$\xspace\fi}
\newcommand{\fescII}{\ifmmode{f_\mathrm{esc,10}^\textsc{ii}}\else $f_\mathrm{esc,10}^\textsc{ii}$\xspace\fi}
\newcommand{\fescIII}{\ifmmode{f_\mathrm{esc,7}^\textsc{ii}}\else $f_\mathrm{esc,7}^\textsc{iii}$\xspace\fi}
\newcommand{\astarII}{\ifmmode{\alpha_\star^\textsc{ii}}\else $\alpha_\star^\textsc{ii}$\xspace\fi}
\newcommand{\astarIII}{\ifmmode{\alpha_\star^\textsc{iii}}\else $\alpha_\star^\textsc{iii}$\xspace\fi}
\newcommand{\fstarII}{\ifmmode{f_{\star,10}^\textsc{ii}}\else $f_{\star,10}^\textsc{ii}$\xspace\fi}
\newcommand{\fstarIII}{\ifmmode{f_{\star,7}^\textsc{iii}}\else $f_{\star,7}^\textsc{iii}$\xspace\fi}
\newcommand{\tstar}{\ifmmode{t_\star}\else $t_\star$\xspace\fi}
\newcommand{\Mturn}{\ifmmode{M_\mathrm{turn}}\else $M_\mathrm{turn}$\xspace\fi}
\newcommand{\LX}{\ifmmode{L_X/{\dot{M}_\star}}\else $L_X/{\dot{M}_\star}$\xspace\fi}
\newcommand{\nuX}{\ifmmode{E_0}\else $E_0$\xspace\fi}
\newcommand{\AVCB}{\ifmmode{A_\mathrm{VCB}}\else $A_\mathrm{VCB}$\xspace\fi}
\newcommand{\ALW}{\ifmmode{A_\mathrm{LW}}\else $A_\mathrm{LW}$\xspace\fi}
\newcommand{\Mpcinv}{\ifmmode{\,\mathrm{Mpc}^{-1}}\else \,Mpc$^{-1}$\xspace\fi} 

\newcommand{\kp}{\ifmmode{k_\textrm{peak}}\else $k_\textrm{peak}$\xspace\fi}
\newcommand{\hp}{\ifmmode{h_\textrm{peak}}\else $h_\textrm{peak}$\xspace\fi}
\newcommand{\hMpc}{\ifmmode{\,h^{-1}\textrm{Mpc}}\else \,$h^{-1}$Mpc\xspace\fi}

\newcommand{\kms}{\,\ifmmode{\mathrm{km}\,\mathrm{s}^{-1}}\else km\,s${}^{-1}$\fi\xspace}
\newcommand{\cm}{\,\ifmmode{\mathrm{cm}}\else cm\fi\xspace}

\begin{document}

   \title{Lyman-alpha emission at the end of reionization: Line strengths and profiles from MMT and JWST observations at $z\sim5-6$}
   \titlerunning{Lyman-alpha emission at the end of reionization}

   \author{
        Gonzalo Prieto-Lyon\fnmsep\thanks{E-mail: gonzalo.p.lyon@gmail.com}\inst{1,2}
        \and Charlotte A. Mason\inst{1,2}
        \and Victoria Strait\inst{1,2}
        \and Gabriel Brammer\inst{1,2}
        \and Rohan P. Naidu\inst{4}
        \and Romain A. Meyer\inst{3}
        \and Pascal Oesch\inst{1,3}
        \and Sandro Tacchella\inst{5,6}
        \and Alba Covelo-Paz\inst{3}
        \and Emma Giovinazzo\inst{3}
        \and Mengyuan Xiao\inst{3}}

    \authorrunning{G. Prieto-Lyon et al.}
        
   \institute{Cosmic Dawn Center (DAWN)
   \and Niels Bohr Institute, University of Copenhagen, Jagtvej 128, DK-2200 Copenhagen N, Denmark
   \and Departement d’Astronomie, University of Geneva, Chemin Pegasi 51, 1290 Versoix, Switzerland
   \and MIT Kavli Institute for Astrophysics and Space Research, 77 Massachusetts Avenue, Cambridge, 02139, Massachusetts, USA
   \and Kavli Institute for Cosmology, University of Cambridge, Madingley Road, Cambridge CB3 0HA, UK 2
   \and Cavendish Laboratory, University of Cambridge, 19 JJ Thomson Avenue, Cambridge CB3 0HE, UK
   }

  \abstract
   {}
   {With JWST, it is now possible to use Lyman-Alpha (\lya) emission from galaxies beyond $z>8$ to trace neutral hydrogen in the intergalactic medium (IGM) as the Universe became reionized. However, observed \lya emission lines are known to be affected by scattering by neutral hydrogen in the IGM and the interstellar and circumgalactic medium (CGM), necessitating "baseline" models of \lya properties in the ionized IGM to disentangle their impacts. In this work, we characterize \lya properties at the end of of the epoch of reionization (EoR) at $z\sim5-6$, providing a baseline that can be applied to $z>6$ observations.}
   {We targeted GOODS-N with MMT/Binospec, obtaining R$\sim$4360 rest-frame UV spectra of 236 galaxies at $z\sim5-6$, selected primarily from HST/CANDELS, finding 62 \lya detections. We used JWST observations from JADES and FRESCO for a subset of our sources to characterize \lya properties as a function of the UV continuum and \Ha emission. We present the first statistical measurements of the \lya FWHM distribution at $z\sim5-6$, along with empirical baseline models for the\lya equivalent width (\ewlya) and escape fraction (\fesc) that are conditional on the UV magnitude and slope.}
    {
    We found our \ewlya and \fesc models strongly depend on UV magnitude, inferring that 45$\pm$5\% and $<62\pm8\%$ of \MUV = -19.5 galaxies have \ewlya$>$25\AA\ and $\fesc>0.2$, respectively. We also found a mean \fwhmlya of 245 \kms and median \lya velocity offset of 258 \kms, both displaying a correlation with increasing UV luminosity. Our median observed \lya line profile is broader and has higher velocity offset compared to pre-JWST models based on $z\sim2$ lines, which might reflect both resonant scattering by residual neutral hydrogen in the IGM at $z\sim5$-6 and increasing ISM and CGM densities. Our median line profile thus predicts higher \lya transmission in a fully neutral IGM, providing insights into recent detections of \lya at $z>10$.
    }
    {}


\maketitle
\section{Introduction}
The epoch of reionization (EoR) was the last phase transition of the Universe when the majority of the neutral hydrogen in the intergalactic medium (IGM) became ionized due to an injection of ionizing photons from the first galaxies \citep[e.g.,][]{Stark2016}. The rate and morphology of this transition is generally determined by the galaxies' ionizing properties, thereby constraining the reionization process and offering unique insights into early galaxies, even those below our direct detection limits. Multiple independent studies have reached the consensus that the Universe was fully reionized at z $\sim$ 5 - 6 \citep[e.g.][]{Fan2006,Becker2015a,Eilers2018,Yang2020,Qin2021b,Bosman2022,Gaikwad2023}, but likely to be predominantly neutral at $z\sim7-8$ \citep[e.g.,][]{Stark2010,Treu2012,Pentericci2014,Schenker2014,Mason2018,Davies2018b,Yang2020,Jung2020,Bolan2022,Umeda2024,Nakane2024,Tang2024b,Jones2024,Mason2025}.
The earliest stages of the EoR at $z>9$ are still very poorly constrained, but this knowledge is crucial for understanding the build up of the first galaxies and the higher-than-expected UV luminosity density detected by JWST \citep{Gelli2024,Munoz2024,Whitler2025}.

Lyman-alpha (\lya) emission from galaxies is one of our best probes of the early stages of the EoR. Its high cross-section for damping wing scattering by neutral hydrogen means \lya flux traces the IGM opacity \citep[e.g.,][]{Miralda-Escude1998b,Mesinger2008} and we are now detecting 1000s of $z>8$ galaxies with JWST \citep[e.g.][]{Adams2022,Donnan2024,Whitler2025}, providing large samples to comprehensively trace the IGM at high redshift.
In contrast, the number density of quasars, the typical IGM probes at lower redshifts \citep[e.g.,][]{Jiang2016,Matsuoka2018,Wang2020,Schindler2023}, decreases significantly at z $>$ 7 \citep{Euclid2019}.
The visibility of \lya\ from galaxies drops considerably at $z>6$, implying an increasing neutral IGM \citep[Seen in the declining fraction of galaxies detected with strong (EW$>$25\,\AA) \lya ,][]{Stark2010,Pentericci2014,DeBarros2017,Jung2019,Fuller2020,Laporte2021,Tang2024b}.
JWST has opened a new window on reionization studies, enabling \lya observations in faint galaxies at $z\simgt6$ without the atmospheric and sensitivity limitations of ground-based telescopes.
Early JWST results have confirmed the downturn in \lya\ emission at $z\simgt6$ seen from the ground and demonstrated strong \lya\ is extremely rare at $z>8$ in current samples \citep[][]{Nakane2024,Chen2024,Tang2024b,Jones2024,Kageura2025}. They have also provided the first detections of \lya\ at $z>10$ \citep{Bunker2023,Witstok2024b}. These observations demonstrate how valuable a tool JWST will be in improving our understanding of the earliest stages of the EoR.

However, the interpretation of \lya observations during the EoR relies on grasping \lya transmission on various spatial scales. Overall, \lya photons are subject to scattering not only by neutral hydrogen in the IGM, but also in the interstellar and circumgalactic media (ISM and CGM).
To use \lya to track the reionization history, an understanding of \lya\ emerging from the ISM/CGM is required. In particular, because of the strong wavelength dependence of the IGM \lya damping wing \citep{Miralda-Escude1998b}, both an understanding of the \lya line strength and spectral shape emerging from the ISM/CGM is important to disentangle the impact of the IGM \citep{Dijkstra2011,Mason2018b,Endsley2022}.

To this end, samples of galaxies at the end of the EoR, z $\sim$ 5 - 6, have been commonly used as a baseline for interpreting higher redshift observations \citep{Stark2011,DeBarros2017}. These galaxies are expected to be physically similar to those at earlier epochs, but suffer minimal damping wing attenuation by the neutral IGM. Therefore, these galaxies provide a window into the emergent \lya emission from the ISM and CGM, and have been used to infer the additional impact of the IGM on $z>6$ galaxies' \lya \citep[e.g.,][]{Mason2018,Whitler2020,Jung2020,Bolan2022}.
However, these baseline samples had been limited to the most readily available observables pre-JWST -- \lya EW and UV photometry, with only a handful of systemic redshifts known from strong rest-UV or FIR lines \citep[e.g.,][]{Stark2017,Bradac2017}: our knowledge of the emerging \lya line profile had been severely limited. These limitations, along with others, such as the unknown \lya reionization analyses, have led previous models to rely on empirical relations based on $z\sim0-2$ galaxy samples \citep[e.g.,][]{Mason2018,Mason2019c,Hayes2023}, where the rest-frame optical is accessible from the ground. However, it is unclear whether these samples are good analogs of reionization-era sources. For example, recent works have shown the \lya velocity offsets of z$\sim$6 galaxies appear systematically higher than models derived from $z\sim2$ galaxies \citep{Prieto-Lyon2023b,Tang2024}.

JWST's ability to observe the rest-frame optical is transforming our ability to understand galaxies in this era. 
Thanks to $R\simgt1000$-resolution observations, such as NIRCam grism data, we now have accurate systemic redshifts for 1000s of $z>5$ galaxies \citep[e.g.,][]{Oesch2023,Alba2024,Meyer2024b} and measurements of \lya escape fractions (\fesc), relative to Balmer lines, and velocity offsets (\DV) are now routinely being measured for $z\sim5-9$ galaxies \cite[e.g.,][]{Lin2024,Chen2024,Tang2023,Tang2024,Saxena2024,Chen2025}. 
\citet{Tang2024} recently presented constraints on the emergent \lya EW and escape fraction distributions in bins of \MUV, UV $\beta$ slope and [OIII]+H$\beta$ EW, and velocity offset distributions in $z\sim5-6$ galaxies.
These observations have provided a more robust baseline for predicting emergent \lya\ at $z>6$ and, thus, for interpreting observed \lya\ in the context of the EoR \citep{Tang2024b}.

However, these works have mostly focused on measuring \lya emission distributions in bins of galaxies properties, limiting the information used in the conditional distributions, and less attention has focused on the shape of the \lya line profile at $z\sim5-6$. 
Accurately measuring the emergent lineshape is critical for understanding the transmission of \lya through the IGM \citep{Dijkstra2010,Mason2018,Mason2019c,Endsley2022,Tang2024c,Mukherjee2024}. 
High resolution spectroscopy is required to accurately measure both the asymmetric shape of \lya and its offset from the systemic redshift (\DV). At a resolution below $\sim$100\kms ($R\simgt4000$) is optimal, as the typical full-width half-max (FWHM) in \lya-emitters in the ionized IGM ($z<5$) range from $\sim 100- 400$\kms \citep{Verhamme2018a,Tang2024c} and values of \DV range from 150\kms to 500\kms \citep[e.g.,][]{Prieto-Lyon2023b,Tang2024c}. \citet{Tang2024} demonstrated that $z\sim5-6$ \lya line profiles appear more offset from systemic than profiles for sources with comparable \ewlya at $z\sim2-3$ \citep{Tang2024c}, which they discuss may be due to residual neutral gas in the ionized IGM resonantly scattering \lya close to line center and/or increased scattering in (relatively dust-poor) ISM and CGM. 
However, we currently have no statistical information about the broadness and asymmetry of \lya line profiles at these redshifts, which provides additional information to understand the evolution in line profiles, primarily due to the moderate to low resolution of \lya spectroscopy in fields overlapping with deep JWST spectroscopy (i.e. VLT/MUSE $R\sim3000$). Additionally, JWST spectroscopic observations using NIRSpec/PRISM have an insufficient resolution (R$\sim$100) to accurately measure \lya velocity offsets. In contrast, NIRCam grism spectra, with a higher resolution (R$\sim$1600), is better suited for studies such as this one.

To that end, we have carried out a deep, high-resolution ($R\sim4400$) near-infrared spectroscopic survey of 236 $z\sim5-6.5$ Lyman-break galaxies in the GOODS North field \citep{Giavalisco2004}, selected from deep HST imaging, with MMT/Binospec \citep{Fabricant2019}, with the aim of constraining the \lya strength, and line profiles emerging from the ISM/CGM of galaxies in the post-reionization era. MMT/Binospec, with its high resolution (R$\sim$4360), high throughput, and wide on-sky coverage (two 8$'$ x 15$'$ fields of view), offers great advantage for \lya studies, surpassing the resolution of most z$\sim$5-6 detections in the literature (e.g., LRIS, MUSE, VIMOS, JWST, R$<$3500). Our sample includes 62 high-resolution \lya detections.
Thanks to the excellent ancillary datasets in GOODS-N, we supplement our \lya\ spectroscopy with JWST/NIRCam slitless spectra from FRESCO + CONGRESS \citep{Oesch2023,Egami2023}, and deep HST and JWST photometry from the CANDELS \citep{Giavalisco2004,Grogin2011,Koekemoer2011} and JADES surveys \citep{Jades-Eisenstein2023a} to obtain precise measurements of galaxy properties (UV magnitudes, UV beta slopes, systemic velocities, \lya escape fractions: \fesc).
With these data, we were able to build two new empirical models for \ewlya and \fesc that are conditional on UV observables, \MUV and $\mathrm{\beta}$, which can be used as a baseline for constraining the reionization history. We also explore the properties of \lya lineprofiles (\DV, FWHM, asymmetry) in our sample. 

The paper is structured as follows. In Section~\ref{sec:data}, we describe our target selection, the MMT/Binospec spectroscopy, and ancillary datasets. In Section~\ref{sec:galaxySelect}, we describe our \lya and rest-optical emission lines measurements. We describe our \lya measurements in the context of other galaxy properties in Section~\ref{sec:lya_galaxy_properties}. We present our new empirical model for emergent \ewlya and \fesc in Section~\ref{sec:MakingModel}. We discuss our results in Section~\ref{sec:disc} and present our conclusions in Section~\ref{sec:conc}. We assume a flat $\Lambda\mathrm{CDM}$ cosmology with $\Omega_m=0.3,\,\Omega_\Lambda=0.7,\,h=0.7$. All magnitudes are in the AB system.

\begin{figure*}[!h]
\begin{center}\includegraphics[width=9cm, trim=0.1cm 1.5cm 0cm 0cm]{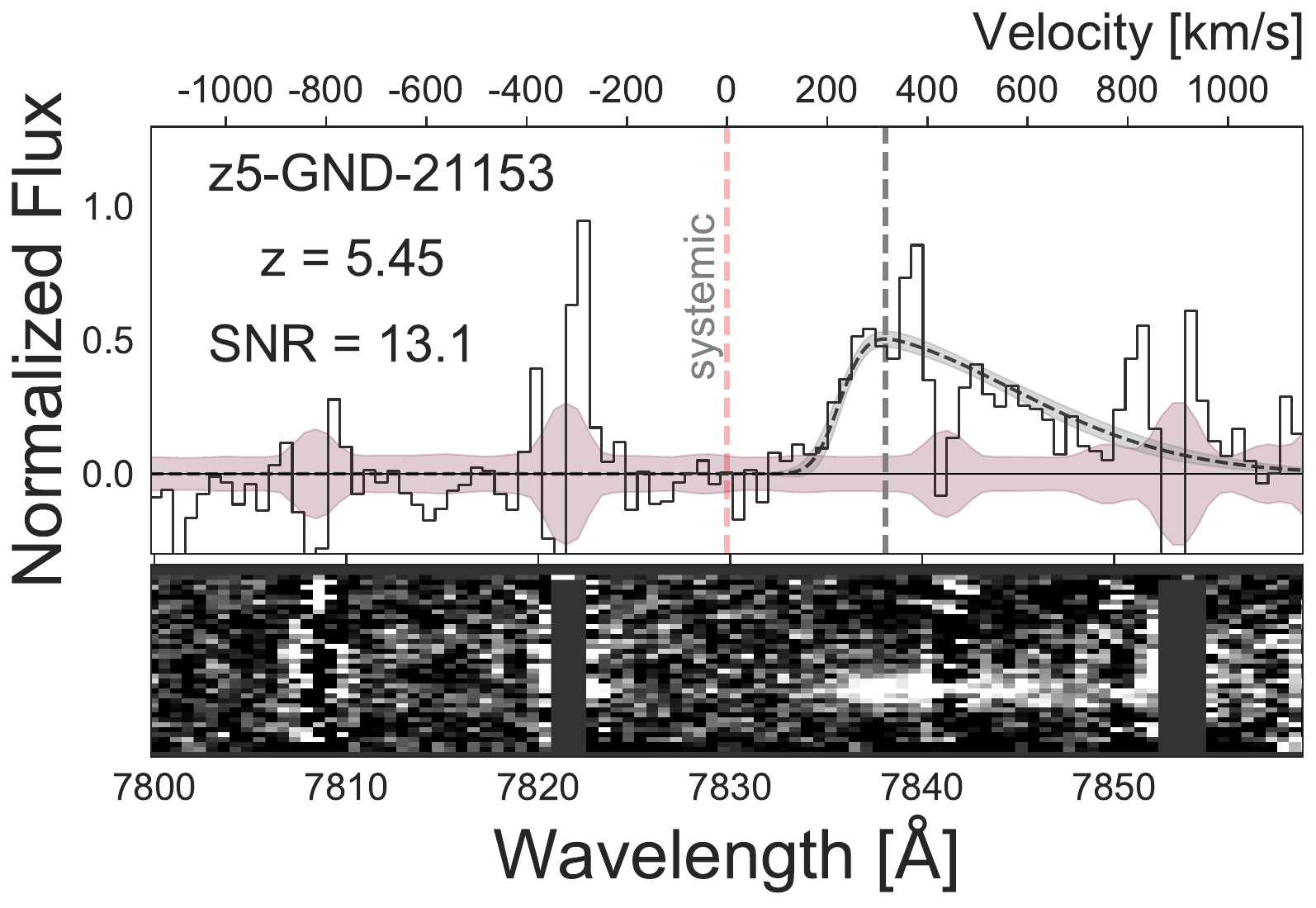}\includegraphics[width=9cm, trim=0.1cm 1.5cm 1cm 0cm]{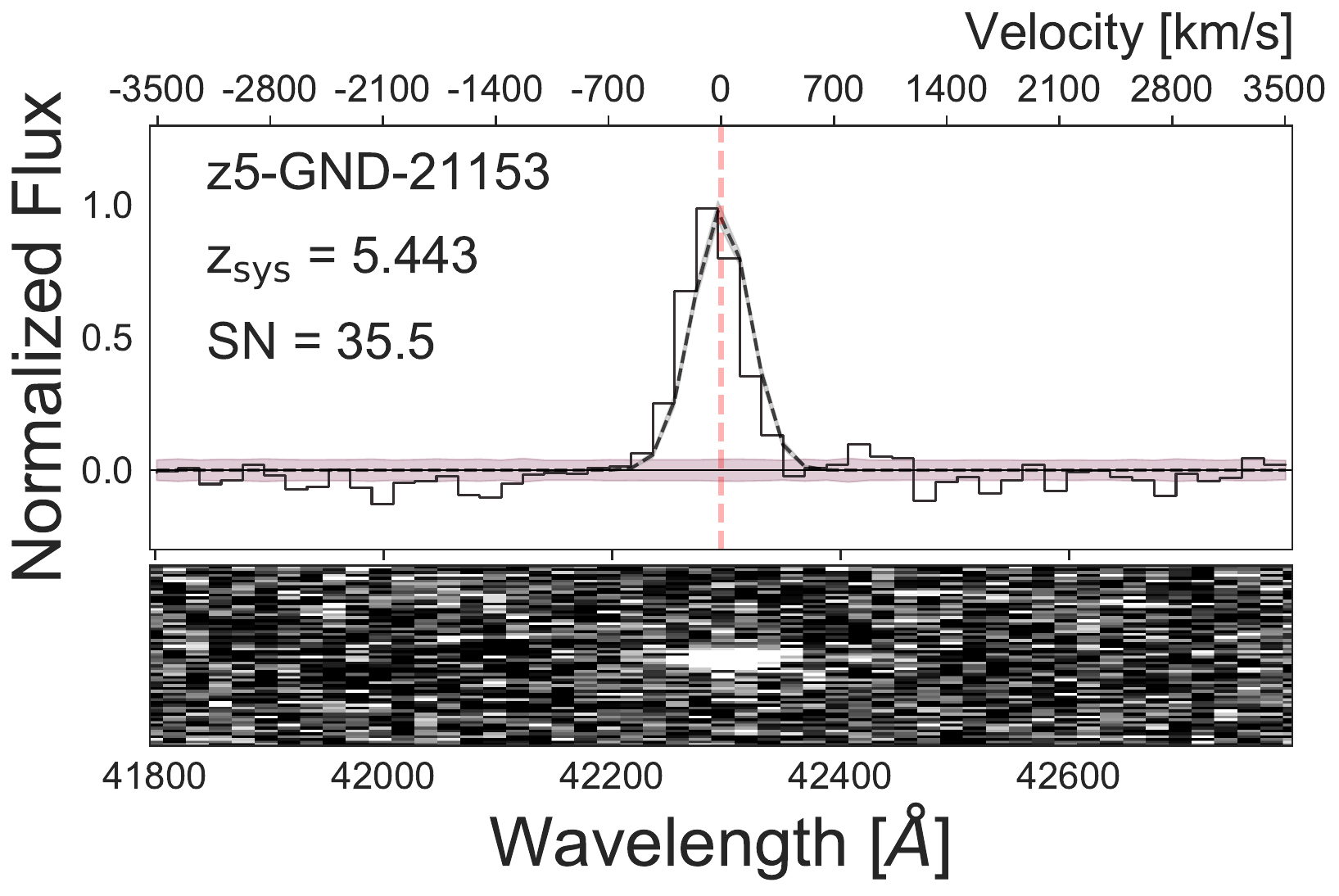}
\end{center}
\caption{Spectra of z5-GND-21153. Left:  MMT/Binospec spectra of \lya. Right: \Ha counterpart from JWST/NIRCam slitless spectra of the same source. We plot the emission line fit with its corresponding 1$\sigma$ uncertainty as a gray shaded area. On the left, the vertical red line shows the systemic/\Ha redshift, in this example there is a 315\kms velocity offset between \lya and \Ha. The upper axis shows velocity centered at the peak of the systemic line. In the bottom are the 2D spectra.}
\label{fig:spectra_example}
\end{figure*}

\section{Data}\label{sec:data}

Here, we describe the observational datasets used in this work. First, we describe our target selection (Section~\ref{sec:data_target}) and the MMT/Binospec spectroscopic observations of z $\sim$ 5 - 7 Lyman-break galaxies in the GOODS-N field (Section~\ref{sec:binospec}). Second, we describe the JWST/NIRCam slitless spectroscopy from the FRESCO survey\footnote{https://jwst-fresco.astro.unige.ch} that provides rest-frame optical spectroscopy of a subset of our targets (Section~\ref{sec:fresco}). Finally, we describe the JWST and HST photometric datasets from the DJA \footnote{https://dawn-cph.github.io/dja/} (The DAWN JWST Archive) photometric catalog of JADES\footnote{https://jades-survey.github.io} \citep{Jades-Eisenstein2023a}, the CANDELS/SHARDS \citep{Giavalisco2004,Grogin2011,Koekemoer2011} catalog from \citet{Barro2019} and 3D-HST\footnote{https://archive.stsci.edu/prepds/3d-hst/} \citep{Skelton2014,Brammer2012} imaging in Section~\ref{sec:data-phot}. Our identification of the emission lines is described in Section~\ref{sec:galaxySelect}.

\subsection{Target selection}\label{sec:data_target}

Our primary targets were composed of 236 $z_\mathrm{phot}\sim5-7$ candidates in GOODS-N \citep{Giavalisco2004} selected by Lyman-break color criteria from \citet{Bouwens2015}. All candidates were detected with a magF160W$<$27.5. ACS/WFC \& WFC3/IR filters from the deep GOODS imaging were used for tracing the Lyman-Break and the rest-frame UV continuum, along with two low wavelength Spitzer/IRAC bands break the degeneracy between the Lyman-Break and 4000\AA\ break. The 236 sources include 24 spectroscopically confirmed \lya-detected galaxies previously reported by \citet{Hu2010}, \citet{Stark2011}, and \citet{Jung2018}, to obtain higher resolution, higher signal-to-noise ratio (S/N) observations of their \lya lineshapes.

\subsection{MMT / Binospec}\label{sec:binospec}

We performed observations of our targets in GOODS-N with MMT/Binospec \citep{Fabricant2019} from 2019 to 2021. The observations were divided into 4 masks with two detectors of 8'x15' field of view, covering a total footprint of 0.27\,deg$^2$, with exposure times between 10 - 16 hours per mask. We used the 600 lines/mm grating, providing R$\sim$4360 and a wavelength range between 7255 - 9750 $\AA$, enabling \lya detections between $z = 4.9$ to 7.0. Thanks to the high resolution per pixel, 0.61 $\AA$/pix and instrumental resolution of $\approx 70$ \kms, we were able to accurately retrieve the profile shape of the observed \lya lines, improving, for instance, on the JWST NIRSpec and VLT/MUSE \lya observations ($R\sim3000$) by a factor of 45\%. In Figure \ref{fig:spectra_example} we show an example of our \lya spectra. We show the 2D spectra of each source in Appendix \ref{appendix:spectra}, noting that due to varying slit lengths on the masks, the targets are not necessarily centered in each slit. Properties of each mask are described in Table~\ref{tab:masks}. All MMT/Binospec spectra discussed in this work is publicly available \footnote{https://sid.erda.dk/sharelink/FCZBDRW6O2}.

\subsection{JWST / NIRCam - Slitless spectra}\label{sec:fresco}

We include NIRCAM/grism observations from FRESCO \citep{Oesch2023} which targeted 62 arcmin$^2$ in GOODS-N, covering 82/236 of our MMT/Binospec targets. The wavelength range with the F444W filter covers 3.8-5$\mu$m, allowing the observation of the most luminous optical lines, \Ha and the \OIII doublet, from z$\sim$4.8-6.6 and z$\sim$6.6-9, respectively, with R$\sim$1600 $\approx 180$\kms. This provides systemic redshift for a subset of our targets, with detected \Ha and \OIII line in FRESCO, with an average 5$\sigma$ flux limit $\gtrsim$2$\cdot$10$^{-18}$erg s$^{-1}$ cm$^{-2}$ (see Section~\ref{sec:lines_fresco} for description of the line detections). \Ha flux measurements also provide an estimate of the intrinsic \lya flux, enabling us to estimate \lya escape fractions. The reduction and spectral extraction of FRESCO data is detailed in \citet{Oesch2023}. The accuracy of the wavelength calibrations is crucial due to \lya velocity offsets near the instrumental resolutions (see Figure \ref{fig:spectra_example}). To verify the calibration, we compared the systemic redshift of FRESCO sources with known ground-based spectra and found no substantial systematic differences, with a scatter in the order of the wavelength pixel resolution. We show the spectra of all galaxies with a \lya and optical line detection in Appendix \ref{appendix:spectra}. 

\subsection{Photometric data}\label{sec:data-phot}

Deep HST and JWST photometry of our candidates is provided by CANDELS \citep{Giavalisco2004,Grogin2011,Koekemoer2011}, 3D-HST \citep{Skelton2014}, and JADES \citep{Jades-Eisenstein2023a} imaging in GOODS-N. These data provide rest-frame UV coverage of our targets. CANDELS and 3D-HST provide imaging in ACS/WFC: F435W, F606W, F775W, F814W, and F850LP; and WFC3/IR: F105W, F125W, F140W and F160W. JADES JWST/NIRCam imaging covers 128/236 of our targets in F090W, F115W, F150W, F200W, F277W, F335M, F356W, F410M, and F444W. 

Inside the JADES FoV, we use the data reduction from DJA processed with \textit{grizli} \citep{Grizli2023a}, which includes both legacy HST ACS/WFC \& WFC3/IR bands and NIRCam imaging from JADES reduced self-consistently \citep{Valentino2023} (hereafter, JADES-DJA). For sources outside of the FoV of JADES, we use ACS/WFC \& WFC3/IR photometry catalogs from \citet{Barro2019}, which provide consistent photometry to the \citet{Valentino2023} catalogs. In cases where we do not find a match in either catalog, we used the 3D-HST catalog \citep{Skelton2014}. For all catalogs we use aperture corrected fluxes.

\subsection{Binospec data reduction}\label{sec:reduction}

In the following section, we describe how we reduced our Binospec data by applying a telluric calibration, a flux calibration, and an additional sky subtraction. The initial reduction and sky subtraction were performed with the Binospec Pipeline \footnote{https://github.com/bjweiner/MMT}. To optimize the reductions for our faint emission line targets we further reduced each individual exposure. None of our targets showed evidence of a UV continuum detection.

To reduce features due to atmospheric emission such as the O2 A\& B bands, we performed a telluric calibration of our sources. We followed a similar method to \citet{Vacca2003}. On each mask, we observed two standard F-type stars. We compared the observed star in each exposure to F-star synthetic models using the \textit{SEDPy}\footnote{https://github.com/bd-j/sedpy} library. We used the synthetic models that best fits the spectrum of our standard stars. We extracted and normalized the spectra to obtain the telluric transmission $T(\lambda)$ by dividing the data by the model (both continuum-normalized). We applied the transmission function to all spectra in each exposure via
\begin{align}\label{eq:transmission}
f_{corr}(\lambda) = \frac{f_{obs}(\lambda)}{T(\lambda-\alpha)^{\beta}}.
\end{align}
The constant $\alpha$ accounts for any wavelength shifts between the spectra and transmission and $\beta$ is used as a scaling factor of the transmission. To calculate $\alpha$ and $\beta$, we minimized the RMS in the two most important telluric regions, O2 A\& B bands in f$_{corr}(\lambda)$ for every exposure for every slit containing a high redshift target.

We performed flux calibration for each exposure by calibrating our standard stars with magnitudes obtained by \textit{SDSS} \citep{SDSS_DR17}. We used the \textit{i} filter of \textit{Sloan} since it covers a similar wavelength range as the MMT/Binospec data. We convolved our standard star spectra with the corresponding \textit{i}$_{sdss}$ filter using SEDPy.

 We performed additional sky subtraction of the exposures for each mask to improve on the pipeline sky subtraction. We generated a global 1D sky spectrum for each individual exposure, $S(\lambda)$, using the standard deviation of flux in all slits containing $z>5$ targets with no clear emission lines. We note that as our targets were faint, we did not expect to detect the continuum, so the "empty" slits were expected to measure the sky. In this procedure, we used a sigma clipping to remove any cosmic rays. We performed an additional sky subtraction to each spectrum per exposure as
\begin{equation}\label{eq:rescale}
    f_\mathrm{skysub} (\lambda) = f_\mathrm{obs}(\lambda) - A\cdot S(\lambda).
\end{equation}

We found $A$ by minimizing the RMS of $f_\mathrm{skysub} (\lambda)$ per each slit's exposure given the S($\lambda$) measured in each mask, so that the distribution of signal in an empty region of the slit is centered around $\frac{S}{N} \sim 0$. For this, we used a nonlinear least-squares minimization (LMFIT).\footnote{https://lmfit.github.io/lmfit-py/}

Finally, we merged the telluric-corrected, flux-calibrated, and sky-subtracted exposures into the final science spectra, using sigma clipping to remove cosmic rays. Exposures with seeing above 1.2$^{\prime\prime}$ were excluded. Table~\ref{tab:masks} provides an overview of the exposure times, average seeing and targets for each of our four masks.

\begin{table*}
\caption{\label{tab:masks} Overview of Binospec masks and $z\sim5-7$ targets}
\renewcommand{\arraystretch}{1.45}
\centering
\begin{tabular}{l|ll|l|cc}
\hline\hline
 & Exposure Time & Median Seeing$^\dagger$ & Flux Limit (5$\sigma$) & Targets & \lya detections \\
 & [hr] & & [erg s$^{-1}$ cm$^{-2}$] & & \\
\hline
Mask 1 & 15.5 & 0.90$^{\prime\prime}$ & $1.26\times10^{-17}$  & 93 & 30 \\
Mask 2 & 13.0 & 0.85$^{\prime\prime}$ & $1.26\times10^{-17}$ & 80 & 17 \\
Mask 3 & 13.3 & 0.92$^{\prime\prime}$ & $1.35\times10^{-17}$ & 60 & 12 \\
Mask 4 & 7.8 & 0.97$^{\prime\prime}$  & $1.63\times10^{-17}$ & 50 & 7 \\
\hline
\hline
\end{tabular}
\centering
\tablefoot{After excluding exposures with seeing $>1.2^{\prime\prime}$.}
\end{table*}

\section{Lyman-alpha and rest-frame optical line identification}\label{sec:galaxySelect}

In the following section, we describe our search for \lya emission in our MMT/Binospec spectra (Section~\ref{sec:lya_selection}), and \Ha and/ or \OIII emission in the FRESCO JWST/NIRCam slitless spectra (Section~\ref{sec:lines_fresco}).

\subsection{Lyman-alpha emission line detection}\label{sec:lya_selection}

For the case of spectroscopically confirmed galaxies (either from rest-optical lines or existing \lya spectroscopy), extracting the \lya profile is straightforward since the redshift is already known. For sources without spectroscopic confirmation, we performed a systematic search for the emission line given the photometric redshift. 

We first collapsed the 2D spectra into 1D spectra for each target by extracting within the median spatial extension of our \lya detections (= 1.7$''$ = 7 pixels). We maximized the signal captured in the extraction, with more extended extractions the S/N plateaus adding mostly noise. Starting at the position of the UV continuum, we repeated this procedure for every spatial position of the slit within 2.4$''$ of the UV coordinates to account for any spatial offset between the UV continuum and \lya emission \citep{Hoag2019b,Lemaux2021}. We then scanned through the 1D spectra to search for high S/N peaks, scanning all wavelengths. We flagged line candidates where S/N$\geq$3 is found in three consecutive wavelength pixels; essentially comprising an integrated S/N$\geq5$ detection. Finally, we manually inspected the candidate lines in the 1D and 2D spectra to remove any false positives near the sky lines or from randomly high peaked background. For galaxies where we had a systemic redshift from the rest-frame optical lines, we carried out a manual search of \lya within $\sim$2000\kms of the systemic redshift. For the detected emission lines, we measured \lya fluxes from a fitted asymmetric Gaussian (Section~\ref{sec:lya_fluxlimit}) as this allowed for more accurate measurements in cases of strong skylines near the emission line, rather than obtaining the flux directly from the 1D spectra. 

\subsubsection{Completeness}\label{sec:completeness}
\begin{figure}[]
\centering\includegraphics[width=9cm]{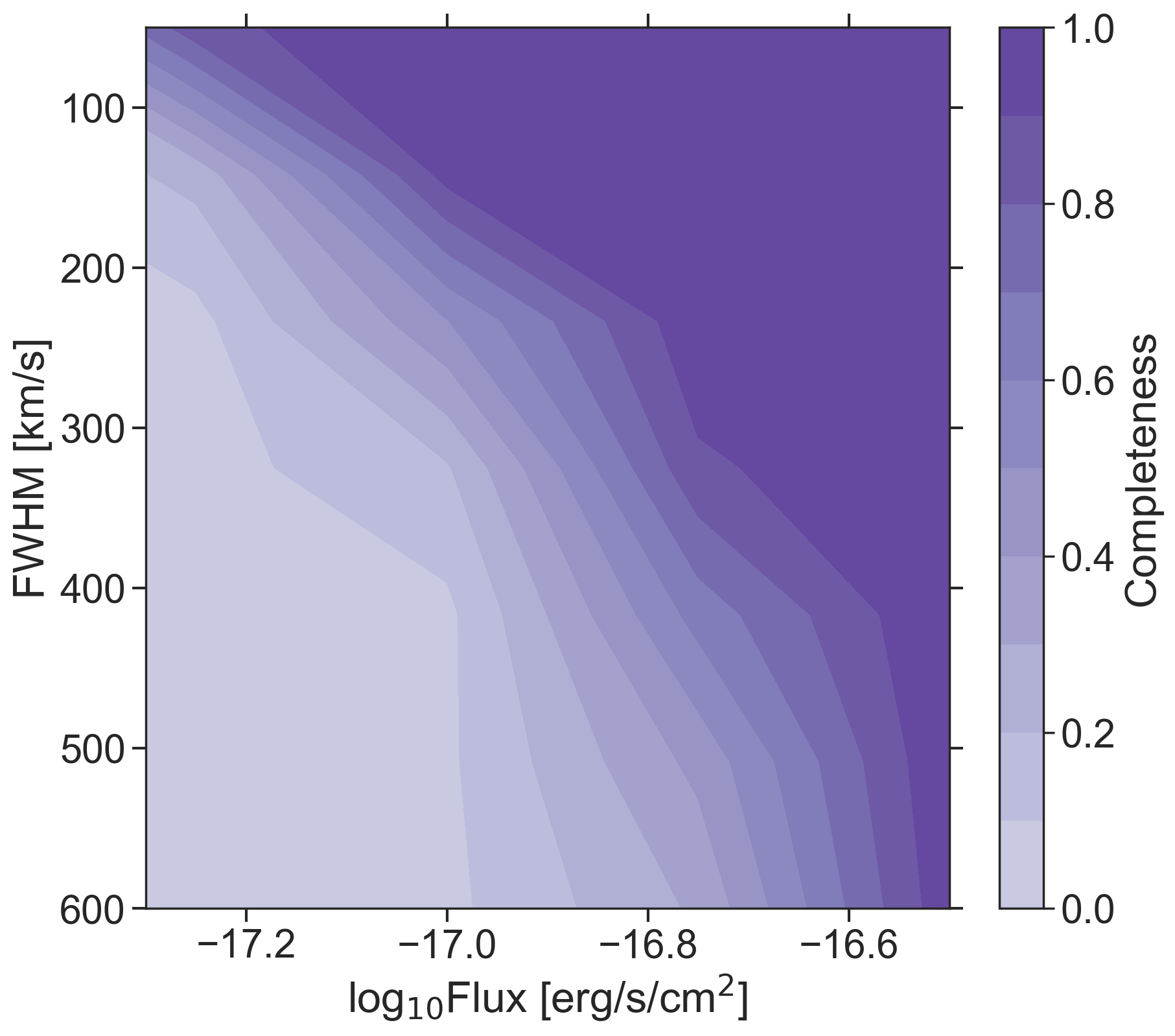}
\caption{
\label{fig:Completeness} Completeness of our \lya search as a function of observed FWHM and flux. The fraction of recovered detections drops with decreasing flux and increased line broadness.}
\end{figure}
To evaluate the completeness of our emission line search, we performed Monte Carlo simulations. We added simulated Gaussian \lya profiles at random wavelengths into the slits of nondetected targets and then attempted to recover them with the method described above. We varied the line flux and FWHM as these are the most important factors in determining the detection of a line, and show the resulting completeness in Figure~\ref{fig:Completeness}. The asymmetry of the profile is sub-dominant to the flux and FWHM, so we used a Gaussian profile for simplicity. The presence of sky lines plays an important role, as a retrieval is generally unlikely if an emission line falls at the same wavelength, leading to high redshift line recovery to become even more difficult due to the increased sky line emission at longer wavelengths. 

Figure~\ref{fig:Completeness} shows that \lya line flux and FWHM play a crucial role in completeness. We achieved $>50\%$ completeness for lines with FWHM=100\kms and flux $\simgt10^{-17.2}$\,erg\,s\,cm$^{-2}$, while lines broader than 500\kms need a flux of $\simgt10^{-16.7}$\,erg\,s\,cm$^{-2}$ to reach the same completeness.

\subsubsection{Slit losses}\label{sec:slit-loss}

As \lya can be spatially extended beyond the width of our extraction window (1.7'')  we corrected our \lya fluxes for slit losses following the approach of \citet{Tang2024}. We first produced a 2D \lya surface brightness profile following for typical z$\sim$5-6 galaxies as measured by MUSE \citep{Leclercq2017}: consisting of two declining exponential functions to describe a \lya core and an extended \lya halo, with 0.3\,kpc and 3.8\,kpc exponential decay constants, respectively. We allocated 35\% and 65\% of the total flux to the core and halo, respectively. We convolved the resulting model with the PSF of each mask, and quantify the flux that would be obtained following the \lya extraction (7pix $\sim$ 1.7$''$). 

For all sources, we estimated 65-70\% of the total \lya flux was captured by our extractions in the Binospec slits. We took this into account for any calculations involving these quantities. We calculated the slit loss directly for individual sources and used the median of 67\% slit loss for our nondetection upper limits. We note that we assumed each galaxy has the same surface brightness profile; therefore, the only difference between galaxies will be the redshift and following angular size conversion.

\subsubsection{\lya fluxes and upper limits}\label{sec:lya_fluxlimit}

For galaxies with \lya detections, we obtained line fluxes by fitting the lines with an asymmetric Gaussian profile. This enabled us to account for the presence of sky lines. Thanks to the $R\sim4000$ resolution of our spectra, we were able to observe detailed \lya line profiles for our detections. We fit a skewed Gaussian model to our \lya lines,
\begin{equation}\label{eq:SkewedGaussian}
    f(x; A, \xi, \omega, \alpha) = \frac{A}{\omega\sqrt{2\pi}}
e^{[{-{(x-\xi)^2}/{{2\omega}^2}}]} \Bigl\{ 1 +
{\operatorname{erf}}\bigl[
\frac{{\alpha}(x-\xi)}{\omega\sqrt{2}}
\bigr] \Bigr\}.
\end{equation}
The location is represented by $\xi$, he skewness is parameterized by $\alpha$, and the scale is $\omega$, which describes the standard deviation with a factor of skewness. For example, a symmetrical Gaussian profile will have $\alpha=0$, whereas for a completely red-sided asymmetric profile, we would have $\alpha \rightarrow +\infty$. During the fitting, we fixed $\alpha \geq 0$ as we expected only profiles that are truncated on the blue side due to the resonant scattering of \lya photons \citep{Laursen2011}. The flux of the line is then obtained by integrating the resulting line profile. To be consistent with the literature, we extracted the FWHM by its definition, by measuring the width of the profile at half its maximum and subtracted the instrumental resolution in quadrature.

We fit the line profiles using an Markov chain Monte Carlo (MCMC) approach with the Python library \textit{emcee}\footnote{https://emcee.readthedocs.io/} \citep{Foreman-Mackey2013}. 
As the skewness, is most sensitive to $\alpha \sim 0-3$, we set a log uniform prior on the skewness between $\alpha \in [0,15]$, allowing perfectly Gaussian to completely asymmetric profiles. For all other parameters of the fit (i.e., amplitude, location, and scale), we used uniform priors with positive values. By setting 30 walkers with 2000 steps, with a burn-in of 750 steps, we obtained converged distributions for our parameters. We used the chains of the MCMC to measure the FWHM posterior distribution. In Appendix \ref{appendix:spectra}, we show some examples of the spectra and the resulting fits.

For the galaxies that did not show a \lya emission feature, we calculated the flux upper limits using the following procedure. We added simulated emission lines to the 2D spectra of our sources, considering a grid of flux, FWHM and redshift. We took uniformly distributed values of flux between $0.1-3.2\times10^{-17}$\,erg/s/cm$^2$ and FWHM values between 100 - 400 km/s, which span the range we find most of our detections (see Section~\ref{sec:lya_galaxy_properties}). To account for the presence of skylines and the photometric redshift, we took uniformly distributed values of redshift within $z_{phot} \pm 0.5 $. We generated fully asymmetric Gaussian profiles and ran the line selection code (Section~\ref{sec:lya_selection}) to detect S/N$>$5 simulated lines.

We simulated 500,000 lines for each galaxy across the above grid of parameters. We calculated the minimum detectable line flux for every sampled FWHM and redshift. We took the median of the resulting flux distribution as the \lya flux limit for the galaxy.

\subsection{Rest-frame optical emission line detection} \label{sec:lines_fresco}

\begin{figure}[]
\begin{center}\includegraphics[width=\columnwidth, trim=0.1cm 0cm 0cm -1cm]{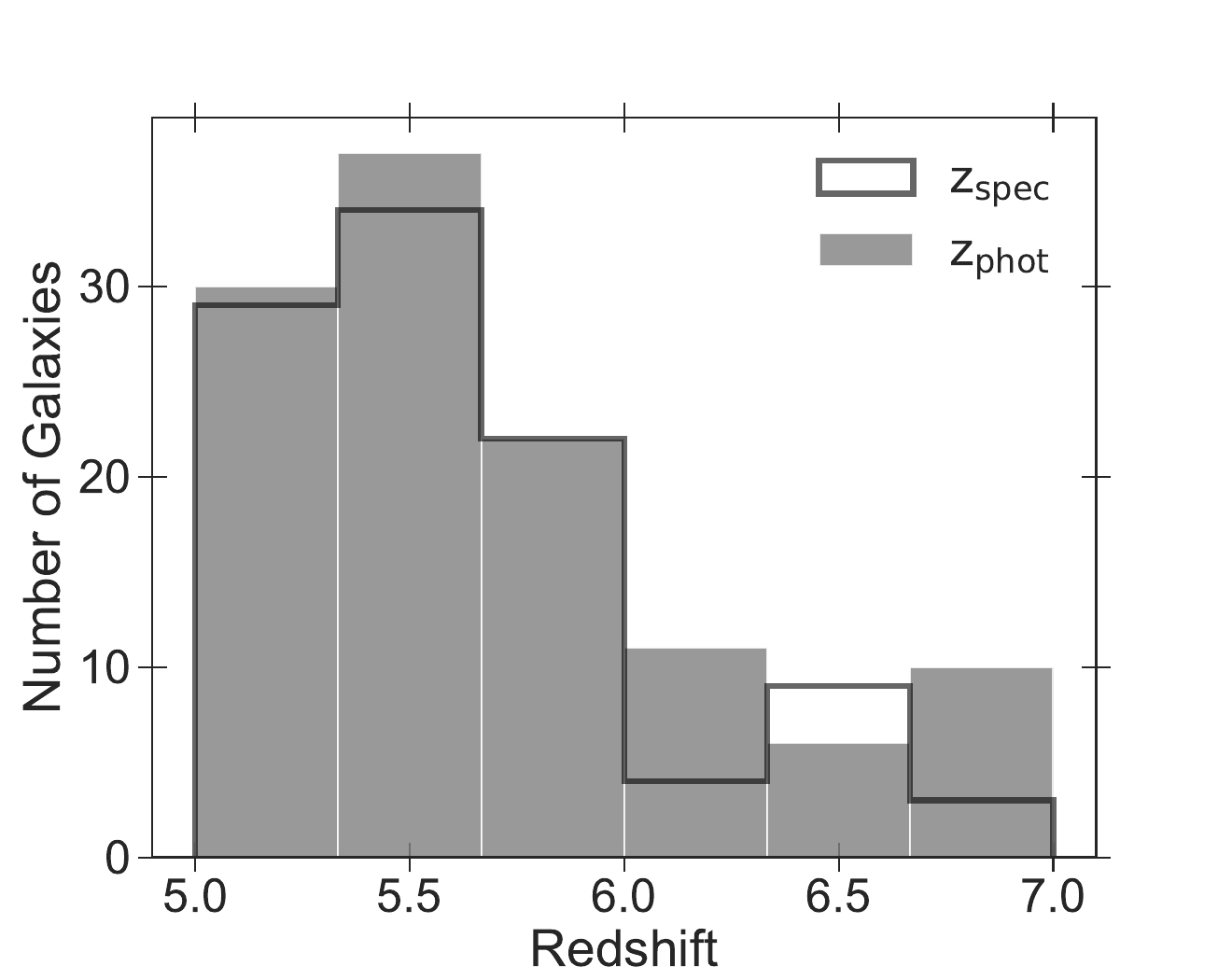}
\includegraphics[width=\columnwidth, trim=0.1cm 0cm 0cm -1cm]{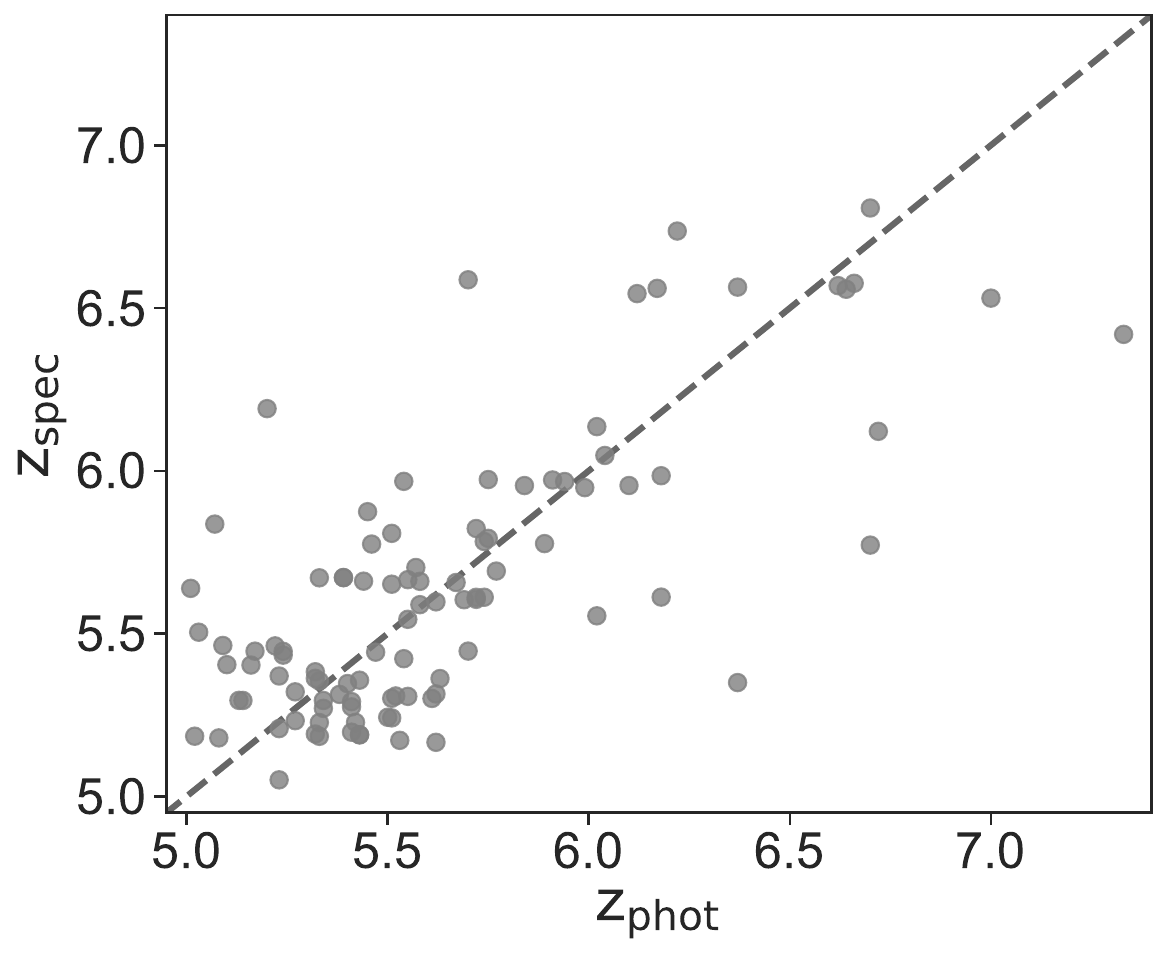}
\end{center}
\caption{Top: Redshift distribution. The final sample consists of 236 galaxies. 105 sources have a spectroscopic redshift, from either \lya and/or \Ha or \OIII (black line). 13 sources have both \lya and \Ha, and 1 source has \lya and \OIII. 131 sources lack spectroscopic confirmation but have $z_{\rm phot}\sim5-7$ and \lya flux limits from Binospec (gray shaded histogram). Bottom: spectroscopic redshift (z$_{\mathrm{\lya}}$ or z$_{\mathrm{sys}}$) and photometric redshift comparison.  }
\label{fig:redshift_distribution}
\end{figure}

82/236 of our Binospec targets fall into the footprint of JWST/NIRCam grism slitless spectroscopy from FRESCO (Section~\ref{sec:fresco}), providing rest-frame optical emission line coverage.
We perform an initial search for emission lines based on the \textit{grizli} extractions \citep{grizli} and refine the search around the \lya redshift for \lya detections or photometric redshift for \lya nondetections. We require a total S/N$>$5 extracted over the median FWHM of the detections (385 \kms) as a detection limit for rest-frame optical lines. Due to the nature of the slitless spectra, contamination from emission lines of overlapping galaxies is present in some of our sources, this effect is mostly taken into account by the extraction as described by \citet{Oesch2023}. In cases where there is a misidentification between the \lya of our target and emission lines of a foreground source, we look for extra emission lines to discard the foreground source, as these do not match with the z$_{\mathrm{spec}}$.

For the 22 galaxies with detected \lya emission which fall in the FRESCO footprint, we do a manual search for S/N$>$5 rest-optical emission lines within $\sim$2000\kms of the \lya\ redshift\footnote{\lya\ velocity offsets are typically $\sim100-500$\kms, with rare instances of offsets up to $1000-2000$\kms \citep[e.g.][]{Erb2014,Steidel2014,Tang2023,Bunker2023,Prieto-Lyon2023b}}. We detect either \Ha or \OIII for 14/22 galaxies. 
For the remaining 60 targets with no \lya detection in Binospec we search over the complete FRESCO wavelength range 3.835 - 5.084 micron for any optical emission lines, focusing mostly within z$_{\mathrm{phot}}\pm\mathrm{0.5}$. We detect \Ha or \OIII for 56/82 galaxies. The \MUV distribution of unconfirmed galaxies is slightly fainter (\MUV = -20.0) than confirmed galaxies (\MUV=-20.2).  Using the systemic redshifts obtained from JWST/FRESCO, we go back to the rest-frame UV data of MMT/Binospec. We replace the photometric redshifts for systemic redshifts and perform a new search for \lya, and find 2 new detections that we failed to recover in our previous searches due as they are on the threshold of our S/N limit ($\sim$5). 
We also discarded three low-S/N ($\sim$ 5) \lya line candidates due high-confidence \Ha detections in FRESCO which made the \lya redshift implausible.
We compared our optical line identification to the catalogs by \citet{Alba2024} using FRESCO \citep{Oesch2023} and CONGRESS \citep{Egami2023} NIRCam grism data.
Using the \citet{Alba2024} FRECSO catalog we confirmed an additional 3 \Ha detections for our sample. 
We did not find any emission lines for our sample in the \citet{Alba2024} CONGRESS catalogs between $3.1-3.8\mathrm{\mu m}$, finding no evidence to suggest lower photometric redshifts (z${\mathrm{phot}}$) for our \lya nondetections, or an incorrect spectroscopic redshift (z${\mathrm{spec}}$) for our low-S/N \lya detections without $>3.8\,\mu$m optical line detections.

In Figure \ref{fig:redshift_distribution}, we show the redshift distribution of the sample. We have a final number of 236 galaxies. We are able to obtain a spectroscopic redshift for 105 galaxies via three approaches: \lya, \Ha and \OIII. Out of the 105 sources, 62 have \lya emission, 50 have \Ha emission and 1 has \OIII. Out of these, 13 have \lya + \Ha and 1 has \lya + \OIII. For 131 targets no emission line was found. We find a median redshift of 5.6 in both the spectroscopic and photometric redshift samples. The median difference $|z_\mathrm{spec}- z_\mathrm{phot}| \approx 0.15$. 

To obtain line fluxes and centroids, we fit a Gaussian line profile to the detected rest-optical emission lines. We show the resulting 1D and 2D spectra for a sub-set of rest-frame optical lines and the fitted models in Appendix \ref{appendix:spectra}. We calculate the flux limits for nondetections with a S/N=5 limit, given the 1D noise array with an extraction the size of the median \Ha FWHM ( = 375\kms) at z$_{\lya}$ or z$_{phot}$, where z$_{\lya}$ is the redshift of \lya at the peak of the fitted skewed Gaussian profile. The observed \Ha FWHM of our galaxies is 2-2.5 times higher than reported with JWST/NIRSpec in \citet{Prieto-Lyon2023b}. These increased line broadening is due to JWST/NIRCam slitless spectra data having the spatial extension of the galaxy convolved with the broadness of the emission lines. Since the spatial extension dominates the observed FWHM of \Ha and \OIII, recovering the intrinsic linewidth is challenging for resolved sources and we leave this aspect to a future work.

\begin{figure*}
\sidecaption
\includegraphics[width=12cm]{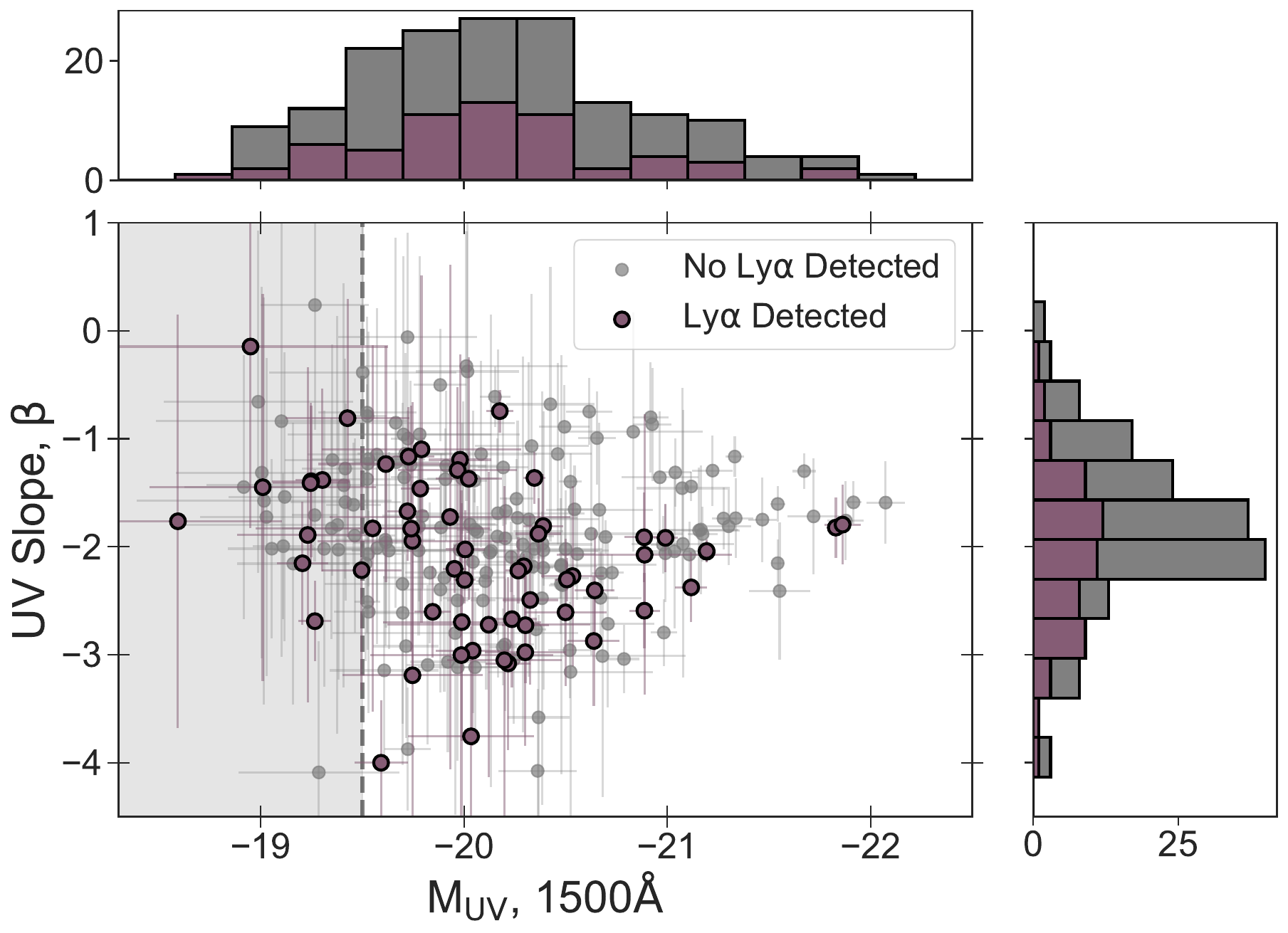}
\caption{\MUV and $\beta$ slope. We show galaxies with \lya detected (purple) and galaxies without detected \lya (gray). These observations are the input into our two empirical models and other results.  On top we show the \MUV distribution, there is no substantial difference between \lya galaxies and \lya non detections, with median M$_{UV}$ of -20.0 and -20.1, respectively. On the right, the UV slope distribution, the two populations show a small difference, with median $\beta$ of -2.0 and -1.8, respectively. The shaded area highlights the range where our sample becomes incomplete due to our target selection (see Section~\ref{sec:discussion_ewlya} for discussion). Extremely blue sources with $\mathrm{\beta<3}$ have large uncertainties due to being constrained by only three photometric data points.\vspace{1.5cm}
}
\label{fig:MUV-slope}
\end{figure*}

\section{Lyman-alpha and galaxy properties}\label{sec:lya_galaxy_properties}

The combination of high-resolution \lya spectroscopy, deep HST and JWST photometry, and rest-frame optical JWST spectroscopy allows us to characterize \lya\ emission in our $z\sim5-6$ sample via measurements of \lya EW, line profiles and velocity offsets, and escape fractions. In the following section we describe these properties in our sample and investigate trends between \lya properties and galaxy properties to better understand \lya transmission in the ISM and CGM in our sample.

\subsection{UV continuum properties}\label{sec:results_uv}

We calculated the UV absolute magnitude (\MUV) and UV slope ($\beta$) for our targets using HST and JWST photometry as described in Section~\ref{sec:data-phot}.
We included only the filters that have their effective width (W$_{eff}$) within the rest-frame UV between 1250-2600\,\AA~ so as to not include bands containing \lya flux. We obtained all filter information from \citet{SVOProfileService}. 
For the redshift, we used z${_\mathrm{sys}}$ when available. If z${_\mathrm{sys}}$ is unavailable but \lya was detected, we used z${_\lya}$. If neither were available, we used z${_\mathrm{phot}}$. 
We note the choice of redshift has negligible effect on our \MUV and $\beta$ results. 

We fit a power law, $f_{\lambda} \propto \lambda^{\beta}$, to the selected photometric filters that capture the UV continuum. We retrieved \MUV by evaluating the fit at 1500 \AA. For sources with JWST NIRCam data, the \MUV and $\beta$ estimates are consistent with the HST estimates, but we obtained lower uncertainties. We fit the continuum using an MCMC method assuming a Gaussian likelihood for the flux densities. 

We plot the resulting \MUV and $\beta$ for our sample in Figure \ref{fig:MUV-slope}, showing the median and 1$\sigma$ uncertainties.
Similarly to previously reported trends in the literature \citep[e.g.][]{Bouwens2012a,Bouwens2014a} we see $\beta$ increase as galaxies are brighter in UV magnitude, but this correlation becomes unclear for fainter sources. We find our F160W selection is incomplete for \MUV>-19.5 sources with blue ($\beta$<-2) UV slopes, which we discuss later in Section~\ref{sec:discussion_ewlya}.

Overall, we see no significant difference in the \MUV distributions between galaxies with and without \lya detection, with median values of $\MUV=-20.0$ and $-20.1$, respectively. We find a slight difference in the $\beta$ slope distributions for galaxies that had a \lya detection and galaxies that do not: galaxies with \lya detection show bluer $\beta$ slopes, with a median value of $\mathrm{\beta}$ = -2.0 , compared to $\mathrm{\beta}$ = -1.8 for galaxies without \lya detections. Performing a KS test shows a p-value>0.1, implying a weak difference between the $\beta$ slope distribution of galaxies with and without detected \lya.

\begin{figure*}[!h]
\sidecaption
\centering\includegraphics[width=12cm]{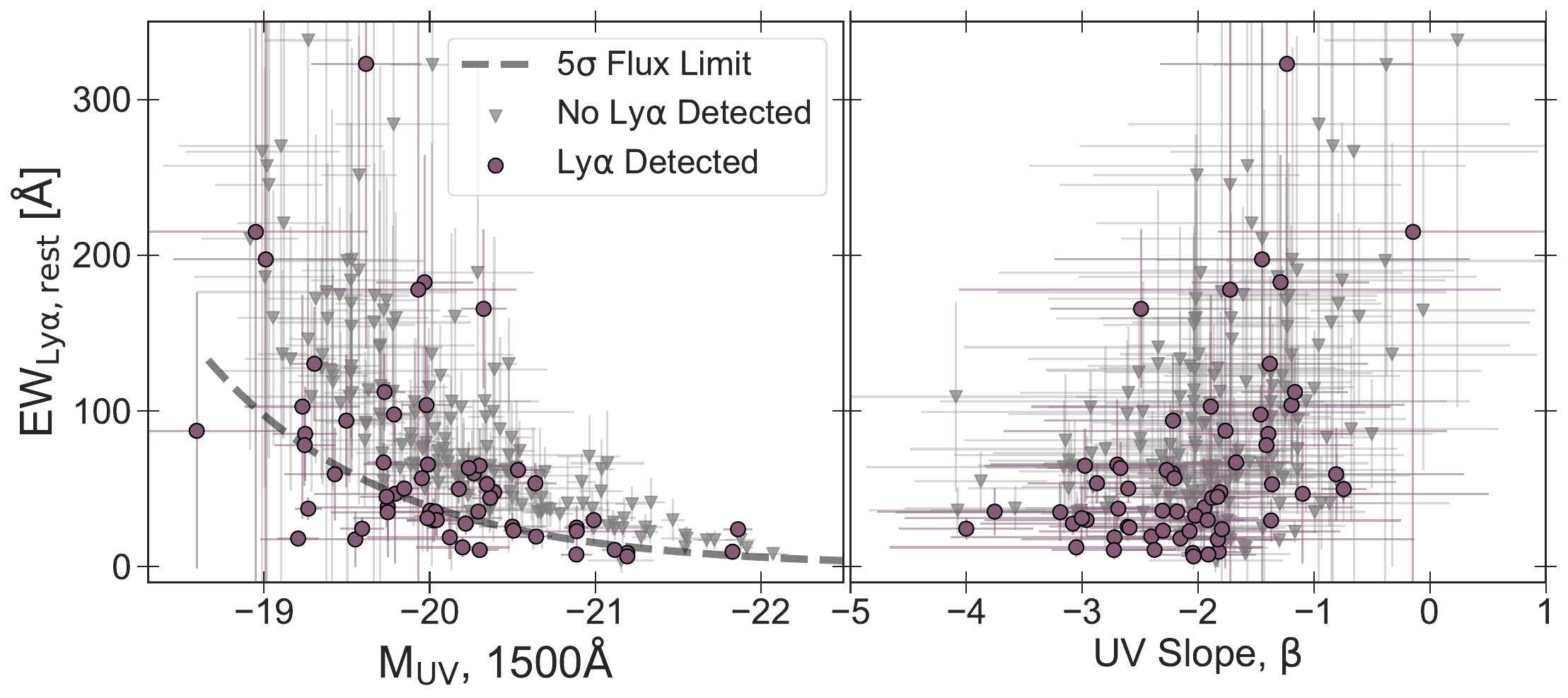}
\caption{Left: \MUV and rest-frame EW$_{\lya}$. Right: $\beta$ slope and rest-frame EW$_{\lya}$. We show galaxies with \lya detection (purple) and galaxies without \lya detection (gray) as upper limits. We find increased \ewlya at fainter \MUV, similar to previous works \citep[e.g.][]{Shapley2003,Stark2011,Oyarzun2017,Tang2024}, and an unclear correlation with UV-slope. We show the median 5$\sigma$ flux limit of the survey in dashed line.\vspace{1.5cm}
\label{fig:EW-MUV-slope}
}
\end{figure*}

\subsection{Lyman-alpha equivalent widths}\label{sec:methods_EW}
To calculate \lya rest-frame equivalent widths (EW), we compared the \lya flux, or upper limit, with the flux density of the UV continuum. We estimated the continuum flux density, $f_{\lambda,UV}$, at the \lya wavelength, 1215.6\,\AA, using our power-law fit to the UV continuum (Section~\ref{sec:results_uv}). The rest-frame EW is
\begin{equation}
    \mathrm{EW_{rest} = \frac{F_{Ly\alpha}}{f_{\lambda,UV} (1+z)}},
\end{equation}
where F$_{\lya}$ is the total flux of the \lya line in erg s$^{-1}$ cm$^{-2}$ (or upper limit) and f$\mathrm{_{\lambda,UV}}$ the flux density [erg s$^{-1}$ cm$^{-2}$ $\AA^{-1}$] of the UV continuum at 1215.6$\AA$. As our sources are faint and the continuum is not detected in the spectra, we do not perform continuum subtraction. For sources with only photometric redshifts, we used z$_{phot}$ as the redshift.

In Figure \ref{fig:EW-MUV-slope} we show our \ewlya measurements against the 2 UV observables. We mainly found enhanced \ewlya in UV faint galaxies, while a trend with UV slope is not evidently present. We also show the average 5$\sigma$ flux limit of our observations. Since the line flux limit is related to \lya broadness, it is possible to observe lines below this limit if they are narrower than the median \fwhmlya of our sample.

\subsection{Lyman-alpha line profiles} \label{sec:results_profiles}

Thanks to the $R\sim4000$ ($\approx 70$\kms) resolution of the Binospec spectra and high S/N of many of our detections, we are able to observe detailed \lya line profile shapes, which offer insights into the scattering of \lya in the ISM and CGM \citep[e.g.,][]{Neufeld1991,Verhamme2006a,Gronke2015a}. At $z\sim0.2-2$, the strongest \lya emitters tend to show narrow \lya emission lines with flux close to the systemic velocity, implying \lya is scattering in low column density neutral gas and thus not diffusing significantly in frequency space \citep[e.g.,][]{Erb2014,Hashimoto2015,Henry2015a,Yang2017c,Matthee2021,Naidu2022a,Tang2024c}. 
At low redshifts ($z=0.05-0.44$), \lya line widths and velocity offsets are seen to correlate with properties, e.g. \MUV, FWHM of rest-frame optical lines, which likely trace gas mass \citep{Hayes2023}.
The sample of published spectra with sufficient S/N and resolution to measure \lya lineshapes at $z\sim5-6$ is small (e.g. MUSE resolution is $\sim$100\kms). Our sample thus provides new insights into the lineprofiles at $z\sim5-6$. We are particularly interested in understanding if \lya lineshapes at $z\sim5-6$ are significantly different from lower-$z$ samples in ways which may impact the transmission of \lya through the IGM.

As discussed in Section~\ref{sec:lya_fluxlimit} we fit our 62 detected \lya lines with an asymmetric Gaussian profile, enabling us to measure the broadness and skewness of the lines, which we discuss in more detail below.
A strong asymmetric \lya lineshape is expected at $z\simgt5$ due to resonant scattering by residual neutral gas in the ionized IGM \citep[e.g.,][]{Laursen2011,Mason2020}. To explore the asymmetry in our data, we choose a binary classification of asymmetric and symmetric. We set profiles with a cut $\alpha> 3$ in our asymmetric Gaussian model (see Equation~\ref{eq:SkewedGaussian}) as asymmetric profiles and sample from our posteriors in our error analysis. We find that 69$\pm$7$\%$ of our \lya sample have asymmetric line profiles, suggesting the IGM is already playing a significant role in shaping \lya emission at $z\sim5-6$, consistent with previous analyses of stacked lines without and with systemic redshifts by \citet{Pentericci2018,Hayes2021} and \citet{Tang2024}, respectively.

The shape of the profile emerging from the ISM/CGM affects the fraction of \lya transmitted through the IGM \citep[e.g.,][]{Dijkstra2007,Dijkstra2011,Mason2018b,Yuan2024}, but can also reveal the impact of the IGM at $z\simgt6$ as the IGM damping wing should produce smooth attenuation as a function of wavelength \citep{Miralda-Escude1998b}, making profiles more symmetric at higher redshifts. 
To test this, we explored the redshift evolution of asymmetry splitting our samples in two bins.
We assumed a threshold $\alpha > 3$ to classify a line as asymmetric (see Appendix \ref{appendix:asymmetry}). Using a different threshold within $\alpha \sim 2-5$ did not alter our conclusions.
We find the asymmetric fraction is 72$\pm$6$\%$ between $z = 5.2 - 5.8$, and 60$\pm$9$\%$ at $z=5.8-6.4$. With the large uncertainties, we cannot draw any conclusions with respect to whether profiles become more symmetric at $z>5.8$, with a stronger impact of the IGM damping wing. Larger samples of \lya line profiles would provide more insight into this effect. 

\begin{figure}
\centering\includegraphics[width=8cm]{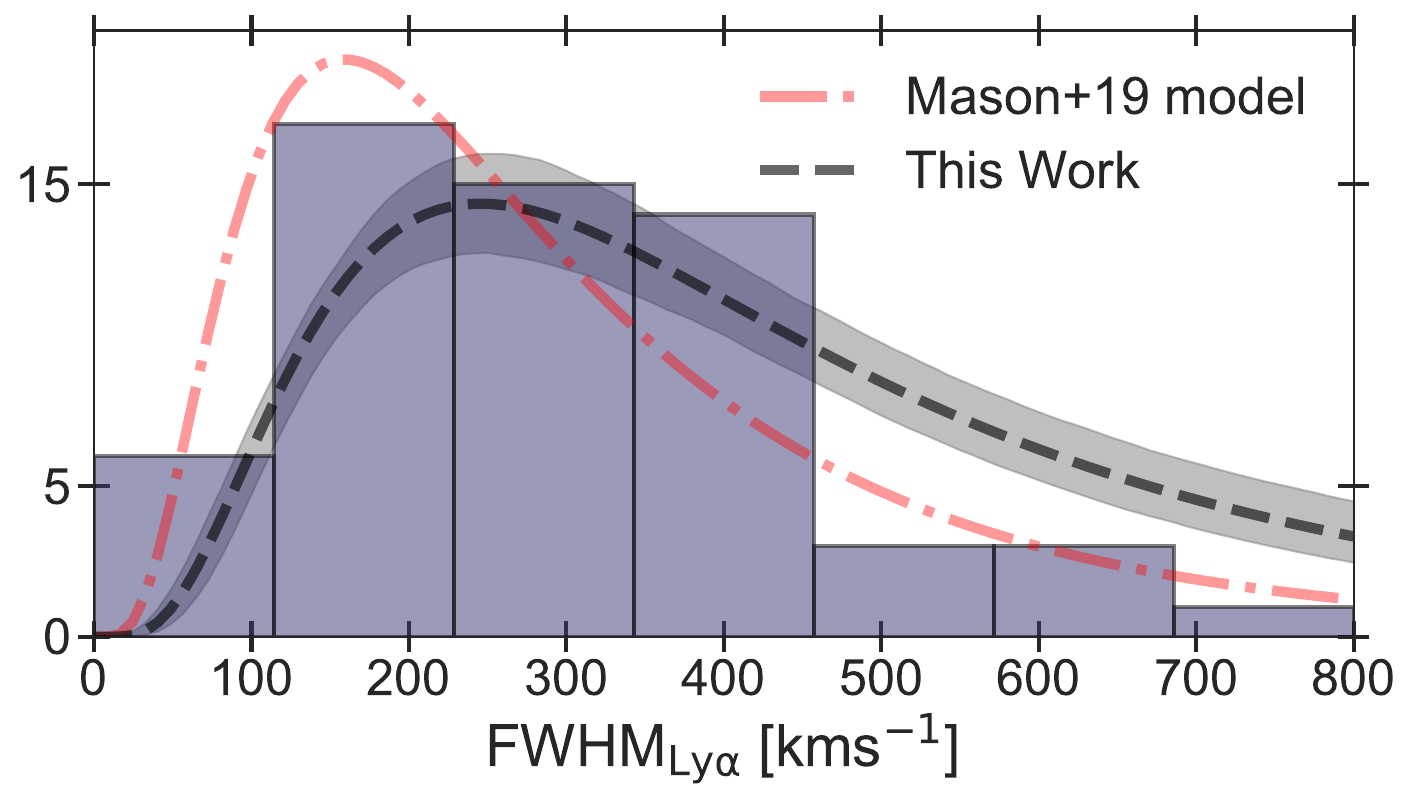}
\caption{
\label{fig:FWHM-MUV-slope} \fwhmlya distribution of our sample. We also plot our fitted log-normal distribution: median as a black dashed line, and the 16th and 84th percentile range in the shaded region. Our log-normal distribution is similar in shape but peaked at higher values than the prediction of \citet{Mason2019c} (red dashed line) evaluated at the same median \MUV and redshift as our sample.}
\end{figure}

In addition to the asymmetry, the distribution of \fwhmlya is a key uncertainty in reionization inferences \citep{Mason2019c}, as it is an important factor in understanding nondetections; namely, we consider whether \lya detection rates lower at $z>6$ because \lya is intrinsically broader and thus more difficult to detect (Section~\ref{sec:completeness}).
We can address this in our sample by inferring the distribution of \fwhmlya at $z\sim5-6$. 
In Figure \ref{fig:FWHM-MUV-slope}, we show the distribution of \fwhmlya our sample.
As the observed distribution looks log-normal, we fit the \fwhmlya sample with a log-normal distribution. The resulting distribution is shown in Figure~\ref{fig:FWHM-MUV-slope}, and is described by
\begin{equation} \label{eqn:FWHM}
\quad P(\log_{10}\fwhmlya)=
\mathcal{N}(\mathrm{ \mu=2.39^{+0.04}_{-0.04},\sigma=0.30^{+0.04}_{-0.03})},  
\end{equation}
where $\mu$ is the mean and $\sigma$ the standard deviation of the normal distribution of log$_{10}$(\fwhmlya). The mean of the log-normal distribution is 245\kms on a linear scale. In Figure~\ref{fig:FWHM-MUV-slope} we show this distribution evaluated at the median \MUV and redshift of our \lya-detected galaxies. For comparison, we plot the model FWHM distribution predicted by \citet{Mason2019c}, based on a model for \lya velocity offsets as a function of \MUV, and an empirical relation between \fwhmlya and \DV derived by \citet{Verhamme2018a} from the \lya peaks redward of systemic. Our inferred distribution peaks at higher \fwhmlya than the \citet{Mason2019c} model, which peaks at 155~\kms. We discuss the implications of this in Section~\ref{sec:disc}.

\begin{figure*}[!h]
\centering\includegraphics[width=0.95\textwidth]{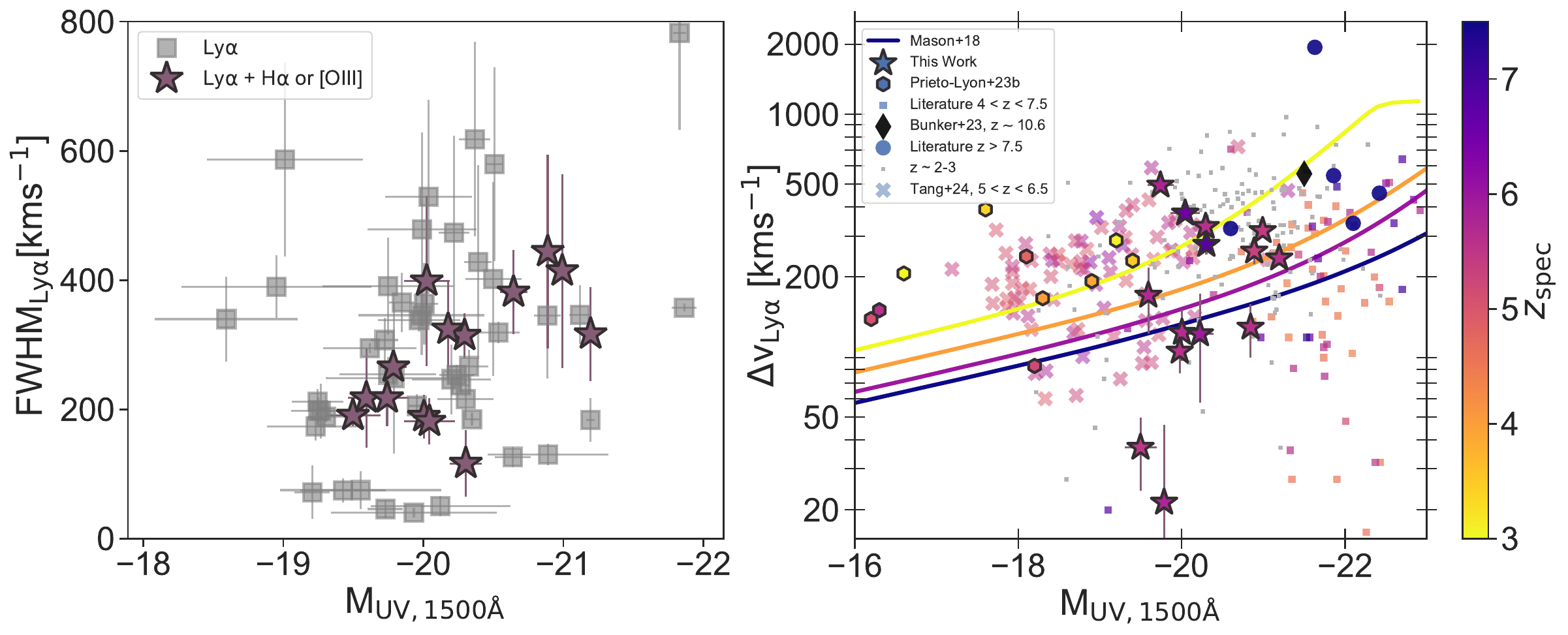}
\caption{
\label{fig:fwhmlya_MUV_FWHMHa} Left: \fwhmlya and \MUV. We show galaxies with only a \lya emission (gray) and with \lya+\Ha emission (purple). We find a mild (p-value $\sim$ 0.1)  correlation between \fwhmlya and \MUV. Right: \lya velocity offset versus UV magnitude, with redshift as a color bar.  We show the data from this work as stars, and the UV-faint (z$\sim$3-5) sample of \citet{Prieto-Lyon2023b} as circles. For reference we show a high redshift sample \citep[i.e.,][z>7.5]{Bunker2023,Tang2023}, mid redshift (i.e. \citep[i.e.,][4<z<7.5]{Bradac2017,Stark2015,Willott2015,Inoue2016,Pentericci2016,Stark2016,Mainali2018a,Cassata2020,Endsley2022,Tang2024}, and low redshift \citep[i.e.,][2<z<3]{Steidel2014,Erb2014}. We include the median semi-empirical model from \citet{Mason2018}. Combining our our data with the literature samples above $z>3$, we find a strong correlation of \DV with increasing \MUV (p-value$<<$0.01).
}
\end{figure*}

To better understand what drives \lya line shapes in our sample, we compared \fwhmlya with galaxy properties. In Figure \ref{fig:fwhmlya_MUV_FWHMHa}, we plot \fwhmlya versus \MUV. Using the Pearson correlation coefficient, we find a mild correlation (p-value $\sim$ 0.1) of increasing \fwhmlya with \MUV, which we would expect if \MUV traces galaxy mass and/or size \citep[e.g.,][]{Shibuya2015,Roper2022,Allen2024,Morishita2024}. Higher galaxy mass and size may indicate enhanced resonant scattering of \lya due to higher N$_\mathrm{HI}$.

\subsection{Lyman-alpha velocity offsets}\label{sec:lya_offset_measure}

For the 14 galaxies in our sample with both \lya and systemic redshifts, we can measure the velocity offset of \lya from systemic, \DV, which provides additional insight into the scattering of \lya in the ISM and CGM \citep{Neufeld1991,Verhamme2008a,Erb2014,Yang2017c}, expressed as
\BE \label{eqn:DV}
\qquad\qquad\qquad\qquad\DV = c \left(\frac{z_{\mathrm{Ly}\alpha} - z_\mathrm{sys}}{1+z_\mathrm{sys}}\right)
\EE
We measured z$_{\mathrm{lya}}$ from the peak of the Skewed Gaussian profile. To measure z$_{\mathrm{sys}}$, we followed the procedure described in Section~\ref{sec:escape_fraction}. The uncertainty in \DV is mostly dominated by the uncertainties of the \lya profile caused by the presence of strong sky-lines, though a few cases of weak \Ha also add to the \DV uncertainty. For most measurements, the error bars remain small but, nonetheless, we are limited by the spectral resolution of both Binospec and the NIRCam grism ($\sim$100\kms).

We show \DV as a function of UV magnitude for our sample in  Figure \ref{fig:fwhmlya_MUV_FWHMHa}. We compare our results to those in the EoR \citep[i.e.,][z>7.5]{Bunker2023,Tang2023,Tang2024b}, mid redshift \citep[i.e.,][4<z<7.5]{Bradac2017,Stark2015,Willott2015,Inoue2016,Pentericci2016,Stark2016,Mainali2018a,Cassata2020,Endsley2022,Prieto-Lyon2023b}, and low redshift \citep[i.e.,][2<z<3]{Steidel2014,Erb2014}. We find a median and standard deviation ($\sigma$) \DV of 258$\pm$144\kms. By excluding \DV<50\kms results, we find a median and $\sigma$, \DV of 315$\pm$125 \kms. Both results are higher than the 205$\pm$75 measured in \MUV>-19 galaxies in \citet{Prieto-Lyon2023b} at $z\sim3-5$. Our results are consistent with similar measurements at $z\sim5-6$ by \citet{Tang2024}, who find a median and $\sigma$,\DV=250$\pm$156\kms for galaxies in the same \MUV range as our sample (\MUV$<-19.5$). Our new data span a relatively narrow \MUV range, but combining with other samples at $z>3$ referenced in the previous paragraph, we see a strong correlation of increasing \DV for higher UV luminosities (p-value<<0.01). We also compare with the semi-empirical model from \citet{Mason2018}. We see most of the $z>5$ data lie above the model, which was derived from a sample of $\MUV<-19$ and z$\sim$2 galaxies \citep{Steidel2014,Erb2014}. 

Our sample includes two galaxies with extremely low \DV<50\kms, z5-GND-17752 ($z=5.77$) and z5-GND-39445 ($z=5.50$), implying only minimal scattering in the ISM/CGM. We find that z5-GND-17752, our lowest \DV (20\kms) measurement, also shows low \fwhmha (120\kms). Looking at their \lya line profiles (Appendix \ref{appendix:spectra}) we find significant emission bluewards of the \lya peak, implying a highly ionized and/or low density sightline through the IGM enabling \lya transmission around systemic velocity \citep{Mason2020}. However, such low \DV is at the limit of the spectral resolution, making them susceptible to systematic errors. We also note our high resolution reveals several sources have low \DV but show an asymmetric profiles with an extended tail to higher velocities ($>300$\,km/s). Such profiles are hard to explain with a simple shell model, but are often well-described by radiative transfer through a clumpy or anisotropic medium (e.g., \citealt{Neufeld1991,Hansen2006,Gronke2017b,Chang2023}). We further discuss implications of our \DV results on \lya transmission in Section~\ref{sec:disc}.

\subsection{Lyman-alpha escape fraction}\label{sec:escape_fraction}

\lya and \Ha are both emitted predominantly in photoionized nebulae in a series of hydrogen line transitions, or recombination cascades, therefore linking the production of both types of photons. The ratio of products of the recombination cascade depends on the electron temperature (T$_{\mathrm{e}}$) and electron number density (n$\mathrm{_e}$), but most importantly the optical depth of the medium \citep{Dijkstra2014}. In this work, we assume a Case-B recombination scenario, where the surrounding medium is optical thick to the Lyman series, as expected of HII regions, with n$_\mathrm{e}$ = 250cm$\mathrm{^{-3}}$, T$\mathrm{_e}$ = 10$\mathrm{^4}$ \citep{Dijkstra2014}. This yields an intrinsic \lya flux of $8.7 \times$ the dust-corrected \Ha flux.

The \lya escape fraction is the ratio of the observed to intrinsic \lya flux,
\begin{equation}\label{eq:fesc}
\quad\quad\quad\quad\quad\quad  f_{\mathrm{esc,ly\alpha}} = \frac{F_\lya}{8.7 F_{\text{dust corrected,}\Ha}}.
\end{equation}
Here, both terms refer to the total flux of the emission lines (observed in the case of \lya, dust-corrected in the case of \Ha), and the 8.7 is the factor derived from case-B recombination. Recovering $\fesc<1$ implies not all \lya escapes from the galaxy, due to dust absorption or scattering by high column densities of neutral hydrogen, effectively removing \lya from the line of sight. 

We corrected all \Ha fluxes for dust attenuation following \citet{Lam2019}. We assume a Small Magellanic Cloud (SMC) dust curve \citep{Prevot1984}, which is expected to be similar to that in $z\sim4-6$ galaxies based on the infrared excess (IRx) - $\beta$ slope relationship \citep{Bouwens2016b}. The dust correction increases \Ha fluxes by $\sim 1-10 \%$. We propagated the uncertainty of $\beta$ into the dust correction, although we did find the error bars of \fesc are dominated by the uncertainty of the line fluxes in all cases. As a point of reference for our \fesc results, we calculated the expected \lya escape fraction assuming the SMC dust curve \citep{Gordon2003}, where $\tau_{\rm dust, Ly\alpha}/ \tau_{\rm dust, H\alpha} \sim 8$. Given a modest $\tau_{\rm dust, H\alpha} = 0.2$, we find \fesc $\sim 0.2$.

\begin{figure*}[!h]
\sidecaption
\centering\includegraphics[width=12cm]{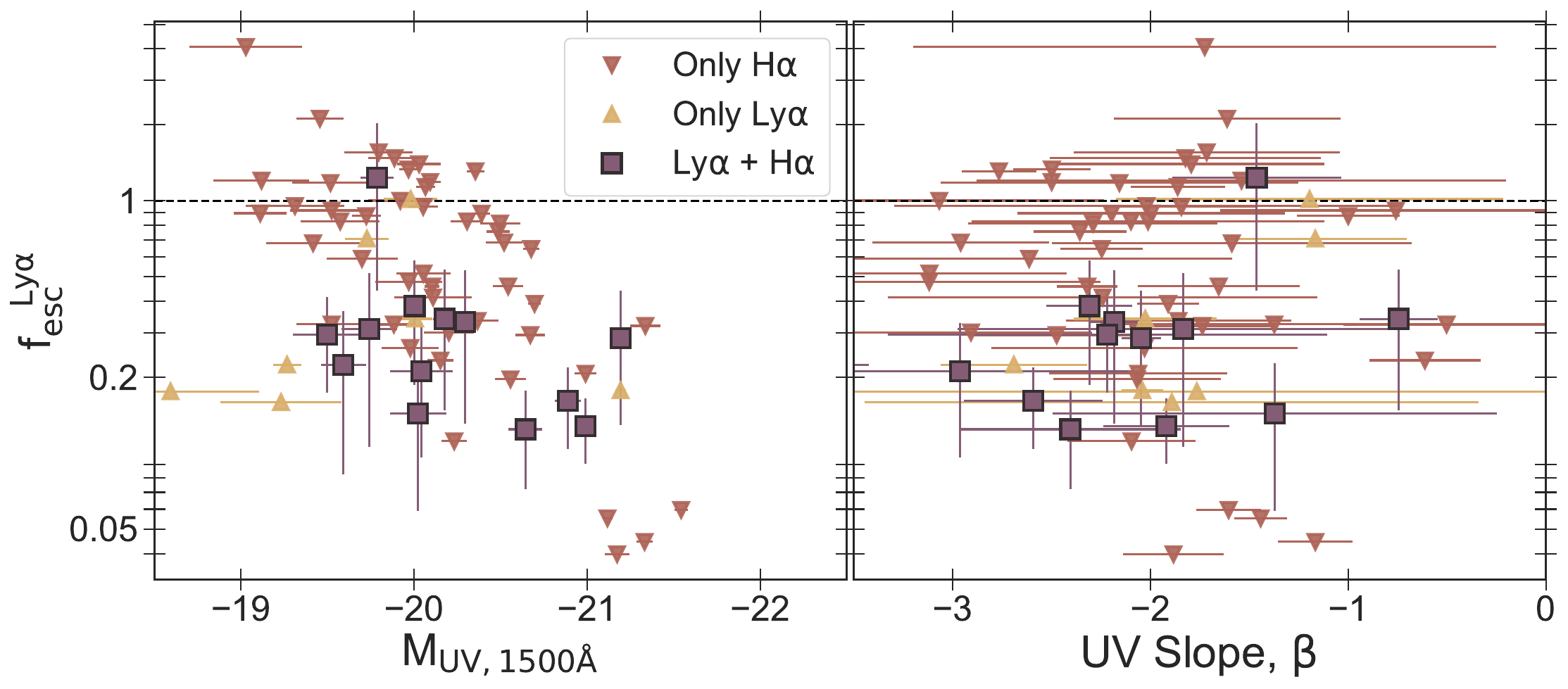}
\caption{ Left: \MUV and \fesc. Right : $\beta$ and \fesc. The sample is divided into three types of measurements; galaxies with \lya and \Ha detections (purple), galaxies with only \Ha detection as upper limits (red), and galaxies with only \lya detection as lower limits (yellow). We find a strong trend (p-value<0.01) where fainter galaxies have increased \fesc. For UV-slope, as in Figure \ref{fig:EW-MUV-slope}, we do not find any clear trends. \vspace{1.5cm}
\label{fig:fesc_MUV_beta}
}
\end{figure*}

\begin{figure*}[!h]
\sidecaption
\centering\includegraphics[width=12cm]{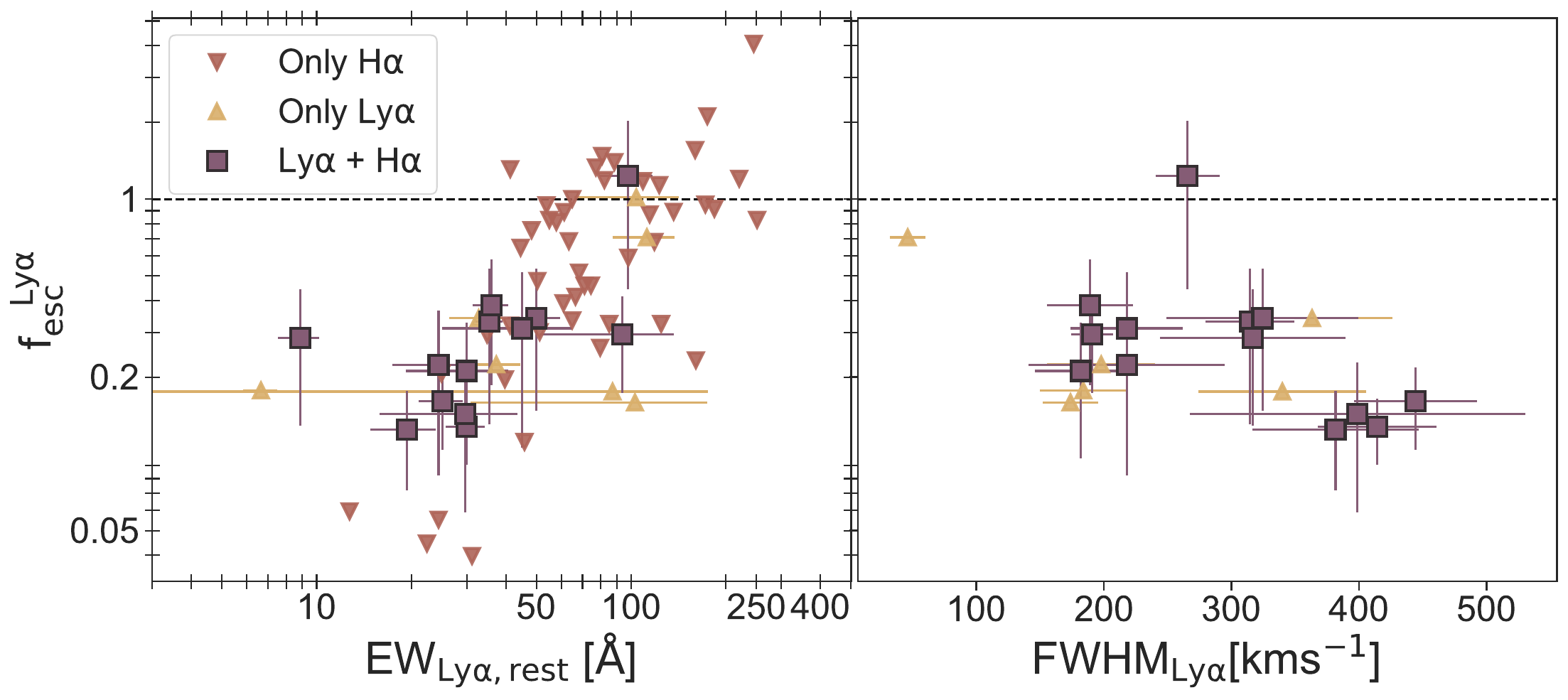}
\caption{Escape fraction of \lya against other \lya properties. Left: \fesc and \ewlya. Right: \fesc and \fwhmlya. The data shows an increase in \ewlya with higher \fesc, and an interesting population of weak \lya with relatively high leakage (\fesc>0.2). We find a mild correlation (p-value<0.05) of higher \fesc with low \fwhmlya.\vspace{1.5cm} 
\label{fig:fesc_lya}
}
\end{figure*}

In Figure \ref{fig:fesc_MUV_beta}, we show the resulting \lya escape fractions as a function of \MUV and $\beta$. We have three classes of measurements, galaxies with \Ha emission only, galaxies with \lya emission only, and galaxies with both \lya and \Ha. These classes produce upper limits, lower limits and measurements of \fesc, respectively. For galaxies with only \Ha, we use the \lya flux limit as presented in Section~\ref{sec:lya_fluxlimit} to obtain an upper limit on \fesc. When only \lya is present, we calculate \Ha upper limits as the 5$\sigma$ flux limit at the expected location of \Ha, assuming z$_{\lya}$, integrated in a width randomly drawn from our FWHM$_\Ha$ distribution. We use the median flux limit of 10,000 realizations for this method. Uncertainties in \fesc resulting from unknown \lya velocity offsets are negligible. Our results in Figure \ref{fig:fesc_MUV_beta} are dominated by upper limits and lower limits, to measure any correlations, we use a Pearson correlation coefficient, and do random draws from a uniform distribution between [0,upper-limit] and [lower-limit,1]. With 1000 iterations we find a strong increasing correlation between \fesc and \MUV (p-value<0.01), while \fesc and UV slope show no trend. We also find that \fesc$> 0.2$ in 13/20 of our measurements and lower limits, indicating \lya escape with low dust attenuation. \\

In Figure \ref{fig:fesc_lya}, we compare \fesc with \ewlya, and \fwhmlya, using the same classification explained in the previous paragraph. Applying the random sampling method outlined in the previous paragraph, we identify mild correlations and anti-correlations (p-value < 0.05): \fesc increases with higher \ewlya but decreases with larger \fwhmlya. Beyond the positive correlation between \fesc and \ewlya, we find a small population of sources with weak \ewlya<10\AA$ $ and \fesc>0.1, indicating a possible ISM configuration that allows weak \lya to still escape from the galaxy. For \lya broadness, we find a mild correlation of high \fesc in sources with narrow \lya profiles, including two of the narrowest profiles in our sample, which exhibit \fesc$\sim$1. Spatially resolved \lya studies (e.g. \citealt{Leclercq2020}) have shown that \fwhmlya can vary between the core and halo of a galaxy, with the \lya profile in the extended halo typically being broader. As a result, slit spectroscopy could miss some of the extended halo emission, potentially leading to narrower \fwhmlya values (see Section~\ref{sec:slit-loss}), however this does not significantly impact our conclusion. We further discuss the implications of these correlations and anti-correlations in Section~\ref{sec:disc}.\\

One important caveat is that our dataset is inherently biased towards detecting sources with high \lya luminosity. As a result, our sample exhibits a Malmquist bias towards the UV-faint regime, where detecting weaker \lya becomes increasingly challenging (see Fig. \ref{fig:EW-MUV-slope} and Fig. \ref{fig:fesc_MUV_beta}). This bias may enhance the strength of the apparent correlations between \ewlya and \fesc with \MUV, but we note that we account for upper limits in our empirical models in the following section.

\section{An empirical model for emergent Lyman-alpha}\label{sec:MakingModel}

Next, we sought to build an empirical model to predict emergent \lya properties based on easily observable galaxy properties, \MUV and UV slope $\beta$. In the following section we describe the approach we used to produce model distributions for both \lya EW and \fesc.\\ 

To estimate the \lya EW and \fesc distributions, we used a Bayesian approach \citep[e.g.,][]{Treu2012,Schenker2014,Oyarzun2017,Mason2018,Tang2024} which allowed us to obtain the posterior distribution of the parameters that describe the \ewlya and \fesc distributions.  
We followed \citet{Mason2018} and assumed the distributions for EW and \fesc are described by an exponential with a potential peak at zero for nonemitters, with two parameters $\theta = (A,X_0)$. We note that previous works have shown the exact form of the distributions (exponential, log-normal, etc.) does not significantly impact results \citep{Schenker2014,Oyarzun2017,Tang2024}. In the following, we use $X$ to denote the `data' EW or \fesc as we model both distributions in the same way. The model distribution is
\begin{equation}\label{eq:Likelihood}
p(X | \theta=\{A,X_0\}) =  B\frac{A}{X_0}e^{-X/X_0} \mathrm{H}(X) + (1-A)\delta(X).
\end{equation}
Here, the parameter $X_0$ represents the scale length of the exponential and $A$ is a normalization term which represents the fraction of \lya emitters (i.e., the fraction of galaxies which have EW or \fesc $>0$). Then, $B$ is a normalization term for the distribution given the limits of EW and \fesc ($B=1$ for EW, and $B=1/(1-\exp(-1/X_0))$ for \fesc). $H(X)$ is the Heaviside function and $\delta(X)$ is the Dirac-delta function.\\
We assumed Gaussian measurement uncertainties with standard deviation $\sigma$ on our data $X_\mathrm{obs}$ for each galaxy, such that
\begin{equation}\label{eq:integralLikelihood}
p(X_{\mathrm{obs}} | X ) = \frac{1}{\sqrt{2\pi}\sigma} \exp\left[{-\frac{(X - X_{\mathrm{obs}})^2}{2\sigma^2}}\right].
\end{equation}
Thus the likelihood for an individual observation is
\begin{equation}\label{eq:integralLikelihood}
\quad\quad\quad p(X_{\mathrm{obs}} | \theta ) = \int dX \, p(X_{\mathrm{obs}} | X) p(X | \theta)
\end{equation}
Where the limits of integration are [$0,\infty$] for \ewlya and [0,1] for \fesc.

Convolving our likelihood with the Gaussian distribution and solving the integral, we obtain the likelihood for our observations,
\begin{equation} \label{eq:ConvolvedLikelihood}
 \begin{split}
 p(X_{\mathrm{obs}} | \theta ) & =  B\frac{A}{2X_0} e^{\frac{\sigma^2 - 2 X_0 X_{\mathrm{obs}}}{2X_0^2}}  \left[\mathrm{erf}\left( Z+\frac{X_\mathrm{max}}{\sqrt{2}\sigma}\right) - \mathrm{erf}(Z)\right] \\
    & + \frac{(1-A)}{\sqrt{2\pi}\sigma}e^{-\frac{X_{\mathrm{obs}}^2}{2\sigma^2}}.\\ 
\mathrm{where} \; Z   &  = \frac{1}{\sqrt{2}\sigma}\left( \frac{\sigma^2}{X_0} - X_{\mathrm{obs}} \right).
 \end{split}
 \end{equation}
 In addition, \textit{erf} is the error function and $X_\mathrm{max}=\infty$ for EW and 1 for \fesc.\\
 
For galaxies with \lya nondetections (i.e. upper limits in \ewlya and \fesc) the likelihood is
\begin{equation} \label{eq:NonDetect}
p(X_{i,\mathrm{obs}} | \theta)  = p(X_{i,\mathrm{obs}} < X_{\mathrm{uplim}} | \theta) + p(X_{i,\mathrm{obs}} > X_{\mathrm{uplim}} | \theta) \cdot (1-C), 
\end{equation}
where the cumulative distribution function is obtained by integrating $p(X_{i,\mathrm{obs}} | \theta )$ (Equation~\ref{eq:integralLikelihood}) from $x_{i,\mathrm{obs}}=-\infty$ to $x_{\mathrm{uplim}}$.We incorporate the second term using the methods from \citet{Schenker2014}, accounting for the probability of an emission line above the 5$\sigma$ flux limit being undetected due to a skyline. The completeness ($C$) is calculated for individual slits over $\Delta \mathrm{z_{phot}}$ = 0.5. If \Ha or \OIII are available, we determine completeness for z$_{sys}$ within a range of typical \DV from 0 to 500 \kms.\\

For galaxies with no \Ha detections, we only had a lower limit on \fesc; thus, in those cases, the likelihood is
\begin{equation} \label{eq:NonDetect-lower}
p(X_{i,\mathrm{obs}} | \theta)  = p(X_{i,\mathrm{obs}} > X_{\mathrm{lowlim}} | \theta), 
\end{equation}
where we integrate $p(X_{i,\mathrm{obs}} | \theta )$ from $X_{\mathrm{lowlim}}$ to $1$.\\

The final posterior is the product of all individual galaxy posteriors, $p(\theta | \{X_\mathrm{obs}\} \propto p(\theta) \prod_i p(X_{i,\mathrm{obs}} | \theta)$. We assume flat priors on A and X$_\mathrm{0}$: $A\in[0,1]$, $W_{0}\in[0,500]$ and $f^{\mathrm{Ly}\alpha}_{\mathrm{esc,0}}\in[0,1]$.

\subsection{Dependence of the distributions on galaxy properties}

To better understand the dependence of \lya emission on galaxy properties, we parameterized our model as a function of \MUV and $\beta$. Previous works, similarly to our results in Section~\ref{sec:lya_galaxy_properties}, have demonstrated that \lya EW and escape may be enhanced in UV-faint, bluer, galaxies \citep[e.g.,][]{Tang2024,Oyarzun2017}.

We followed the approach done by \citet{Oyarzun2017} and parameterized our model parameters $A$ and X$_\mathrm{0}$ as a linear combination of \MUV and $\beta$,
\begin{equation} \label{eq:Params}
\begin{split}
\qquad\qquad\qquad A & = A_{\MUV}\MUV + A_{\beta}\beta + A_c;\\
\qquad\qquad\qquad X_0 & = X_{\MUV}\MUV + X_{\beta}\beta + X_c.
\end{split}
\end{equation}
To compute the posterior distributions, we followed the same method as before, assuming the same priors on $A$ and $X_0$. 

We performed tests on mock datasets to assess whether our sample size is sufficient to robustly recover these parameters. We find that our \lya EW sample (236 galaxies, including 62 \lya detections) is large enough that we could expect to recover the parameters in Equation~\ref{eq:Params} within 5-30\% of the true parameters. We find our \fesc sample (72 galaxies, including 10 \lya+\Ha detections) is not large enough to fit the linear model robustly, so we assume constant $A$ and $X_0$, and find we can recover these robustly when splitting our sample into two bins. Both $A$ and $X_0$ are recovered with similar accuracy and precision in our tests.

\subsection{Lyman-alpha equivalent width model}\label{sec:EW_results}

\begin{figure}[]
\centering\includegraphics[width=9cm]{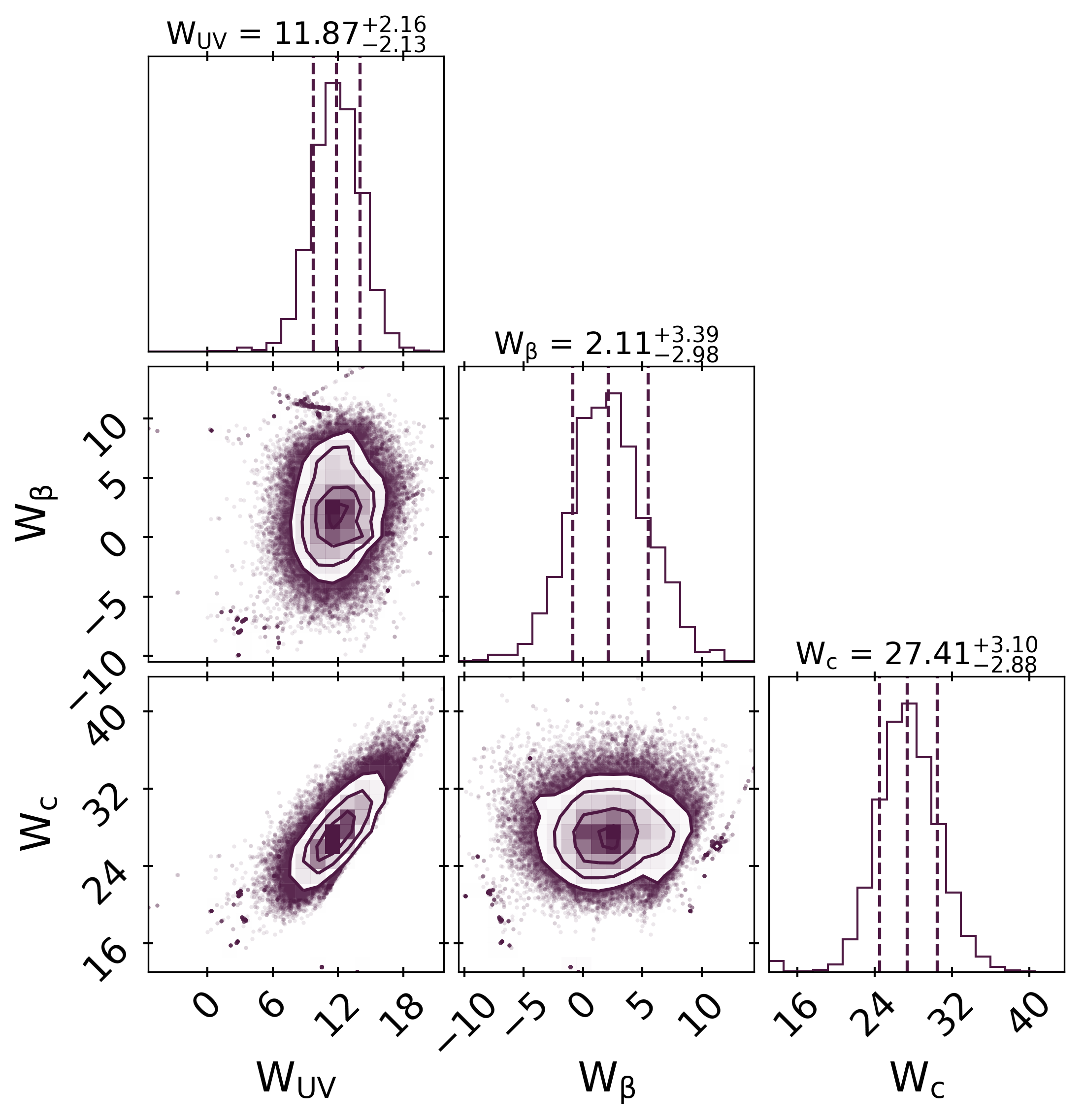}
\caption{
\label{fig:EW-corner} Posteriors for the \lya EW distribution model (Section~\ref{sec:EW_results}). The posteriors show the coefficients for the linear model for EW scale length, W$_\mathrm{0}$ as a function of \MUV ($W_\mathrm{UV}$), UV-slope ($W_\beta$), and a constant ($W_\mathrm{c}$). We see a strong correlation between W$_{\mathrm{UV}}$ and W$_{\mathrm{c}}$, and that there is no significant dependence on UV slope ($W_\beta \approx 0$).}
\end{figure}

\begin{figure}[]
\centering\includegraphics[width=9cm]{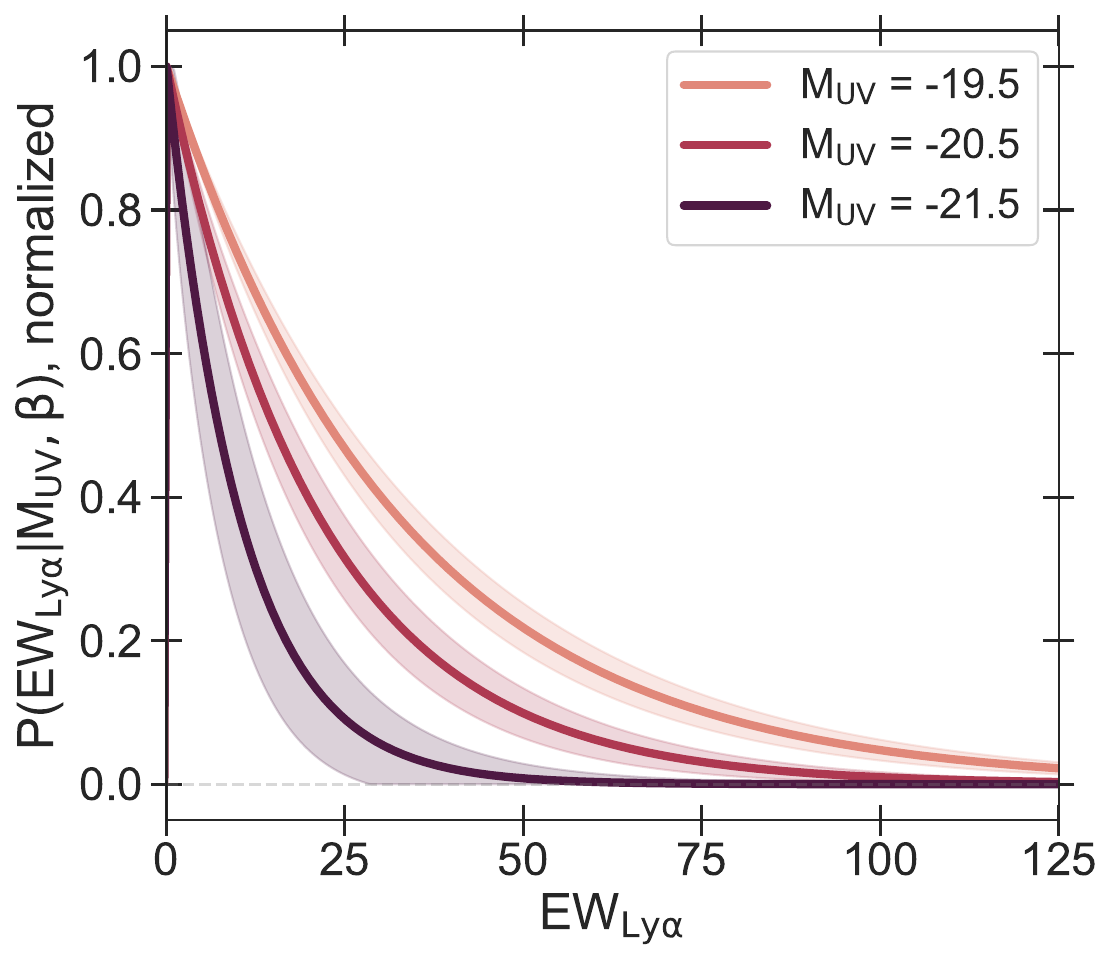}
\caption{
\label{fig:PEW} Probability distribution of \ewlya. We plot 3 distributions for different \MUV within our data range, showing UV faint galaxies are more likely to be stronger \lya emitters.}
\end{figure}

As described above, we fit for the \lya EW distribution assuming an exponential model with parameters described by a linear model in \MUV and $\beta$ (Equation~\ref{eq:Params}).
We used the measured EW, uncertainties and 5$\sigma$ upper limits, as described in Section~\ref{sec:methods_EW}, in our likelihood (Equation~\ref{eq:ConvolvedLikelihood}), accounting for uncertainties in \MUV and $\beta$ for each galaxy by sampling from their posteriors (Section~\ref{sec:results_uv}).
We performed the fitting using an MCMC with \texttt{emcee} \citep{Foreman-Mackey2013} with 10,000 steps and 25 walkers, enough to reach convergence. We find that our model clearly returns $A\approx1$; namely, a purely exponential distribution with no peak at EW$=0$\,\AA. Thus, we fit only for the scale-length $X_0 \equiv EW_0$.\\

We show the resulting corner plots for the $EW_0$ parameters in Figure~\ref{fig:EW-corner}. We find that our model can be completely described by \MUV, with no significant dependence on the UV slope -- as the linear coefficient, $W_\beta$, is consistent with zero within 1$\sigma$. This lack of correlation with the UV slope is likely due to our source selection, as discussed further in Section~\ref{sec:discussion_ewlya}. Consequently, we opt to re-run the model excluding the UV slope. The inferred scale-length is:
\begin{equation} \label{eq:Model_results}
\begin{split}
 \qquad\qquad EW_0 & = 11.2^{+2.0}_{-2.0}\cdot(\MUV+20)+ 27.2^{+3.1}_{-2.8},
\end{split}
\end{equation}
where we give the median and 68\% credibile intervals for each parameter given their posterior distributions.

In Figure \ref{fig:PEW}, we visualize the normalized \ewlya distribution. We show the resulting distribution for 3 examples with \MUV= -19.5, -20.5, -21.5. We plot the median and 68\% credible interval of the distributions obtained by sampling from the posterior predictive distributions for the $EW_0$ parameters. We see that faint galaxies have a broader EW distribution than bright galaxies, and therefore much higher likelihood of having a strong \ewlya as discussed in the next paragraph.

Using our EW distribution model we can also calculate the `\lya fraction': the fraction of Lyman-break galaxies with EW$>25$\,\AA$ $ \citep[e.g.,][]{Stark2010}. We obtain this by integrating Equation \ref{eq:Likelihood} given our inferred parameters in Equation \ref{eq:Model_results}. For the three cases shown in Figure~\ref{fig:PEW} the \lya fraction ranges from 8 to 45\% ($\pm$5\%) for $\MUV=-21.5$ to \MUV = -19.5.
Assuming the most commonly used UV magnitude range in the literature ($-18.75 < \MUV < -20.25$, i.e. median $\MUV=-19.5$) we find a \lya fraction of 45\%$\pm$5\%, consistent with recent findings by \citet{Tang2024} of 35$\pm$7$\%$. Based on our distributions, we expect 21$\pm$3$\%$ of \MUV = -19.5 galaxies to show \ewlya>50\AA, fully consistent with \citet{Tang2024} who found $22\pm3$\%, and 5$\pm$2$\%$ to show  \ewlya>100\AA, $\approx2\sigma$ lower than \citet{Tang2024}. Our sample is $\sim3\times$ smaller than that of \citet{Tang2024}, and less complete at $\MUV \sim -19.5$, so it is likely that we are missing some of the high EW tail of the distribution, which are likely more common towards the fainter end of the \MUV range.
Overall, we should expect a non-negligible fraction of very strong emitters as we move deep into the EoR, where $\MUV > -19.5$ galaxies become more common. 

We also attempted to calculate tighter upper limits for \lya nondetections where \Ha is detected in FRESCO. We estimate the \lya upper limit from the dust-corrected \Ha flux, adopting a Case B recombination scenario and assuming 100\% \lya escape fraction (See Section~\ref{sec:escape_fraction}). Using this approach, we find a median reduction of 27\% in the \lya upper limits for 10\% of the total \lya nondetection sample. We re-ran our analysis with these upper limits but find no significant changes in the resulting model posteriors.

\subsection{Lyman-alpha escape fraction model}\label{sec:fesc_results}

\begin{table}
\caption{\label{tab:fesc_results} Exponential scale length ($\fesc_{0}$) and delta (A) parameter results}
\renewcommand{\arraystretch}{1.45}
\centering
\begin{tabular}{l|rr}
\hline\hline
Sub-sample & $\fesc_0$ & $A$ \\
\hline
All             &  $0.27^{+0.16}_{-0.08}$ & $>$ 0.99*  \\ 
\MUV > -20.1      &  $0.52^{+0.23}_{-0.22}$ & $>$ 0.98*  \\
\MUV < -20.1      &  $0.20^{+0.09}_{-0.05}$ & $>$ 0.98*  \\
$\beta$ > -2.0  &  $0.40^{+0.27}_{-0.17}$ & $>$ 0.97*  \\
$\beta$ < -2.0  &  $0.38^{+0.13}_{-0.08}$ &  $>$ 0.99*  \\
\hline\hline
\end{tabular}
\tablefoot{ We state the median, along with the 16th and 84th percentile values from the posterior for each parameter. *For the lower limits, we give the 68\% limit. We note that due to the large number of upper limits, the recovered $\fesc_0$ should be viewed as upper limits.}
\end{table}

\begin{figure}[]
\centering\includegraphics[width=8cm]{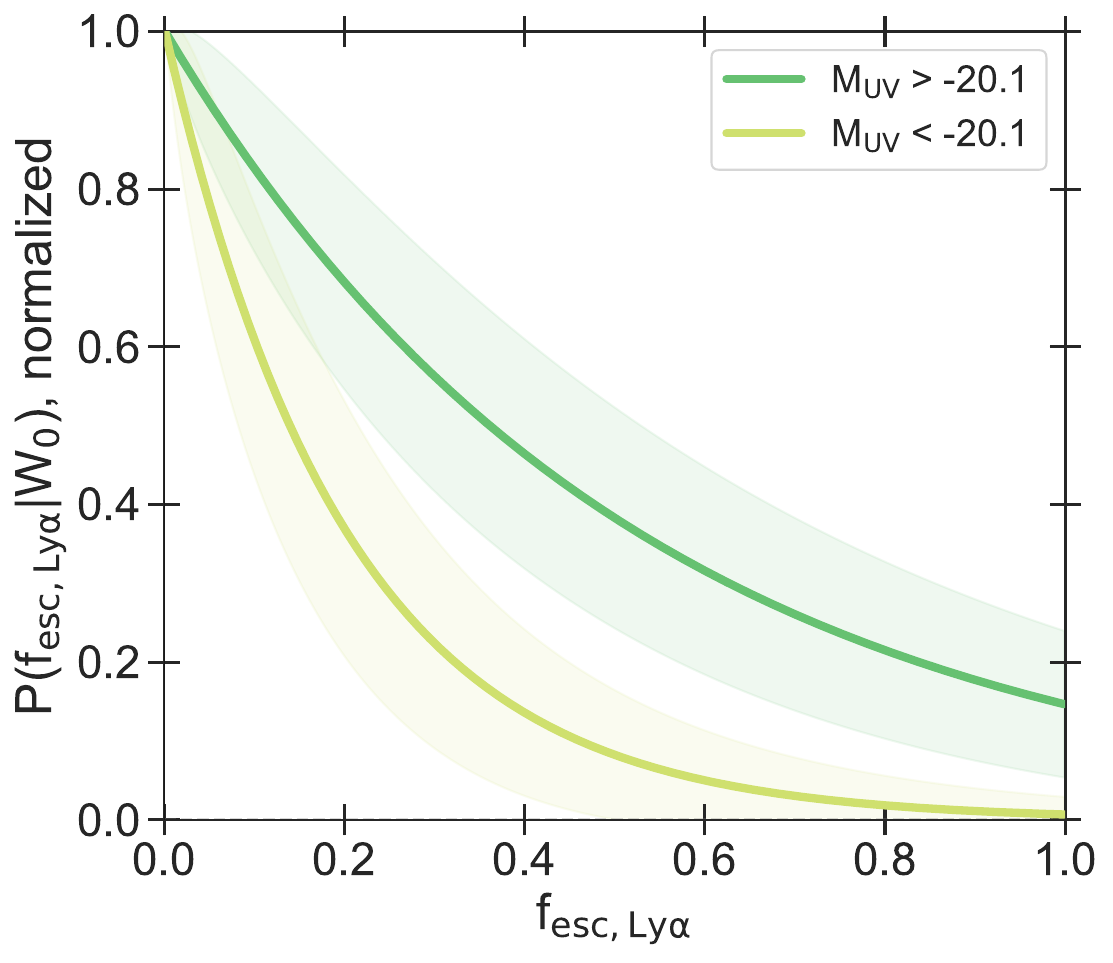}
\centering\includegraphics[width=8cm]{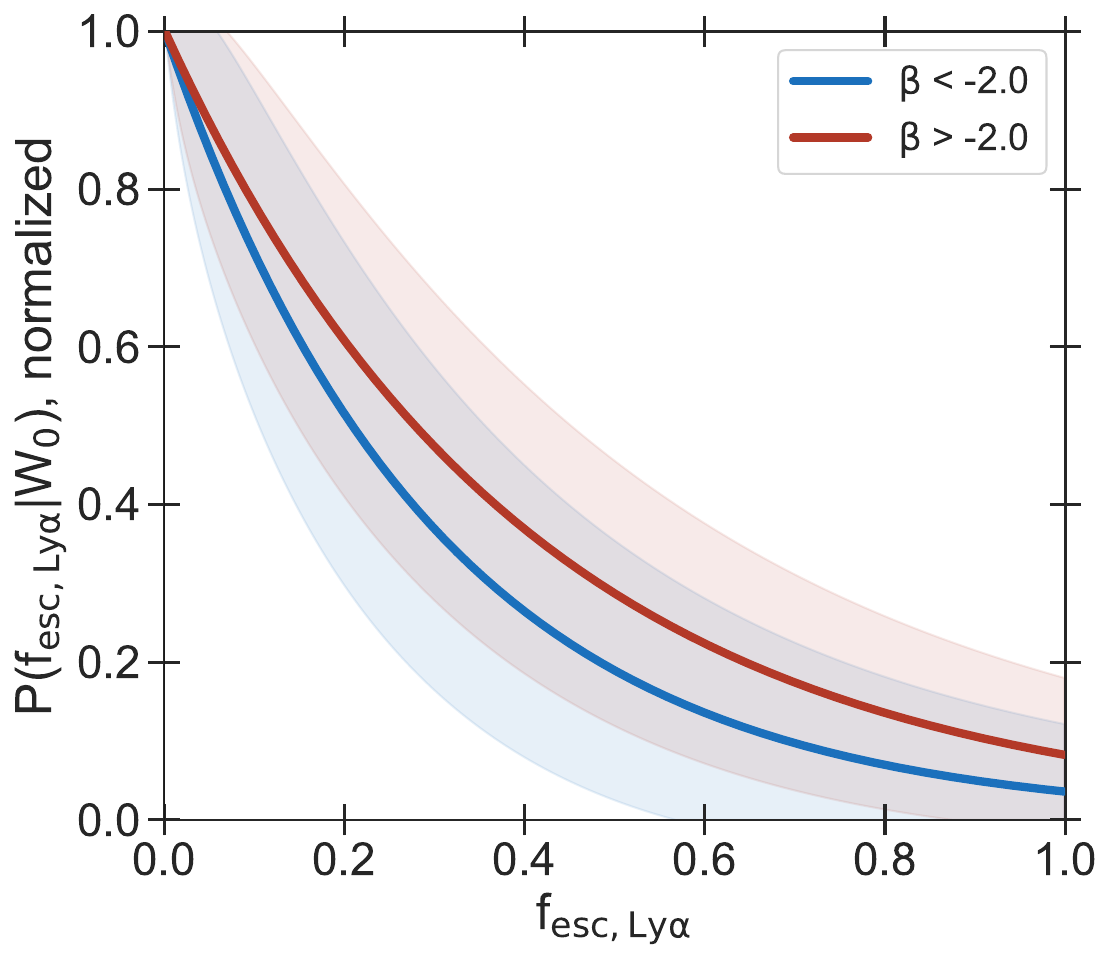}
\caption{Probability distributions of observed \fesc. Results are obtained by binning the dataset in \MUV and UV slope separately. As in our \ewlya results, we find that the model is dominated by \MUV, while the two UV slope scenarios are not statistically different.  
\label{fig:Pfesc1} }
\end{figure}

As our sample of sources with \fesc measurements is not large enough to fit the linear model (Equation~\ref{eq:Params}), we fit the \fesc distribution assuming constant parameters $A$ and $X_0$ in two bins. We create two UV magnitude bins: UV-faint (\MUV>-20.1), UV-bright (\MUV<-20.1); and two UV slope bins: red ($\beta$>-2.0) and blue ($\beta$<2.0). These bins are defined so each contains the same number of sources.

We inferred the parameters $A$ and $X_0$ in our four sub-samples, with the resulting inferred parameters given in Table \ref{tab:fesc_results}. We show the inferred distributions in Figure \ref{fig:Pfesc1}. We recover median $\fesc=0.18\pm 0.06$ from our distribution fit to the full sample. Similarly to our EW fits, we find a delta function at \fesc = 0 is disfavored.

We find the probability of high \fesc is much higher in our fainter \MUV bin. We find median $\fesc=0.13\pm 0.05$ in our UV-bright bin compared to 0.28$\pm 0.08$ in our UV-faint bin. 
From our UV-faint bin we obtain that 62$\pm 8 \%$ of \MUV>-20.1 galaxies should have \fesc $> 0.2$. This is considerably higher than the 30$\pm 6 \%$ found by \citet{Tang2024} for \MUV$\sim -19.5$ galaxies. We attribute this difference to our high upper limits on \fesc which do not significantly constrain our models and the use of a uniform prior on \fesc. If we decrease our \lya flux limits by a factor 2(4), we obtain 50$\pm12\%$ (35$\pm10\%$) of \MUV>-20.1 galaxies have \fesc $> 0.2$, demonstrating the sensitivity of the inference to upper limits. Thus we consider our reported distribution an upper limit on the underlying distribution.

We do not find significant difference between the \fesc distributions binned by UV slope. Previous works at z$\sim$5-6 \citep{Tang2024,Chen2024} have shown that \fesc strongly correlates with UV slope. However, similar to z$\sim$2 \lya surveys \citep[i.e.][]{Naidu2022a}, we do not see this correlation in our sample (see also Figure~\ref{fig:fesc_MUV_beta}). The lack of trends with UV slope might also be driven by our target selection which is incomplete for blue ($\beta \simlt -2$) \MUV$<$-19.5 galaxies (see discussion in Section~\ref{sec:discussion_ewlya}).

\section{Discussion}\label{sec:disc}

In this paper, we present new measurements and models of \lya for Lyman-Break selected galaxies at $z\sim5-6$. In Sections \ref{sec:lya_galaxy_properties} and \ref{sec:MakingModel}, we connect the shape and strength of the \lya line with other physical properties, such as \MUV. In the following discussion, we explore the implications of our findings for predicting \lya emission emerging from the ISM/CGM (Section~\ref{sec:discussion_ewlya}), as well as the impact of this transmission on its observability during the EoR (Section~\ref{sec:discussion_observability}),.

\subsection{Predicting \lya\ transmission from the ISM/CGM at $z\sim5-6$}\label{sec:discussion_ewlya}

The main goal of this paper is to build a predictive model for \lya at the end of the EoR, providing a basis to better interpret \lya observations at $z>6$. These baseline observations are important due to the degeneracy between scattering of \lya by neutral hydrogen in the IGM and ISM/CGM during the EoR. In particular, we focused on \lya line shape and strength properties, including \fwhmlya, \DV, \ewlya and \fesc. In this section, we discuss their trends with physical galaxy properties, \MUV, UV slope. Finally, we explore the possible physical drivers behind these trends and how they can be used to better predict \lya at $z>6$.

In Section~\ref{sec:results_profiles}, we show that \lya lineshape properties: \DV and \fwhmlya, have tentative correlations with \MUV. In Figure \ref{fig:fwhmlya_MUV_FWHMHa}, combining our \DV results with other samples from the literature works at $z>3$, we see a strong correlation of increasing \DV with higher UV luminosity. 
We also found a mild correlation of increased \fwhmlya in UV bright galaxies. It has been shown in theoretical works \citep[e.g.][]{Neufeld1991,Verhamme2006a} that broader \lya profiles and higher \DV are byproducts of scattering in high column density HI gas, increasing \lya resonant scattering events. The trends we find of increasing \DV and \fwhmlya with UV luminosity are thus consistent with the physical picture that UV bright galaxies generally trace more massive and spatially extended galaxies, favoring higher HI column densities.
This scenario is supported by observations showing a strong correlation between galaxy size and UV luminosity \citep[e.g.,][]{Shibuya2015,Morishita2024}. Furthermore, results from hydrodynamical simulations predict high HI column densities are more prevalent in UV bright galaxies, where supernova feedback is less efficient in disrupting the neutral ISM and opening channels of low HI column density \citep[e.g.,][albeit with large sightline variance]{Rosdahl2022,Gelli2025}. 

A key goal of this work was to better understand what shapes \lya line profiles at $z\sim5-6$, to help predict emergent \lya during the EoR at $z>6$, as pre-JWST reionization inferences have primarily been based on line profiles from $z\sim2$ galaxies \citep[e.g.,][]{Mason2018}. Recently, \citet{Tang2024} demonstrated the median \lya line profile at $z\sim5-6$ is shifted to higher velocity offsets relative to a $z\sim2$ comparison sample with similar \lya EW \citep{Tang2024c}, and discussed this may reflect enhanced scattering in the ISM/CGM at these redshifts and/or resonant scattering by infalling residual neutral gas in the IGM with mean neutral fraction $\overline{x}_\mathrm{HI} \simgt 10^{-4}$ \citep{Dijkstra2007,Laursen2011}.
Resonant scattering by infalling IGM gas should produce a sharp cut-off in the line profile, at the velocity of the infalling gas, resulting in observed \lya lines which are narrower and with more redshifted peaks compared to lines transmitted through the highly ionized IGM \citep{Santos2004a,Mason2018,Park2021,Tang2024c}. 
We find that $\approx70\%$ of \lya lines in our $z\sim5-6$ sample are highly asymmetric \citep[see also,][]{Pentericci2018,Hayes2021}, and our median \fwhmlya=245\,\kms is lower than that of the $z\sim2-3$ sample presented by \citet{Tang2024c} (=290\,km/s), which were selected as reionization-era analogs, with \OIII+H$\beta$ EW comparable to $z\simgt6$ galaxies. These observations are qualitatively consistent with the predicted sharp flux cut-off due to resonant scattering by the IGM. 

However, in Section~\ref{sec:results_profiles} we also demonstrated our $z\sim5-6$ \fwhmlya distribution peaks at higher values than the model predictions by \citet{Mason2018,Mason2019c}, which were based on \lya red peaks in $z\sim2$ Lyman-break galaxies. We also found the majority of our velocity offset measurements exceed this model (also seen among other $z>5$ sources in the literature). 
In the \citet{Mason2018} model, \lya velocity offsets and \fwhmlya are empirically estimated to scale with halo mass, $ \sim M_h^{1/3}$, with redshift evolution incorporated by a shift in star formation to lower mass halos at fixed \MUV at higher redshifts \citep[due to more rapidly rising star formation histories, e.g.,][]{Behroozi2013c,Mason2015}. 
One possible factor missing from the \citet{Mason2018} model is cosmological density evolution, which would increase the HI column density in the ISM/CGM at fixed halo mass with increasing redshift. 
This would be expected to produce a redshift scaling of both \DV and \fwhmlya $\sim (1+z)^{2/3}$ at fixed halo mass, implying $\sim 1.6\times$ higher \DV and \fwhmlya at $z\sim5$ compared to the original model. 
Such a scaling appears in good agreement with the ratio between our median observed line profile and the original model predictions: median observed \DV=258(\fwhmlya=245)\,\kms, compared to \DV=150(\fwhmlya=160)\kms predicted by the \citet{Mason2018} model. 
Current samples of sources with both \lya velocity offsets and \fwhmlya measurements at $z\sim5-6$ are still small, but larger samples will be important for better determining the impact of IGM infall relative to ISM/CGM density evolution on \lya line profiles at these redshifts.

In Section~\ref{sec:EW_results}, we showed that \ewlya and \fesc generally increase with decreasing UV luminosity. Additionally, we identified a strong correlation between \fesc and \ewlya. We also showed find that \fesc generally increases as \lya lines become narrower. These findings are consistent with previous studies, where \ewlya and \fesc have an enhanced probability of being stronger in galaxies that are UV faint and bluer \citep{Oyarzun2017,Tang2024}. However we find no strong trends with UV slope in our sample, as we discuss below. 
As discussed by \citet{Tang2024}, the strong correlation of \ewlya and \fesc suggests that strong \ewlya requires an environment capable of high \fesc. This is consistent with the scenario discussed above, where reduced resonant scattering events are more common in UV faint galaxies, leading to lower probability of \lya being destroyed by dust or scattered out the line of sight \citep{Dijkstra2014}, and therefore increased \fesc. The link between \lya strength and the properties of the ISM/CGM suggest that \ewlya and \fesc can be predicted statistically using non-\lya observables. This is supported by previous work at low redshifts linking \lya emission to rest-frame optical properties \citep[e.g.,][]{Trainor2019a,Hayes2023}, and at $z>6$ by our results, which, consistent with other recent studies using JWST observations of \lya-emitting galaxies at $z\sim5-6$ \citep{Tang2024,Chen2024,Lin2024}, show that \lya strength can be predicted statistically from UV observables at these redshifts. Empirical baselines mapping from UV observables to \lya are crucial for interpreting \lya detections in $z>10$ spectra, where we lose rest-frame optical wavelength coverage with NIRSpec.

Our measurements in Figures \ref{fig:EW-MUV-slope} and \ref{fig:fesc_MUV_beta} show no clear trends of \ewlya and \fesc with UV slope. The same is true of our \ewlya and \fesc empirical models in Sections~\ref{sec:EW_results} and \ref{sec:fesc_results}: we find no significant dependencies of the two models on UV slope. Nonetheless, previous works have found that \lya is strongly enhanced in galaxies with bluer UV continuum \citep{Oyarzun2017,Tang2024,Lin2024}. We attribute this missing trend in our sample to our target selection, magF160W $<$ 27.5 (see Section~\ref{sec:data_target}) which results in a lack of \MUV$>-19.5$, blue ($\beta<$-2) galaxies. We tested this by comparing our sample in \MUV and UV slope against the JADES-DJA photometric catalogs. We find that imposing a magF160W $<$ 27.5 cut for galaxies within $5<z_{\mathrm{phot}}<6.5$ in the JADES-DJA catalog biases against blue ($\beta<$-2) UV slopes for \MUV$\gtrsim -19.5$ galaxies. At \MUV$>-19.5$ we find median $\beta$=-1.6 in our data, while the full JADES-DJA sample finds $\beta$=-2.2. If we apply the magF160W < 27.5 cut to JADES-DJA, the median UV slope also becomes $\beta$=-1.6. As the faintest, bluest galaxies typically have strongest \lya emission \citep{Oyarzun2017,Tang2024,Lin2024} we attribute our lack of trends of \ewlya and \fesc with UV slope to our sample incompleteness. However, omitting sources with \MUV$>-19.5$ from our \ewlya and \fesc models does not significantly impact our conclusions.

\subsection{Implications for \lya observability during reionization}\label{sec:discussion_observability}

\begin{figure}[]
\centering\includegraphics[width=\columnwidth]{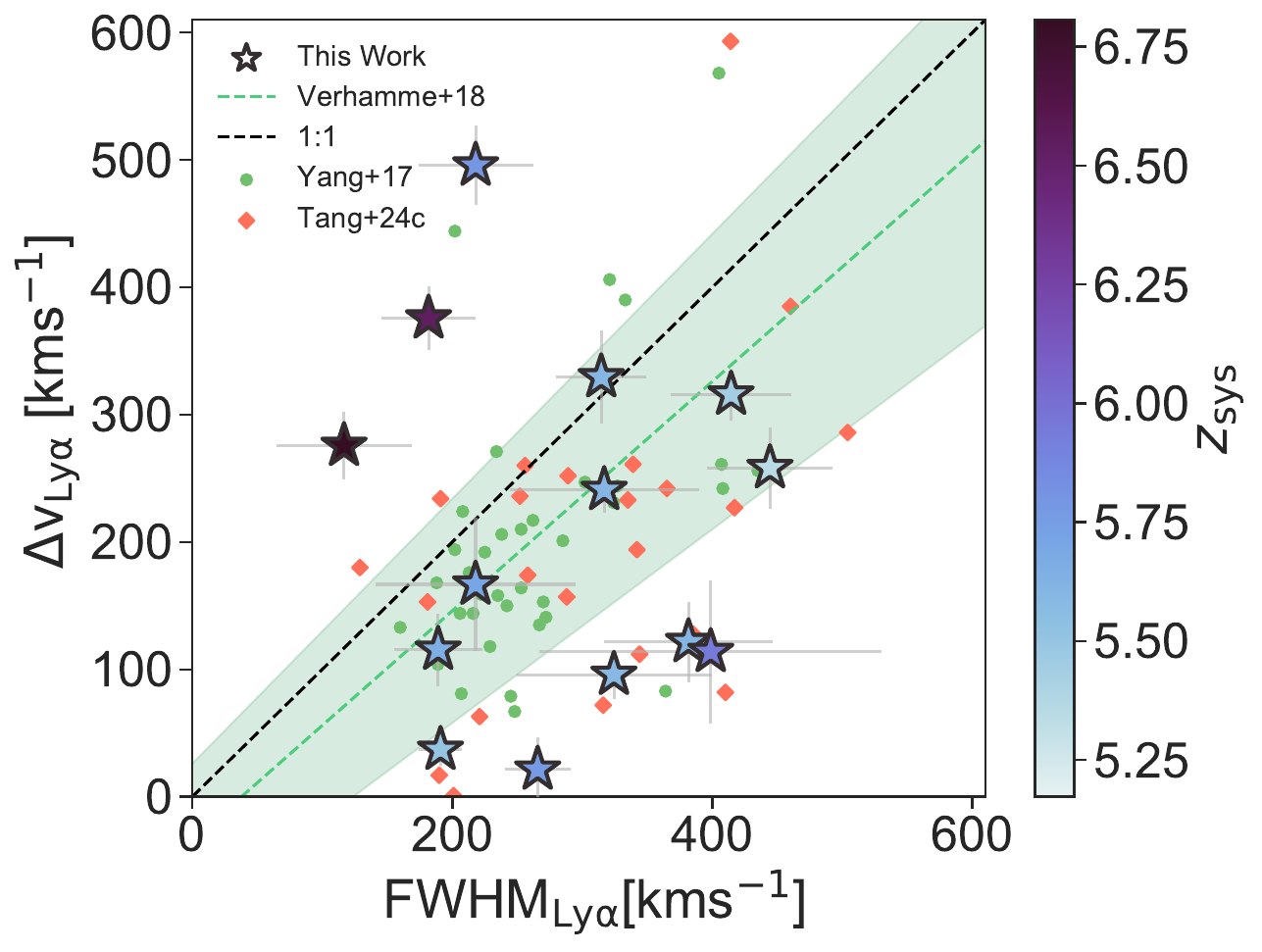}
\caption{\lya FWHM versus \DV for our sub-sample with systemic redshifts measured from FRESCO. We color-code our data points by their systemic redshift. For comparison we show the one-to-one relation, and the relation and 1$\sigma$ range (including intrinsic scatter) derived by \citet{Verhamme2018a} from sample of $z\sim0-6$ galaxies. We add data points from \citet{Yang2017c} (z$\sim$0.1-0.4, Green Peas) and \citet{Tang2024c} (z=2.1-3.4) While the sample is small and there is large scatter, both at low and high redshift, we note the three highest redshift sources in our sample lie above the \citet{Verhamme2018a} relation, suggesting the neutral IGM may be impacting these spectra redwards of the \lya line-center.
\label{fig:fwhm_dv}}
\end{figure}

Our primary goal was to better understand the typical emergent \lya emission as it leaves the ISM/CGM in high-redshift galaxies. As the effects of HI in the IGM and ISM/CGM on \lya are entangled during the EoR, we must separate the emergence of \lya from galaxies and its absorption caused by the damping wing of the IGM \citep{Miralda-Escude1998b,Dijkstra2007,Dijkstra2011,Mason2018}. With this goal, we have produced baseline measurements with observations at $z\sim5-6$. Our observations at the end of the EoR, allow us to isolate the effects of the ISM/CGM over \lya, and to better infer the reionization timeline in the future. In particular, with $R\sim4000$ resolution and observations of up to 15 hours, we have made the first measurement of the \fwhmlya distribution at $z\sim5-6$, previously limited to bright sources and stacked profiles with lower resolution \citep[e.g.,][]{Pentericci2018a,Hayes2021}. 

We found in Section~\ref{sec:results_profiles} that our \fwhmlya distribution is comparable, indeed slightly higher, than the predictions of \citet{Mason2019c}. We also found that velocity offset measurements at $z>5$ tend to lie above these predictions. 
To investigate what may drive this evolution, in Figure~\ref{fig:fwhm_dv} we plot \lya FWHM versus \DV for our sub-sample with systemic redshifts from FRESCO. 
At $z\sim0-6$, a close, almost one-to-one correlation has been observed between these quantities, as expected from \lya radiative transfer simulations \citep{Verhamme2018a}, which we also plot for comparison. 
We compare our sample with samples at lower redshift \citep[$z\sim0-3$]{Yang2017c,Tang2024c}.
We see our sample broadly follows the trends in the literature. While there is large scatter around the \citet{Verhamme2018a} relation both at low and high redshifts, we note that three of the highest redshift sources ($z>6$) in our sample lie above this relation ($\DV \simgt2$FWHM), with \lya lines that are narrower than expected given their velocity offsets. 
One possibility is that these redshifts the IGM is already shaping \lya lineshapes, as the IGM damping wing, and infall of residual neutral IGM, preferentially attenuate \lya\ flux close to line center \citep{Dijkstra2007,Mason2020}.
Our results indicate that larger samples with systemic redshift measurements from JWST and sufficient resolution to measure \lya FWHM will be important for quantifying the impact of the IGM on $z\gtrsim6$ lines in more detail.

The \lya lineshape plays a key role in the \lya detection rate once the IGM becomes increasingly neutral at $z\simgt6$, as the damping wing profile of the IGM is strongly wavelength dependent \citep{Miralda-Escude1998b,Mesinger2015}. Therefore, we must understand to what extent the \lya transmission drop past $z\simgt6$ is purely due to the IGM damping wing, or to the emergent line properties of \lya. The \fwhmlya has two opposing effects on the observability of \lya: First, a broader line profile will more easily transmit through the IGM as it increases the fraction of flux emitted at redder wavelengths, where it experiences lower damping wing optical depths \citep{Prieto-Lyon2023b,Mukherjee2024,Yuan2024}. However, a broader profile will lead to the same flux being spread over a wider wavelength range, hindering its detection in flux-limited observations. Increased \DV will always boost \lya transmission through the IGM by shifting \lya to redder wavelengths with lower damping wing optical depths. 

To assess the impact of the lineshape on reionization analyses, we can compare our lineshape with the assumptions by \citet{Mason2018,Mason2019c} in their neutral fraction inference.
We compare the IGM \lya transmission for an $\MUV = -20$ galaxy  at $z=9$: comparing our median line profile and the model by \citet{Mason2018}.
We assume a completely neutral IGM, in a region without ionized bubbles \citep[using the damping wing approximation by][]{Dijkstra2014}. 
The transmitted \lya using our line shape median values (Gaussian, with \fwhmlya = 245\kms, \DV= 258\kms) is $9\times$ higher than that predicted by \citet{Mason2019c} (Gaussian \fwhmlya = 160\kms \DV= 150\kms, and truncated at 80 km/s due to infalling IGM) -- finding 0.9$\%$ vs 0.1$\%$ of \lya flux is transmitted through the IGM, respectively. We note that for bubbles $R\simgt1$\,pMpc \citep[roughly the expected mean bubble size when the IGM is $\sim70$\% neutral,][]{Lu2023} the difference in transmission becomes negligible, implying the lineshape plays an important role in \lya visibility in the earliest stages of reionization.

We also examine how the strongly asymmetric shape of \lya should facilitate \lya transmission in a mostly neutral IGM. For the 14 sources in our sample with measured velocity offsets we calculate the fraction of flux at $\DV > 400$\,km/s, which can be transmitted even in a neutral IGM at $z\sim13$ \citep{Dijkstra2011,Mason2018b,Yuan2024}. We obtain a median and standard deviation of 22$\pm 18 \%$ from this sample, agreeing closely with recent results from hydrodynamical radiative transfer simulations by \citet{Yuan2024}. This suggests the red tail of \lya emission be important for allowing \lya to be visible even at extremely high redshifts \citep[e.g.,][]{Witstok2024b}.

The higher transmission for \lya, which is both more offset from the systemic redshift and has extended red wings, implies \lya can be visible even in a very neutral IGM \citep{Dijkstra2011,Witstok2024c}. Our results suggest we may require a higher neutral fraction at $z\simgt8$ to explain the \lya decline compared to analyses using $z\sim 2$ lineshapes as a baseline. This highlights the importance of including $z\sim5$-6 samples, such as those used here, when making reionization inferences \citep[see also,][]{Yuan2024}. For instance, \citet{Tang2024b} recently adopted the $z\sim5-6$ \lya line profile from \citet{Tang2024} as a baseline when constraining the IGM neutral fraction at $z\sim6-13$ from JWST observations.

\section{Conclusions}\label{sec:conc}

A detailed understanding of \lya as it emerges from galaxies in the first billion years is critical for interpreting \lya observations during the EoR. Our main goal in this work was to produce empirical baseline models for \lya to enable better constraints on the reionization history of the Universe. We present an analysis of Lyman-break galaxies at $z\sim5-6$, when the IGM is expected to be mostly ionized. With our new ground-based \lya spectra, supplemented by JWST imaging and spectra, we studied the correlations between \lya shape and strength with UV luminosity. Our conclusions are as follows:

\begin{enumerate}
    \item We present high-resolution MMT/Binospec restframe-UV spectroscopy for 236 Lyman-break galaxies. We detected \lya at S/N$>5$ for 62 of these, with a median $5\sigma$ flux limit of 1.34$\times 10^{-17}$erg s$^{-1}$ cm$^{-2}$. Overlapping ancilliary JWST/NIRCam spectra from FRESCO enabled us to measure z$_\mathrm{sys}$ for 56 of our targets, including 14 galaxies detected with \lya.\\

    \item Using $R\sim4600$ spectroscopy, we measured \lya line profiles with a high resolution of $\sim70$\kms. We measured for the first time the \fwhmlya distribution at $z\sim5-6$, a key unknown quantity in interpreting the decline of \lya at $z>6$. We found a mean of \fwhmlya=245\kms, which is higher than models based on $z\sim2$ \lya observations \citep{Mason2018,Mason2019c}, but lower than $z\sim2-3$ observations of reionization-era analogs by \citet{Tang2024c}. We discuss the possibility that this could be due to resonant scattering by infalling IGM with $\overline{x}_\mathrm{HI}\simgt 10^{-4}$ at $z\sim5-6$, as well as increasing ISM and CGM densities at higher redshifts at fixed halo mass. We note that the latter effect was not included in the \citet{Mason2018} model.\\

    \item We obtained 14 new measurements of \lya velocity offsets at $z>5$ through systemic redshifts from \Ha or \OIII from FRESCO NIRCam slitless spectra. We found correlations between physical properties and \lya lineshape: \DV and \fwhmlya, with the majority of galaxies displaying $\DV \simgt 100$\kms, with a median of 258\kms. Combined with $z>3$ data from the literature, we report a strong correlation for increased \DV in UV bright galaxies. We find mild correlations of higher \fwhmlya in galaxies that are UV-bright. Our lineshape trends are consistent with a higher \lya optical depth and numerous scattering events in the ISM and CGM, characterizing increasingly UV-bright galaxies.\\ 
    
    \item We measured the line strength properties \ewlya and \fesc. We found strong correlations of \ewlya and \fesc with increasing UV luminosity and mild correlations of increasing \fesc with lower \fwhmlya. The decline of \fesc with line broadness and UV luminosity is consistent with a scenario of higher \lya optical depth in UV bright galaxies.\\

    \item We built an empirical model for the probability distribution of \ewlya and \fesc at the end of the EoR, as a function of \MUV and UV slope. Our \ewlya and \fesc models strongly depend on \MUV. We found that strong \ewlya ($>25$\,\AA) and \fesc (>0.2) are common at $z\sim5-6$. Our models predict that 45$\pm$5\% and $<62\pm8$\% of \MUV=-19.5 galaxies have \ewlya $>25$\,\AA\ and \fesc$>0.2$.\\
\end{enumerate}

The NIR capabilities of JWST are now enabling the community to unlock new information on emergent \lya line profiles and escape fractions through rest-frame optical emission lines. As $R\simgt 1000$ optical line samples at $z\sim 5 - 9$ grow steadily by the thousands, in large part thanks to NIRCam grism surveys \citep[e.g.][]{Kashino2023,Naidu2024,Alba2024,Meyer2024b}, we can further refine our understanding of emergent \lya at $z\sim 5-6$. The unprecedented detections of \lya at $z\sim10-13$ \citep{Bunker2023,Witstok2024c} have opened a new window on studying the earliest stages of reionization. Baseline \lya empirical correlations and models, such as those we have presented \citep[see also][]{Chen2024,Tang2024}, are crucial for interpreting the growing JWST observations of \lya at $z\sim6-13$ to constrain the reionization process \citep{Tang2024b,Mason2025}.

\section*{}
\fontsize{8}{10}\selectfont
\begin{acknowledgements}
We thank Hakim Atek, Anne Verhamme and Johan Fynbo for useful discussions. We thank Igor Chilingarian, Sean Moran for their initial reductions and advice regarding the Binospec pipeline, and Dan Fabricant, Ben Weiner and Ben Johnson for their advice in planning and reducing Binospec observations. We thank Daniel Eisenstein for helpful discussions on the Binospec survey design and comments on a draft of this manuscript. GPL and CAM acknowledge support by the VILLUM FONDEN under grant 37459. CAM acknowledges support from the Carlsberg Foundation under grant CF22-1322. The Cosmic Dawn Center (DAWN) is funded by the Danish National Research Foundation under grant DNRF140.

All MMT/Binospec spectra and catalogs used in this work are publicly available in the Electronic Research Data Archive at University of Copenhagen at the link: \hyperlink{https://sid.erda.dk/sharelink/FCZBDRW6O2}{https://sid.erda.dk/sharelink/FCZBDRW6O2}.

\end{acknowledgements}

\bibliographystyle{aa}
\bibliography{library}
\onecolumn
\appendix

\section{\lya detections}\label{appendix:tables}
\fontsize{8}{10}\selectfont
\renewcommand{\arraystretch}{1.1} 
\hspace*{-0.7cm}
\vspace*{-3cm} 
\begin{tabular}{|l||ll|ll|ll|ll|l|l|}
\hline\hline
 ID & RA & DEC & z$_{\mathrm{\lya}}$ & z$_{\mathrm{sys}}$ & \MUV & $\mathrm{\beta}$ & \ewlya & \fwhmlya  & \fesc & \DV \\
  &  &  &  &  &  & & [\AA] & [\kms]  & [\kms] & [\kms] \\
\hline
z5-GNW-1503 & 189.139303 & 62.111230 & 5.0508 & -- & -20.9 $\pm$ 0.1 & -1.9 $\pm$ 0.4 & 8 $\pm$ 3 & 345 $\pm$ 97 & -- & --  \\
z5-GNW-12024 & 189.142960 & 62.179790 & 5.1661 & -- & -18.6 $\pm$ 0.5 & -1.8 $\pm$ 1.9 & 87 $\pm$ 89 & 340 $\pm$ 66 & > 0.18 & --  \\
z5-GND-3052 & 189.288742 & 62.173836 & 5.1796 & -- & -20.3 $\pm$ 0.1 & -1.4 $\pm$ 0.2 & 53 $\pm$ 6 & 185 $\pm$ 10 & -- & --  \\
z5-GNW-11071 & 188.972424 & 62.173125 & 5.1848 & -- & -19.3 $\pm$ 0.1 & -1.4 $\pm$ 0.8 & 130 $\pm$ 37 & 189 $\pm$ 8 & -- & --  \\
z5-GND-32413 & 189.357854 & 62.211445 & 5.2326 & -- & -19.2 $\pm$ 0.1 & -2.2 $\pm$ 0.6 & 18 $\pm$ 5 & 72 $\pm$ 41 & -- & --  \\
z5-GNW-20906 & 189.437681 & 62.318150 & 5.2911 & -- & -21.8 $\pm$ 0.0 & -1.8 $\pm$ 0.3 & 10 $\pm$ 1 & 782 $\pm$ 253 & -- & --  \\
z5-GNW-11014 & 189.073073 & 62.172883 & 5.2946 & -- & -20.4 $\pm$ 0.1 & -1.9 $\pm$ 0.3 & 44 $\pm$ 7 & 618 $\pm$ 96 & -- & --  \\
z5-GNW-21219 & 189.310135 & 62.330258 & 5.2946 & -- & -20.3 $\pm$ 0.1 & -2.2 $\pm$ 0.3 & 60 $\pm$ 7 & 237 $\pm$ 11 & -- & --  \\
z5-GNW-22490 & 189.394847 & 62.308059 & 5.3211 & -- & -20.2 $\pm$ 0.1 & -3.1 $\pm$ 0.8 & 28 $\pm$ 7 & 474 $\pm$ 128 & -- & --  \\
z6-GND-35647 & 189.329865 & 62.200844 & 5.3496 & -- & -19.4 $\pm$ 0.3 & -0.8 $\pm$ 1.1 & 59 $\pm$ 30 & 75 $\pm$ 18 & -- & --  \\
z5-GND-27819 & 189.110337 & 62.225441 & 5.3510 & 5.3465 & -20.9 $\pm$ 0.1 & -2.6 $\pm$ 0.3 & 25 $\pm$ 4 & 444 $\pm$ 48 & 0.16 $\pm$ 0.06 & 258 $\pm$ 32  \\
z5-GNW-12482 & 189.079256 & 62.182873 & 5.3543 & -- & -20.5 $\pm$ 0.1 & -2.3 $\pm$ 0.2 & 23 $\pm$ 3 & 580 $\pm$ 113 & -- & --  \\
z5-GNW-2438 & 189.106533 & 62.118483 & 5.3569 & -- & -19.8 $\pm$ 0.1 & -2.6 $\pm$ 0.4 & 50 $\pm$ 9 & 366 $\pm$ 46 & -- & --  \\
z5-GNW-1663 & 189.147548 & 62.112511 & 5.3697 & -- & -19.7 $\pm$ 0.1 & -1.7 $\pm$ 0.5 & 67 $\pm$ 13 & 308 $\pm$ 34 & -- & --  \\
z5-GNW-32760 & 189.451265 & 62.241037 & 5.4033 & -- & -19.7 $\pm$ 0.3 & -3.2 $\pm$ 1.4 & 35 $\pm$ 19 & 254 $\pm$ 27 & -- & --  \\
z5-GNW-21283 & 189.403559 & 62.315688 & 5.4046 & -- & -20.0 $\pm$ 0.3 & -3.8 $\pm$ 1.0 & 35 $\pm$ 15 & 529 $\pm$ 58 & -- & --  \\
z5-GNW-7098 & 189.214160 & 62.149048 & 5.4338 & -- & -20.5 $\pm$ 0.2 & -2.3 $\pm$ 0.8 & 62 $\pm$ 21 & 319 $\pm$ 17 & -- & --  \\
z5-GNW-29609 & 189.436412 & 62.265038 & 5.4455 & -- & -20.4 $\pm$ 0.1 & -1.8 $\pm$ 0.5 & 48 $\pm$ 8 & 429 $\pm$ 30 & -- & --  \\
z5-GND-21153 & 189.049628 & 62.244033 & 5.4483 & 5.4430 & -21.0 $\pm$ 0.1 & -1.9 $\pm$ 0.3 & 30 $\pm$ 4 & 414 $\pm$ 47 & 0.13 $\pm$ 0.04 & 316 $\pm$ 20  \\
z5-GNW-13514 & 188.967311 & 62.189120 & 5.4623 & -- & -20.3 $\pm$ 0.2 & -3.0 $\pm$ 0.9 & 65 $\pm$ 24 & 217 $\pm$ 30 & -- & --  \\
z6-GNW-23350 & 189.414474 & 62.333523 & 5.4633 & -- & -- & -- & -- & 325 $\pm$ 37 & -- & --  \\
z5-GNW-23042 & 189.415009 & 62.333538 & 5.4635 & -- & -20.0 $\pm$ 0.3 & -1.3 $\pm$ 0.8 & 183 $\pm$ 82 & 338 $\pm$ 18 & -- & --  \\
z5-GND-39445 & 189.178643 & 62.187234 & 5.5048 & 5.5043 & -19.5 $\pm$ 0.2 & -2.2 $\pm$ 1.1 & 94 $\pm$ 43 & 191 $\pm$ 16 & 0.29 $\pm$ 0.12 & 37 $\pm$ 13  \\
z6-GNW-14478 & 189.001450 & 62.194984 & 5.5546 & -- & -19.0 $\pm$ 0.7 & -0.1 $\pm$ 1.7 & 215 $\pm$ 447 & 390 $\pm$ 48 & -- & --  \\
z5-GND-7766 & 189.139984 & 62.291809 & 5.5913 & 5.5888 & -20.2 $\pm$ 0.1 & -0.7 $\pm$ 0.2 & 50 $\pm$ 10 & 324 $\pm$ 75 & 0.34 $\pm$ 0.19 & 96 $\pm$ 19  \\
z6-GND-36100 & 189.191274 & 62.199519 & 5.6029 & 5.5975 & -20.3 $\pm$ 0.0 & -2.2 $\pm$ 0.1 & 35 $\pm$ 3 & 315 $\pm$ 35 & 0.33 $\pm$ 0.20 & 329 $\pm$ 37  \\
z6-GND-36553 & 189.156387 & 62.197773 & 5.6109 & 5.6052 & -21.2 $\pm$ 0.0 & -2.0 $\pm$ 0.1 & 9 $\pm$ 1 & 317 $\pm$ 73 & 0.29 $\pm$ 0.16 & 241 $\pm$ 18  \\
Stark11-13066 & 189.156399 & 62.197716 & 5.6115 & -- & -21.2 $\pm$ 0.0 & -2.0 $\pm$ 0.1 & 7 $\pm$ 1 & 184 $\pm$ 34 & > 0.18 & --  \\
z6-GNW-14511 & 189.100538 & 62.195344 & 5.6155 & 5.6117 & -20.6 $\pm$ 0.1 & -2.4 $\pm$ 0.6 & 19 $\pm$ 5 & 382 $\pm$ 65 & 0.12 $\pm$ 0.05 & 122 $\pm$ 32  \\
Hu10-3 & 189.055939 & 62.129990 & 5.6328 & -- & -- & -- & -- & 307 $\pm$ 14 & -- & --  \\
Hu10-7 & 189.032780 & 62.143962 & 5.6389 & -- & -19.6 $\pm$ 0.3 & -1.2 $\pm$ 1.1 & 323 $\pm$ 183 & 295 $\pm$ 8 & -- & --  \\
z6-GND-43125 & 189.189137 & 62.300659 & 5.6585 & 5.6569 & -20.0 $\pm$ 0.1 & -2.3 $\pm$ 0.2 & 36 $\pm$ 5 & 189 $\pm$ 34 & 0.38 $\pm$ 0.20 & 116 $\pm$ 28  \\
Hu10-6 & 189.324563 & 62.299734 & 5.6611 & -- & -20.2 $\pm$ 0.1 & -2.7 $\pm$ 0.6 & 63 $\pm$ 17 & 253 $\pm$ 16 & -- & --  \\
z6-GNW-10822 & 188.995306 & 62.171460 & 5.6657 & -- & -20.5 $\pm$ 0.2 & -2.6 $\pm$ 0.7 & 26 $\pm$ 9 & 402 $\pm$ 47 & -- & --  \\
Hu10-5 & 189.399713 & 62.239490 & 5.6713 & -- & -21.1 $\pm$ 0.1 & -2.4 $\pm$ 0.3 & 11 $\pm$ 2 & 347 $\pm$ 46 & -- & --  \\
Hu10-11 & 189.366007 & 62.196189 & 5.6719 & -- & -19.2 $\pm$ 0.2 & -1.4 $\pm$ 0.7 & 78 $\pm$ 26 & 212 $\pm$ 20 & -- & --  \\
z5-GND-37006 & 189.366007 & 62.196189 & 5.6720 & -- & -19.2 $\pm$ 0.2 & -1.4 $\pm$ 0.7 & 85 $\pm$ 30 & 199 $\pm$ 20 & -- & --  \\
Stark11-22381 & 189.255389 & 62.357754 & 5.6920 & -- & -20.2 $\pm$ 0.3 & -3.1 $\pm$ 1.6 & 12 $\pm$ 6 & 247 $\pm$ 54 & -- & --  \\
Stark11-3982 & 189.039261 & 62.247654 & 5.7080 & 5.7030 & -19.6 $\pm$ 0.1 & -4.0 $\pm$ 0.6 & 24 $\pm$ 7 & 218 $\pm$ 77 & 0.22 $\pm$ 0.14 & 167 $\pm$ 52  \\
z7-GNW-22375 & 189.342596 & 62.308518 & 5.7718 & -- & -20.1 $\pm$ 0.5 & -2.7 $\pm$ 1.4 & 19 $\pm$ 14 & 51 $\pm$ 16 & -- & --  \\
z5-GND-17752 & 189.091798 & 62.253735 & 5.7744 & 5.7750 & -19.8 $\pm$ 0.1 & -1.5 $\pm$ 0.4 & 98 $\pm$ 19 & 266 $\pm$ 25 & 1.23 $\pm$ 0.79 & 22 $\pm$ 25  \\
Stark11-17705 & 189.208207 & 62.232128 & 5.8027 & 5.7925 & -19.7 $\pm$ 0.2 & -1.8 $\pm$ 1.1 & 45 $\pm$ 20 & 218 $\pm$ 44 & 0.31 $\pm$ 0.21 & 496 $\pm$ 31  \\
z6-GND-30340 & 189.388906 & 62.217840 & 5.8079 & -- & -20.0 $\pm$ 0.4 & -2.7 $\pm$ 1.1 & 66 $\pm$ 42 & 479 $\pm$ 53 & -- & --  \\
z6-GNW-21823 & 189.355591 & 62.312592 & 5.8228 & -- & -20.9 $\pm$ 0.4 & -2.1 $\pm$ 1.3 & 23 $\pm$ 16 & 130 $\pm$ 17 & -- & --  \\
z5-GNW-11701 & 189.055409 & 62.177498 & 5.8364 & -- & -19.6 $\pm$ 0.6 & -1.8 $\pm$ 1.7 & 18 $\pm$ 18 & 75 $\pm$ 29 & -- & --  \\
z5-GND-464 & 189.270721 & 62.148424 & 5.8745 & -- & -20.0 $\pm$ 0.1 & -2.2 $\pm$ 0.2 & 57 $\pm$ 6 & 206 $\pm$ 18 & -- & --  \\
Stark11-24923 & 189.284538 & 62.287491 & 5.9498 & 5.9483 & -20.0 $\pm$ 0.2 & -1.4 $\pm$ 1.1 & 30 $\pm$ 14 & 399 $\pm$ 131 & 0.14 $\pm$ 0.08 & 114 $\pm$ 56  \\
Stark11-26902 & 189.307610 & 62.323443 & 5.9547 & -- & -19.0 $\pm$ 0.6 & -1.4 $\pm$ 1.8 & 197 $\pm$ 215 & 587 $\pm$ 63 & -- & --  \\
z6-GNW-25971 & 189.334732 & 62.286125 & 5.9549 & -- & -19.7 $\pm$ 0.5 & -1.9 $\pm$ 1.3 & 38 $\pm$ 32 & 390 $\pm$ 75 & -- & --  \\
z6-GND-14309 & 189.334047 & 62.263058 & 5.9676 & -- & -20.0 $\pm$ 0.2 & -1.2 $\pm$ 1.0 & 104 $\pm$ 38 & -- & > 1.01 & --  \\
Stark11-6706 & 189.079712 & 62.141884 & 5.9719 & -- & -20.0 $\pm$ 0.5 & -3.0 $\pm$ 1.5 & 31 $\pm$ 21 & 346 $\pm$ 62 & -- & --  \\
Stark11-16773 & 189.197796 & 62.199983 & 5.9731 & -- & -19.3 $\pm$ 0.1 & -2.7 $\pm$ 0.4 & 37 $\pm$ 7 & 198 $\pm$ 42 & > 0.22 & --  \\
z6-GND-19165 & 189.347733 & 62.249866 & 6.0471 & -- & -20.0 $\pm$ 0.1 & -2.0 $\pm$ 0.4 & 33 $\pm$ 6 & 363 $\pm$ 63 & > 0.34 & --  \\
z7-GND-43678 & 189.235201 & 62.295597 & 6.1212 & -- & -19.7 $\pm$ 0.1 & -1.2 $\pm$ 0.5 & 112 $\pm$ 25 & 46 $\pm$ 14 & > 0.71 & --  \\
z6-GNW-2993 & 189.127197 & 62.122612 & 6.1357 & -- & -19.8 $\pm$ 0.5 & -1.1 $\pm$ 1.6 & 47 $\pm$ 45 & 249 $\pm$ 117 & -- & --  \\
Jung18-28438 & 189.178023 & 62.223718 & 6.5518 & 6.5442 & -20.0 $\pm$ 0.2 & -3.0 $\pm$ 0.9 & 30 $\pm$ 11 & 182 $\pm$ 36 & 0.21 $\pm$ 0.12 & 376 $\pm$ 25  \\
\hline
\hline
\end{tabular}

\newpage
\fontsize{8}{10}\selectfont
\renewcommand{\arraystretch}{1.1} 
\hspace*{-1cm}
\begin{tabular}{|l||ll|ll|ll|ll|l|l|}
\hline\hline
 ID & RA & DEC & z$_{\mathrm{\lya}}$ & z$_{\mathrm{sys}}$ & M$_{\mathrm{UV,1500}}$ & $\mathrm{\beta}$ & \ewlya & \fwhmlya & \fesc & \DV \\
   &  &  &  &  &  & & [\AA] & [\kms] & [\kms] & [\kms] \\
\hline
Hu10-1 & 189.358260 & 62.207637 & 6.5604 & -- & -20.3 $\pm$ 0.1 & -2.5 $\pm$ 0.8 & 166 $\pm$ 51 & 267 $\pm$ 25 & -- & --  \\
Hu10-2 & 189.356888 & 62.295321 & 6.5760 & -- & -21.9 $\pm$ 0.1 & -1.8 $\pm$ 0.4 & 24 $\pm$ 4 & 358 $\pm$ 6 & -- & --  \\
Jung18-5752 & 189.199585 & 62.320965 & 6.5867 & -- & -19.2 $\pm$ 0.3 & -1.9 $\pm$ 1.6 & 103 $\pm$ 72 & 174 $\pm$ 22 & > 0.16 & --  \\
z6-GND-44831 & 189.175135 & 62.282267 & 6.7365 & -- & -20.6 $\pm$ 0.1 & -2.9 $\pm$ 0.6 & 54 $\pm$ 13 & 127 $\pm$ 16 & -- & --  \\
z7-GND-8358 & 189.155310 & 62.286461 & 6.8135 & 6.8072 & -20.3 $\pm$ 0.1 & -2.7 $\pm$ 0.5 & 11 $\pm$ 5 & 117 $\pm$ 52 & -- & 276 $\pm$ 26  \\
z8-GND-35384 & 189.231989 & 62.202333 & 6.8743 & -- & -19.9 $\pm$ 0.6 & -1.7 $\pm$ 2.3 & 178 $\pm$ 198 & 41 $\pm$ 8 & -- & --  \\
\hline
\hline
\end{tabular}
\\
Notes: *: Not sufficient photometric coverage in catalogs described in Section~\ref{sec:data-phot}.\\

\section{Spectra of \lya-emitting galaxies: MMT/Binospec and NIRCam/JWST}\label{appendix:spectra}

\begin{figure*}[!h]
\begin{center}
\includegraphics[width=4.5cm, trim=0cm 0cm 0cm 0cm]{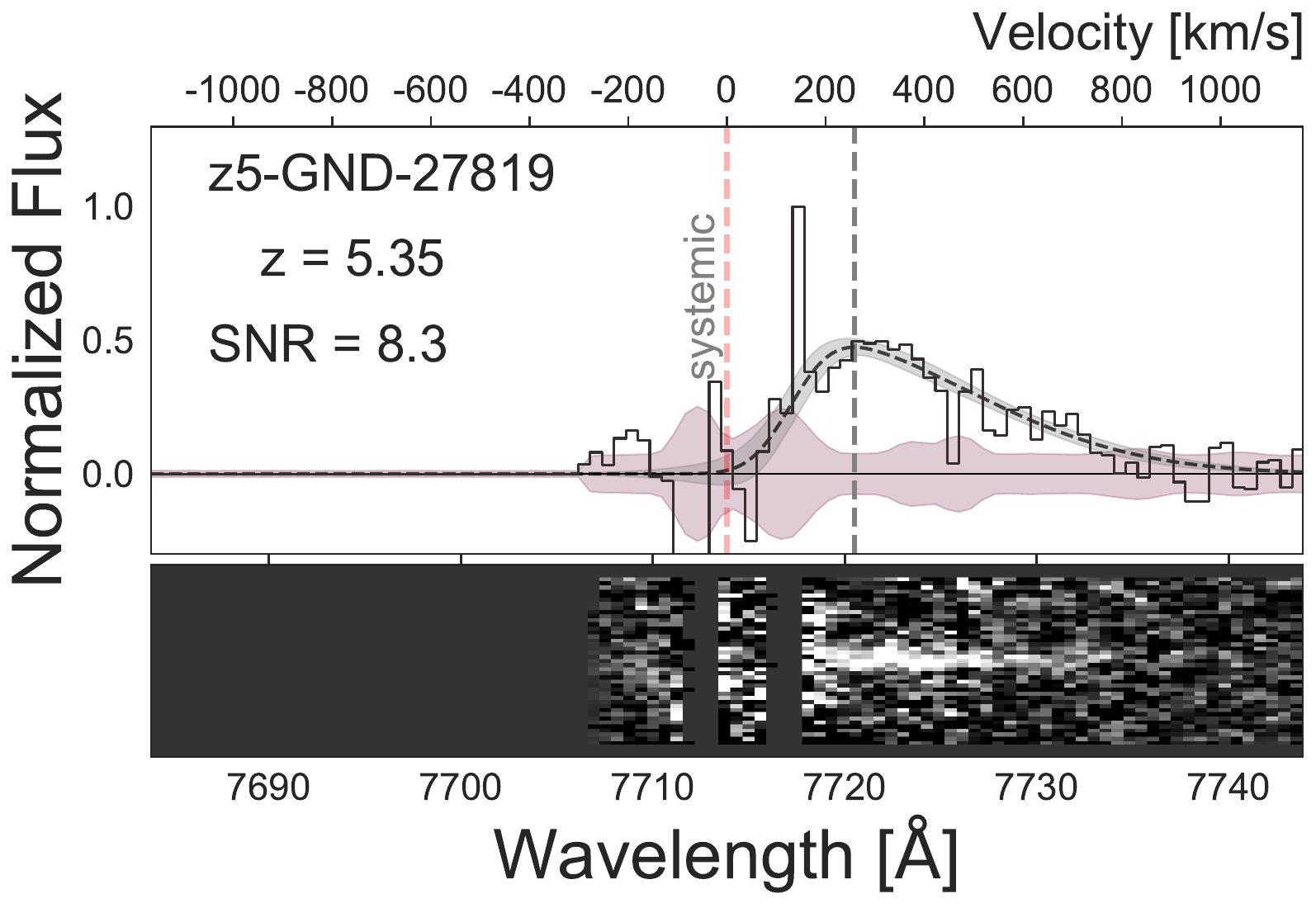}\includegraphics[width=4.5cm, trim=0cm 0cm 1cm 0cm]{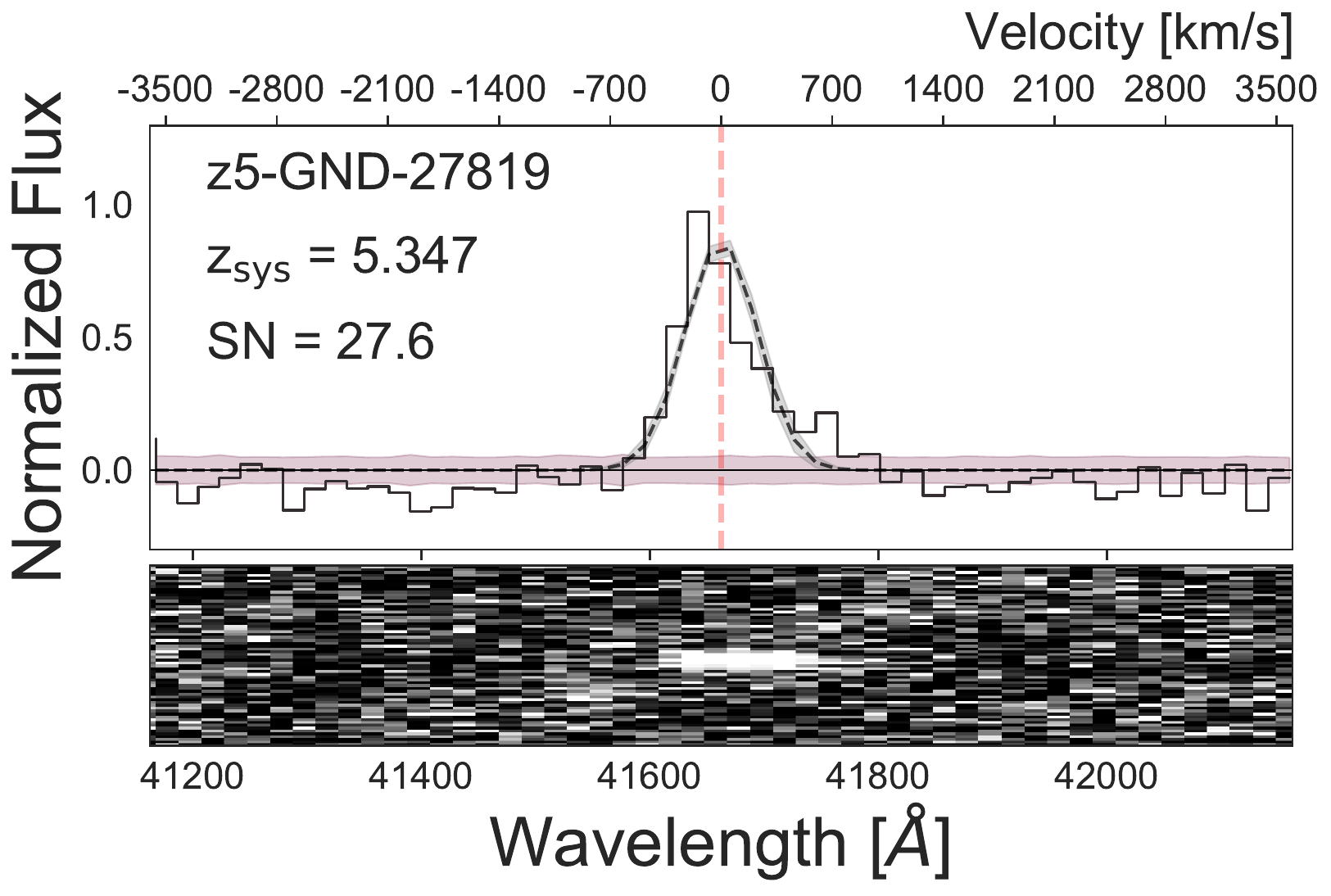}\hspace{0.2cm}\includegraphics[width=4.5cm, trim=0cm 0cm 0cm 0cm]{Spectra/Lya/z5_GND_21153.pdf}
\includegraphics[width=4.5cm, trim=0cm 0cm 0cm 0cm]{Spectra/Optical/z5_GND_21153.pdf}
\end{center}
\end{figure*}

\begin{figure*}[!h]
\begin{center}
\includegraphics[width=4.5cm, trim=0.1cm 1.5cm 0cm 0cm]{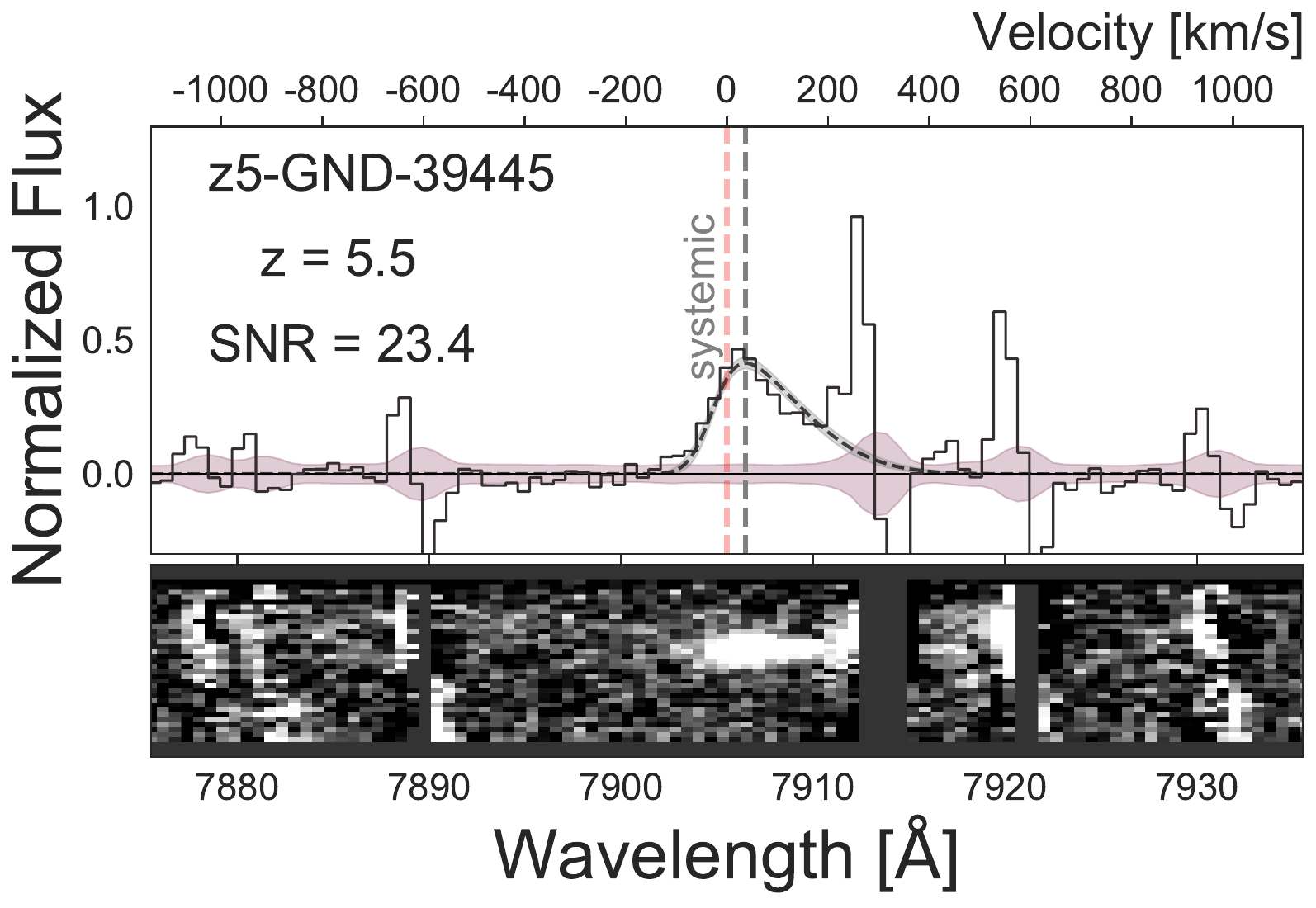}\includegraphics[width=4.5cm, trim=0.1cm 1.5cm 1cm 0cm]{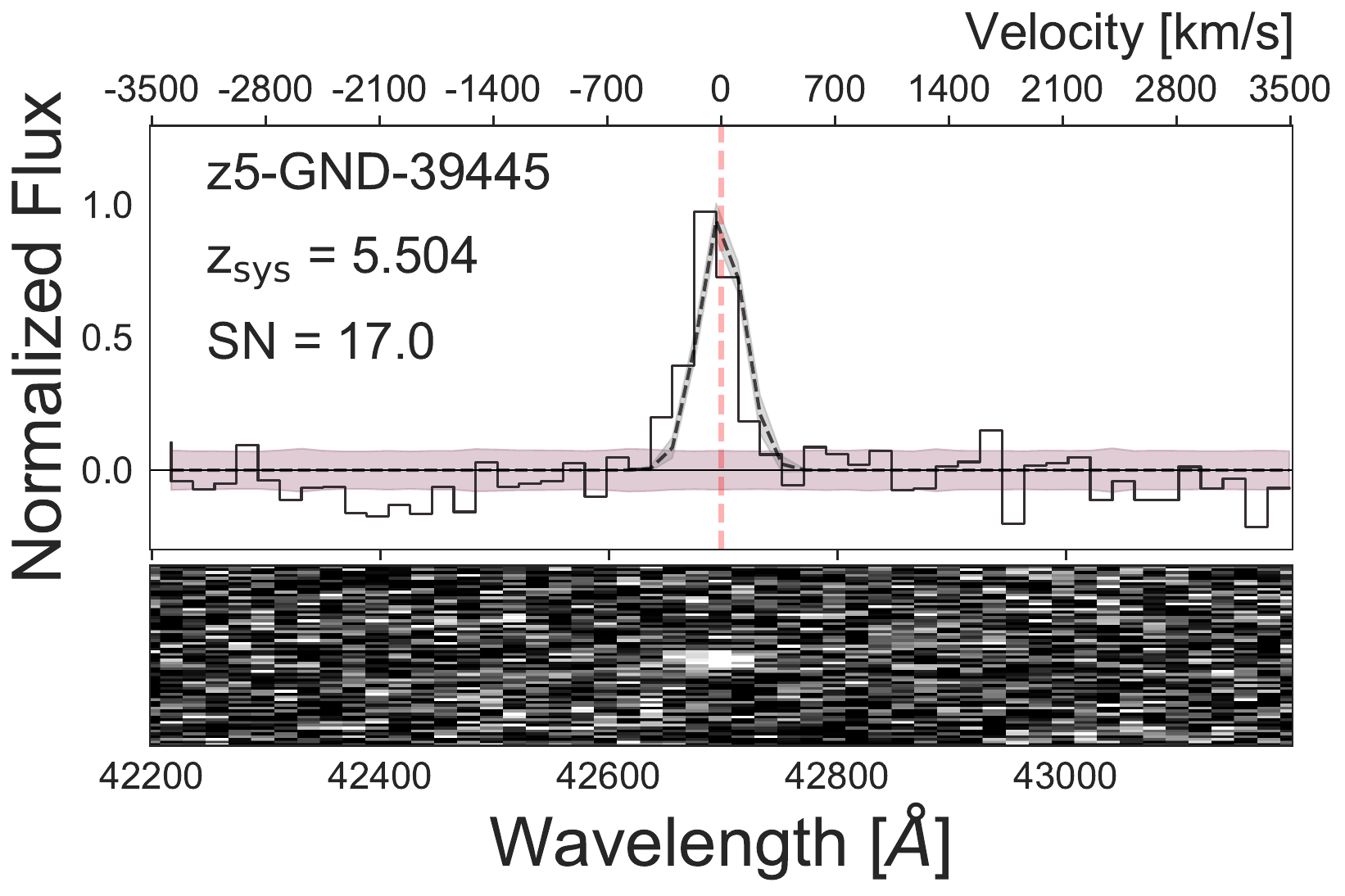}\hspace{0.2cm}\includegraphics[width=4.5cm, trim=0.1cm 1.5cm 0cm 0cm]{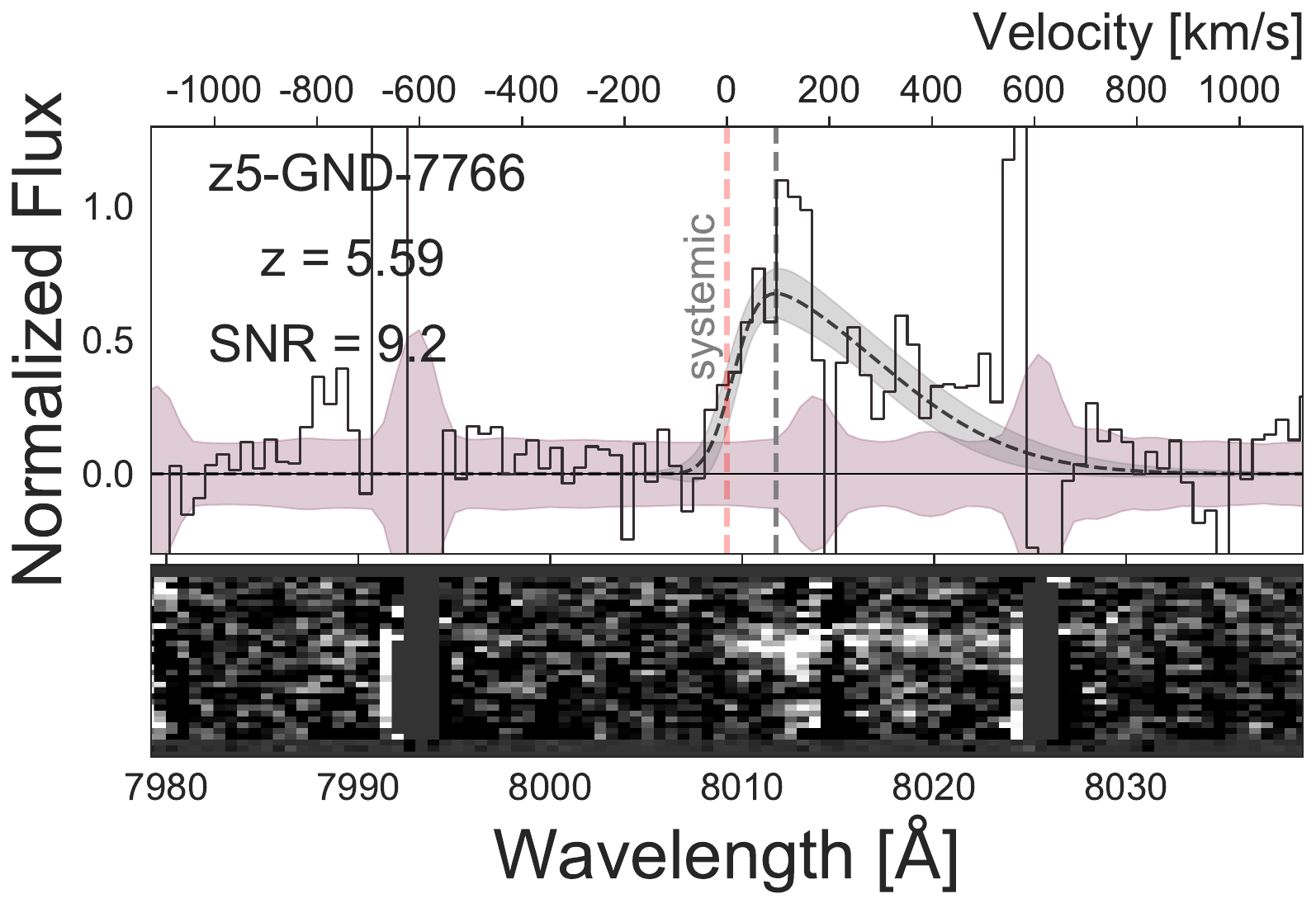}
\includegraphics[width=4.5cm, trim=0.1cm 1.5cm 1cm 0cm]{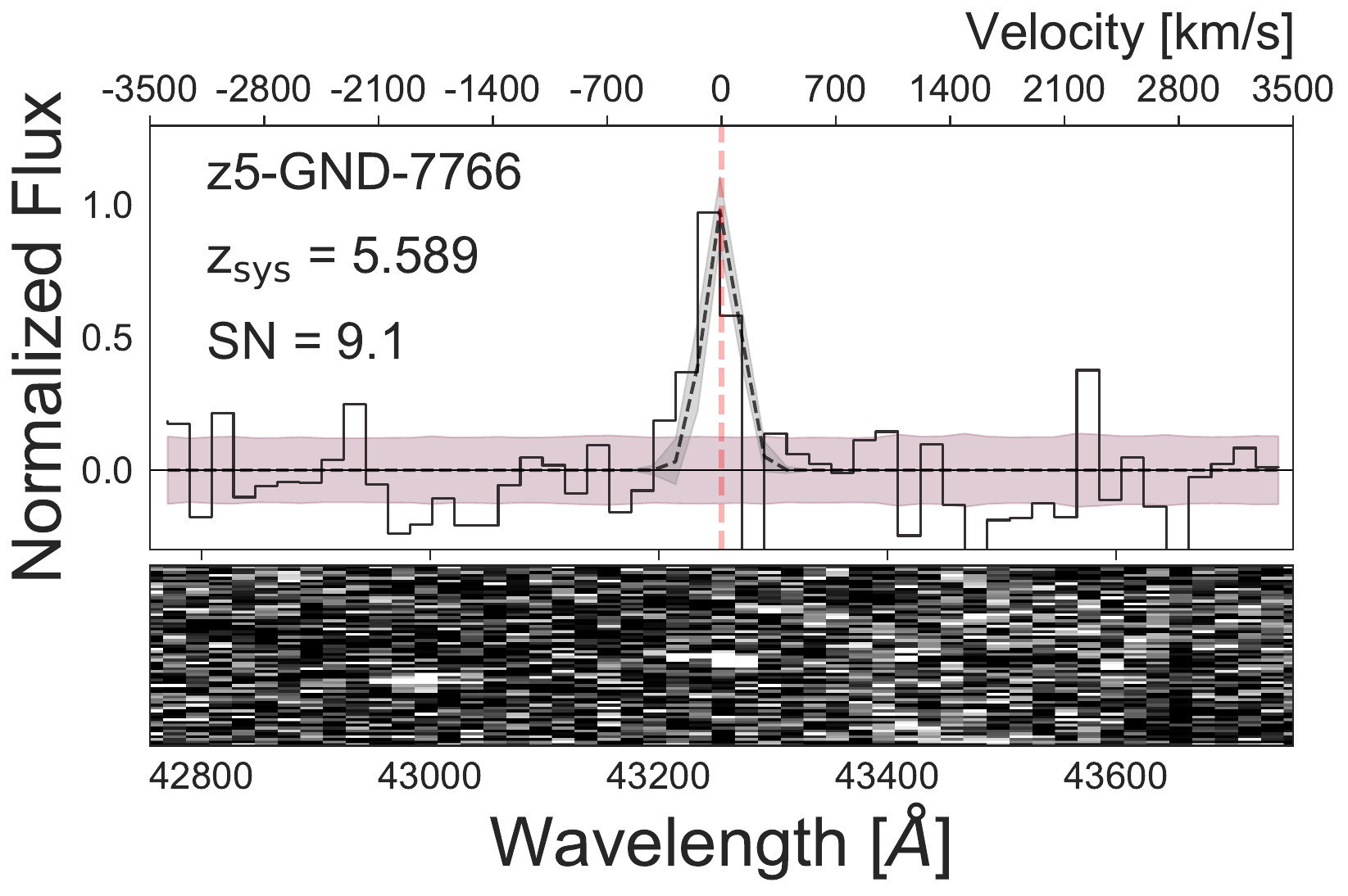}
\end{center}
\end{figure*}

\begin{figure*}[!h]
\begin{center}
\includegraphics[width=4.5cm, trim=0.1cm 1.5cm 0cm 0cm]{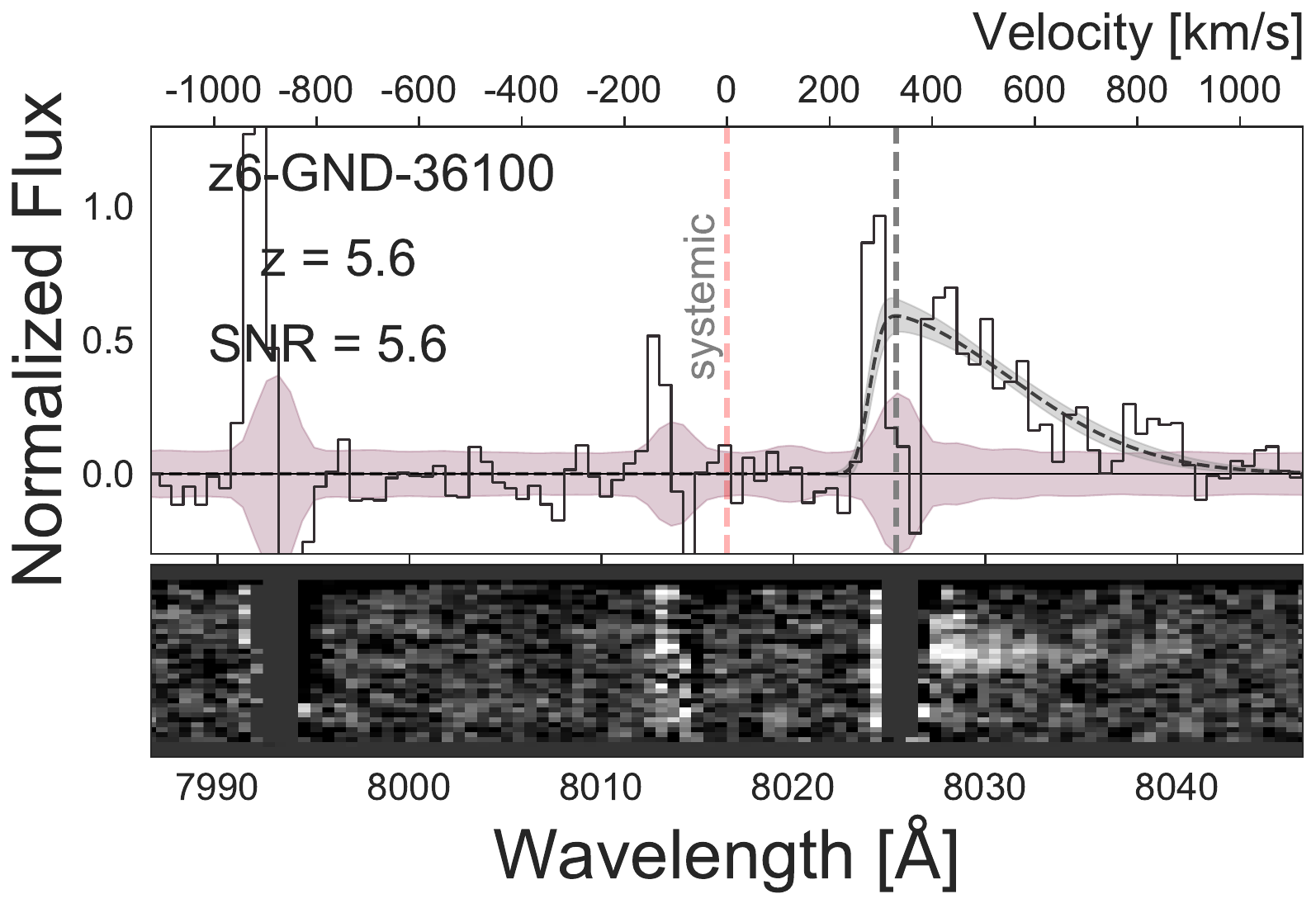}\includegraphics[width=4.5cm, trim=0.1cm 1.5cm 1cm 0cm]{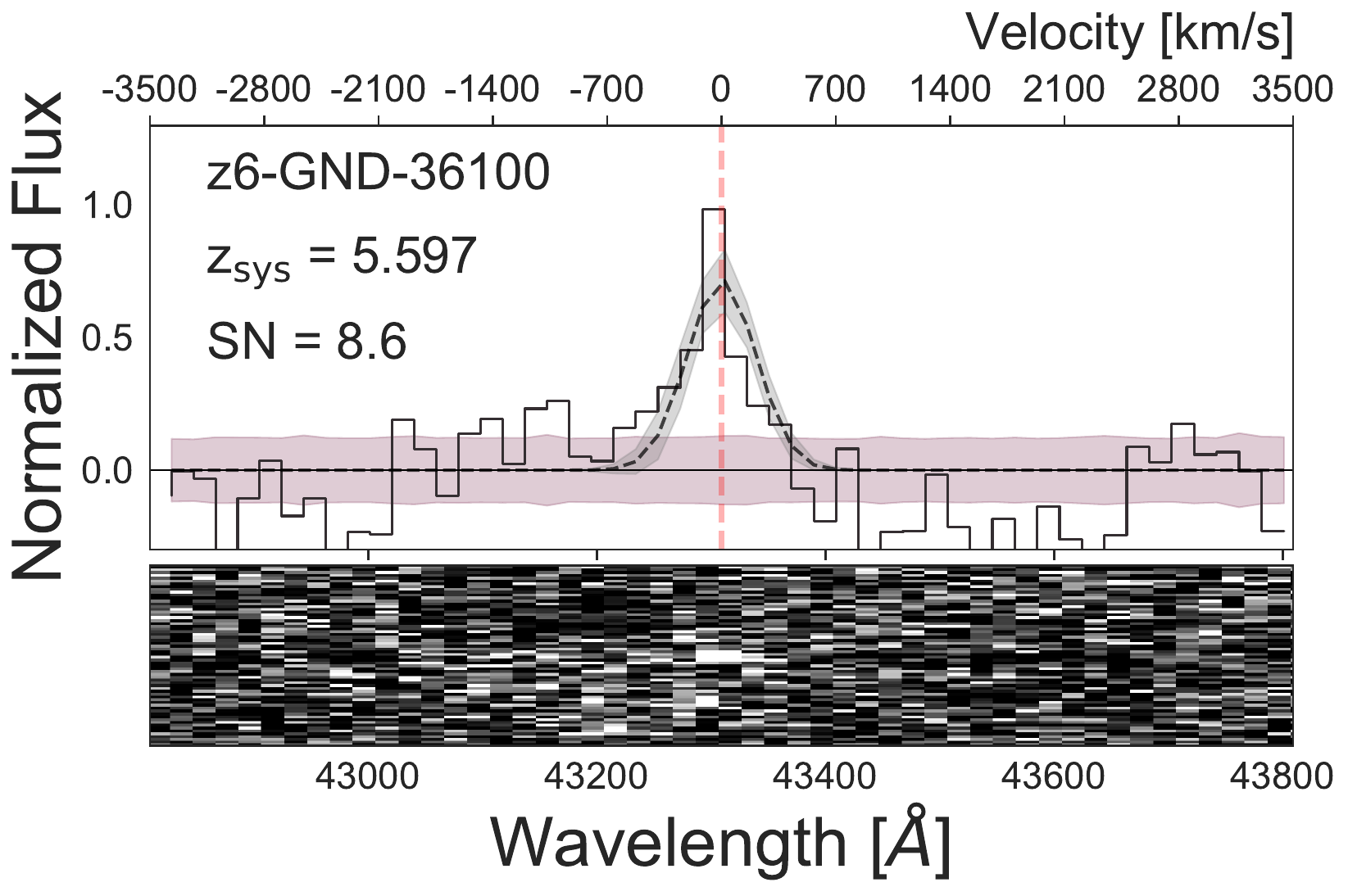}\hspace{0.2cm}\includegraphics[width=4.5cm, trim=0.1cm 1.5cm 0cm 0cm]{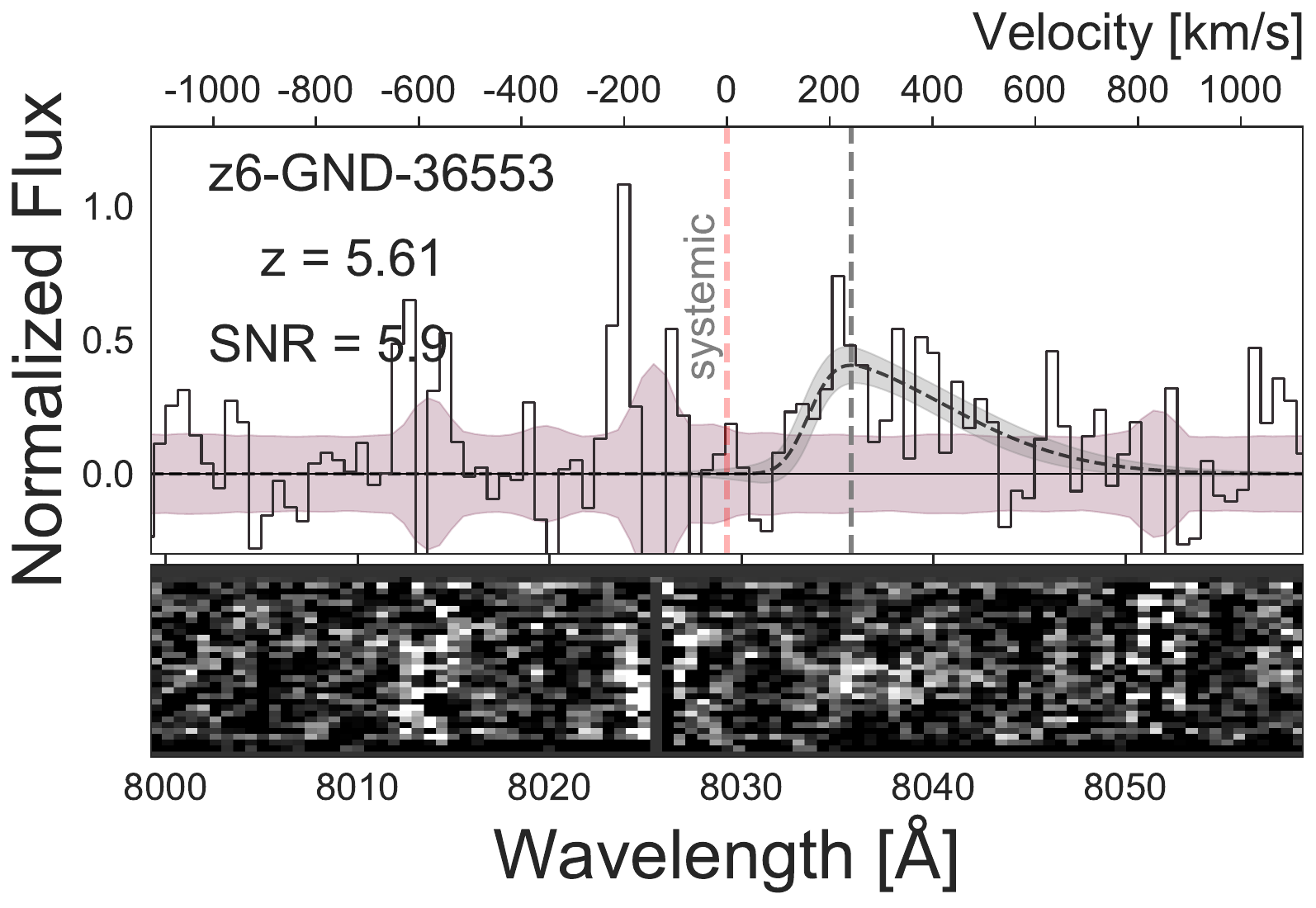}
\includegraphics[width=4.5cm, trim=0.1cm 1.5cm 1cm 0cm]{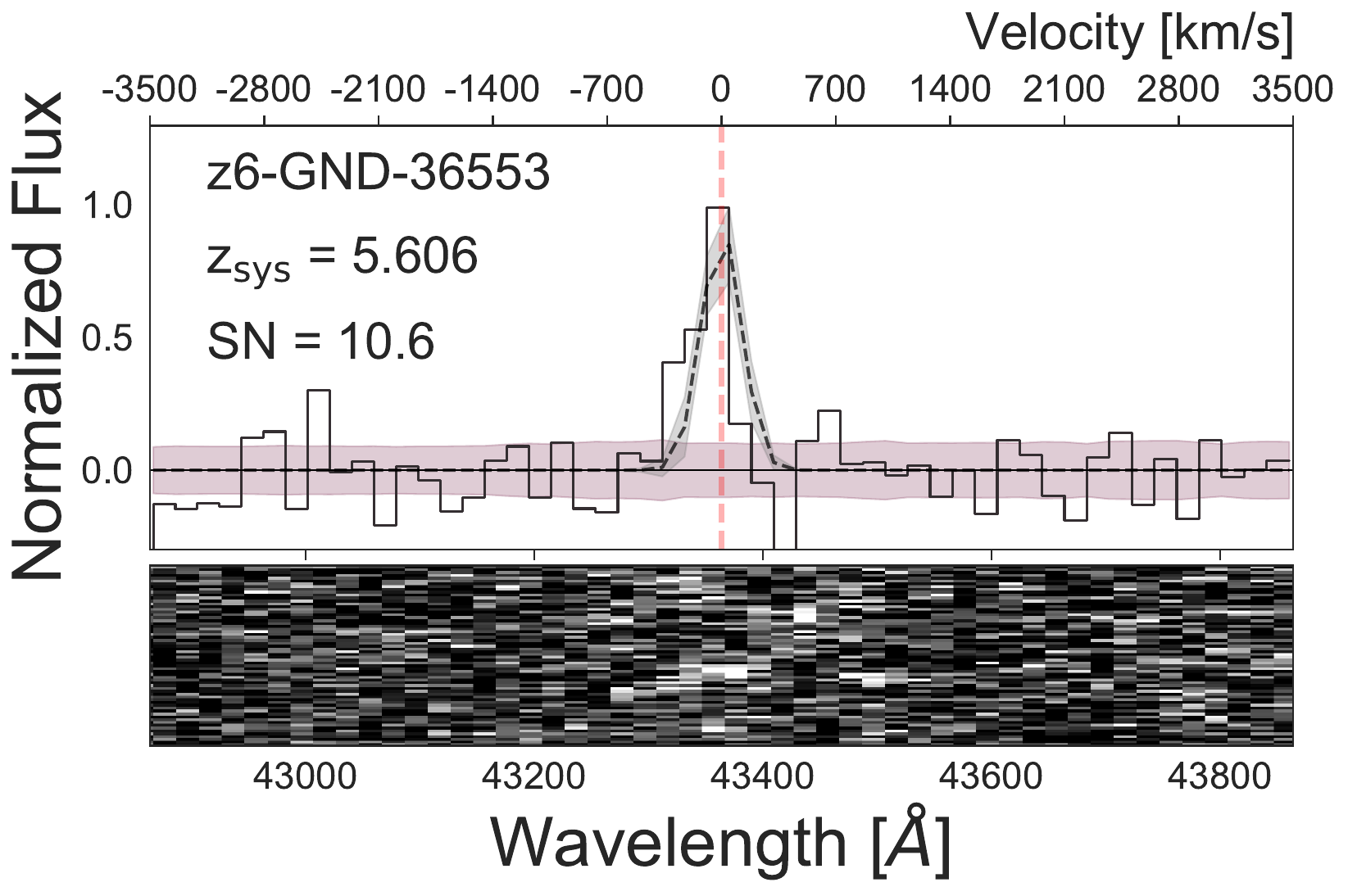}
\end{center}
\end{figure*}

\begin{figure*}[!h]
\begin{center}
\includegraphics[width=4.5cm, trim=0.1cm 1.5cm 0cm 0cm]{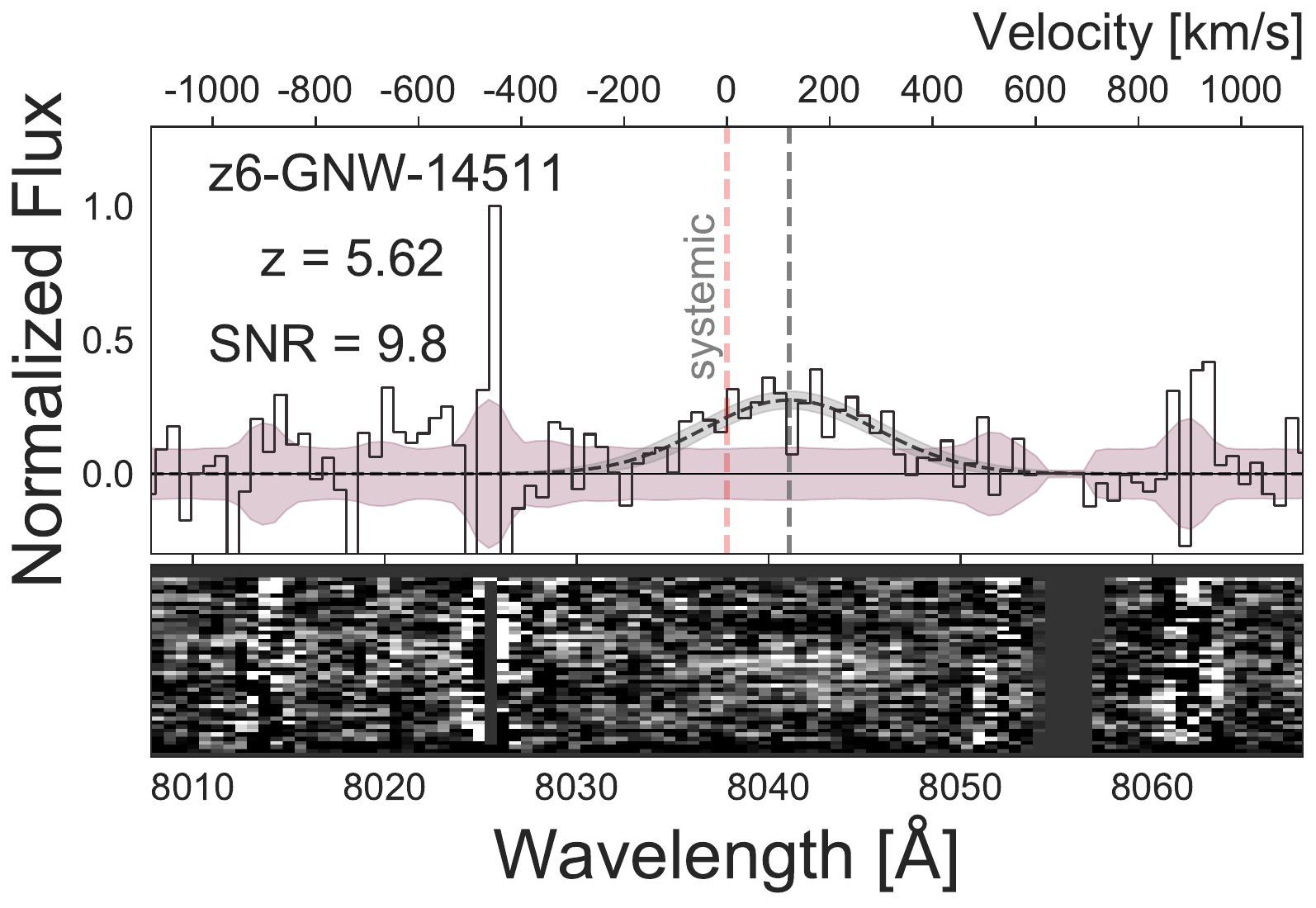}\includegraphics[width=4.5cm, trim=0.1cm 1.5cm 1cm 0cm]{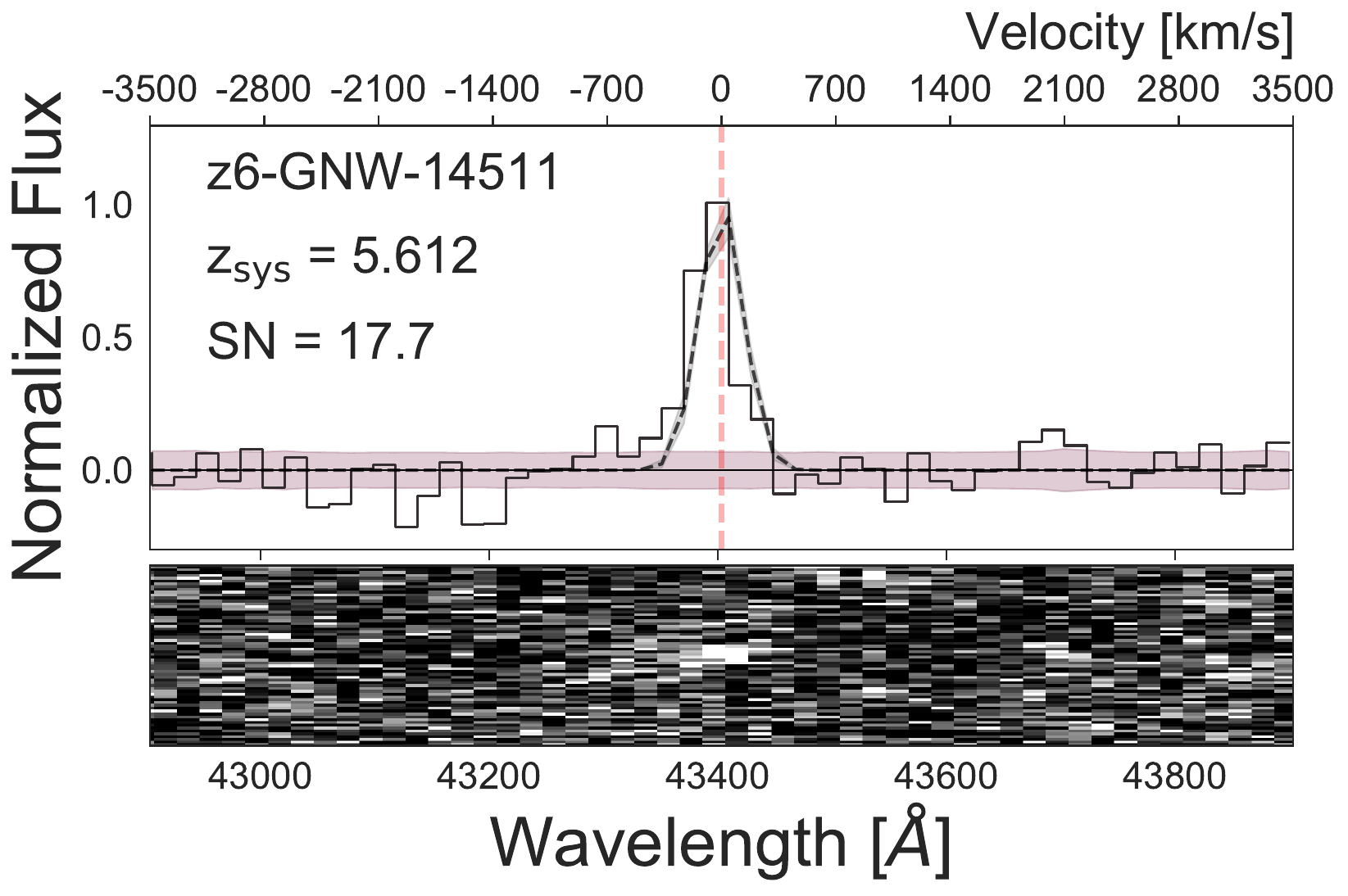}\hspace{0.2cm}\includegraphics[width=4.5cm, trim=0.1cm 1.5cm 0cm 0cm]{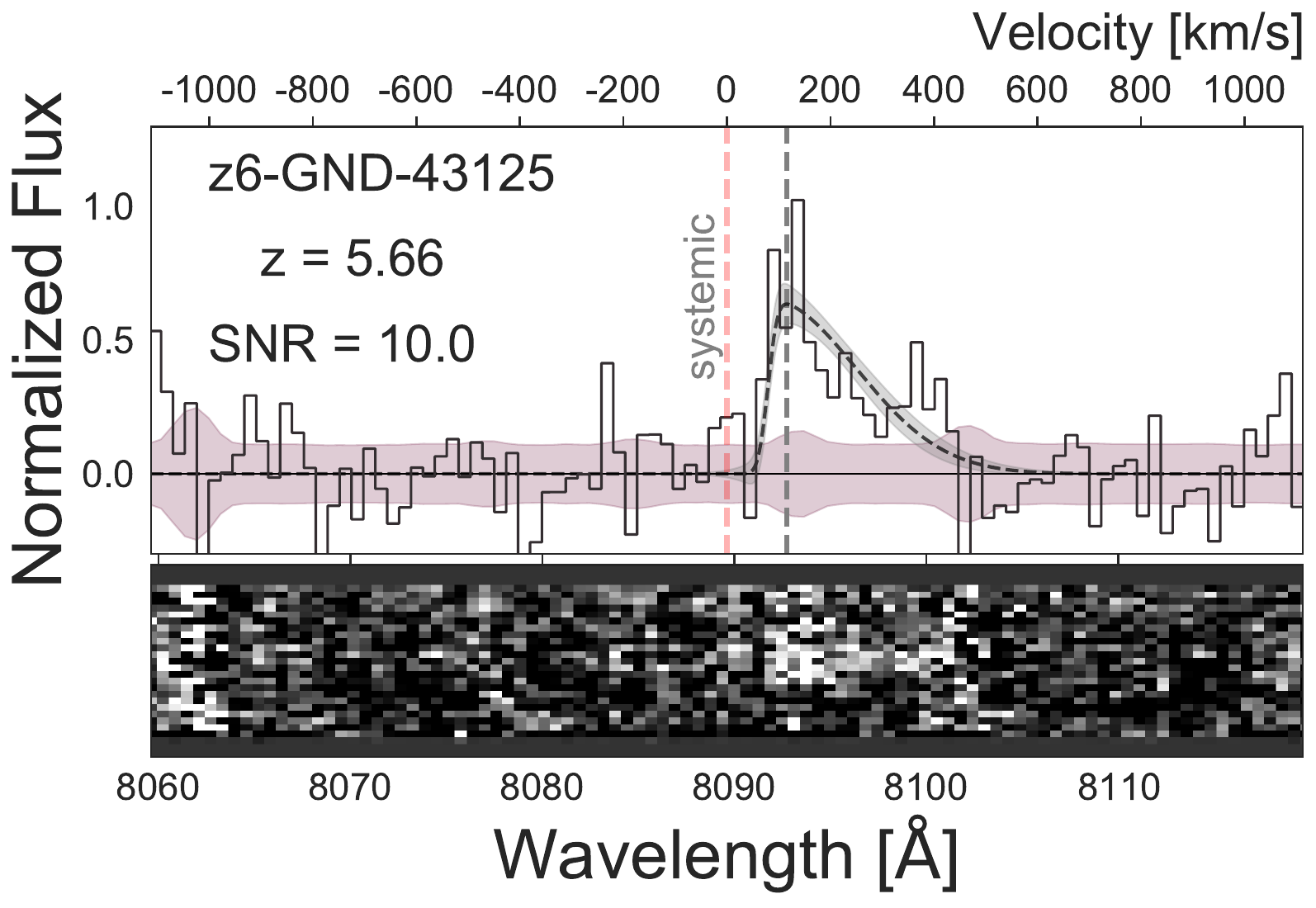}
\includegraphics[width=4.5cm, trim=0.1cm 1.5cm 1cm 0cm]{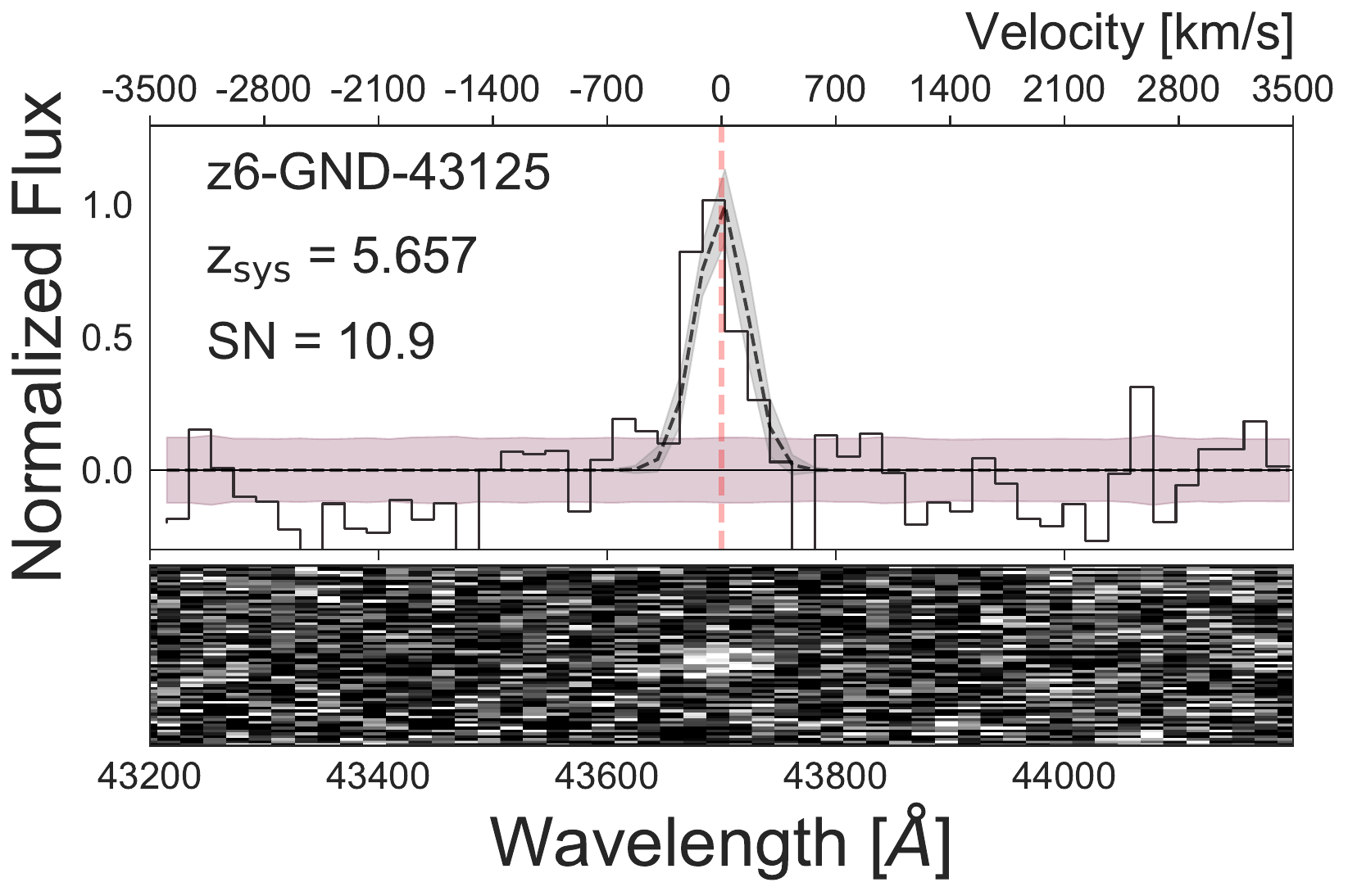}
\end{center}
\end{figure*}

\begin{figure*}[!h]
\begin{center}
\includegraphics[width=4.5cm, trim=0.1cm 1.5cm 0cm 0cm]{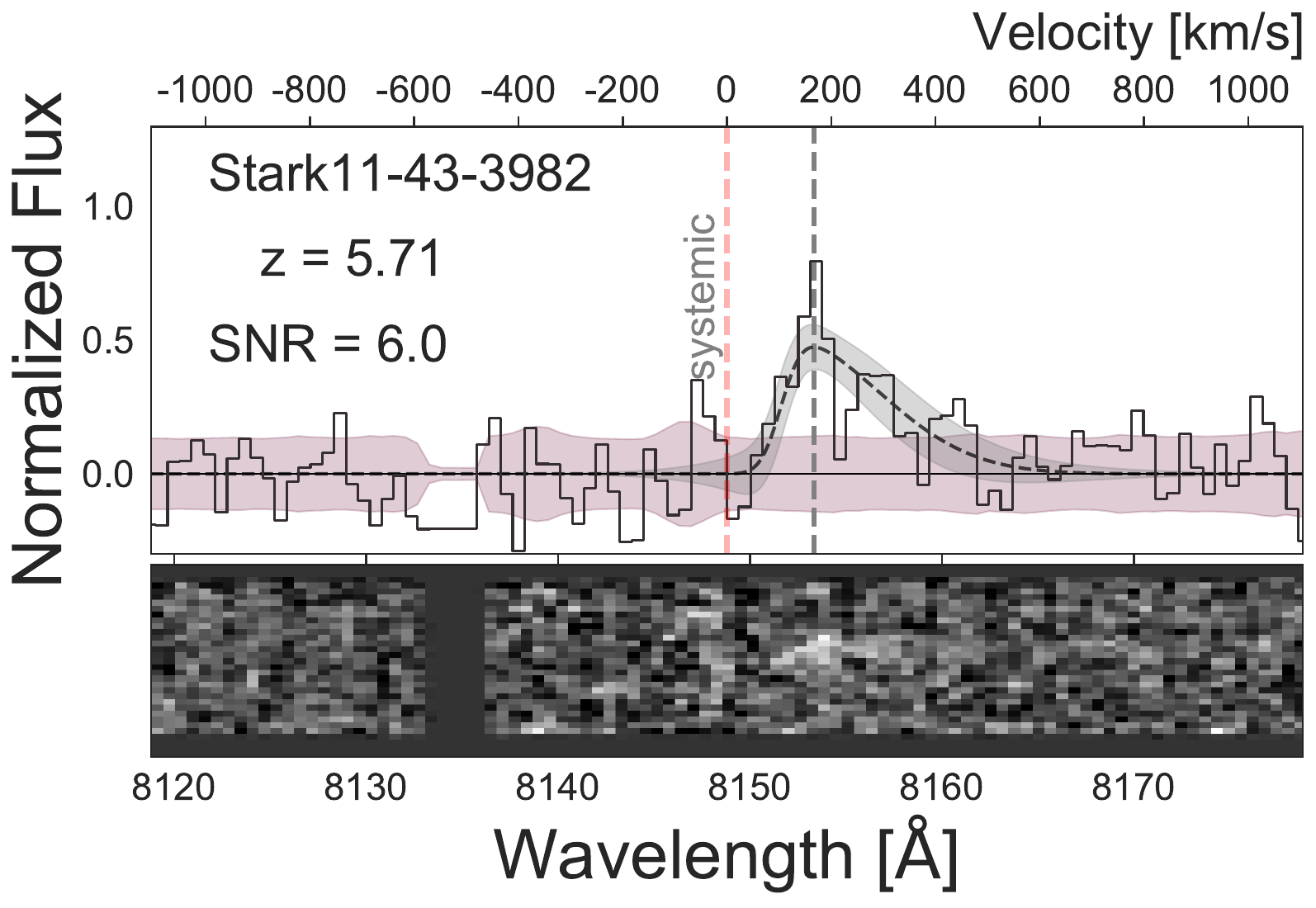}\includegraphics[width=4.5cm, trim=0.1cm 1.5cm 1cm 0cm]{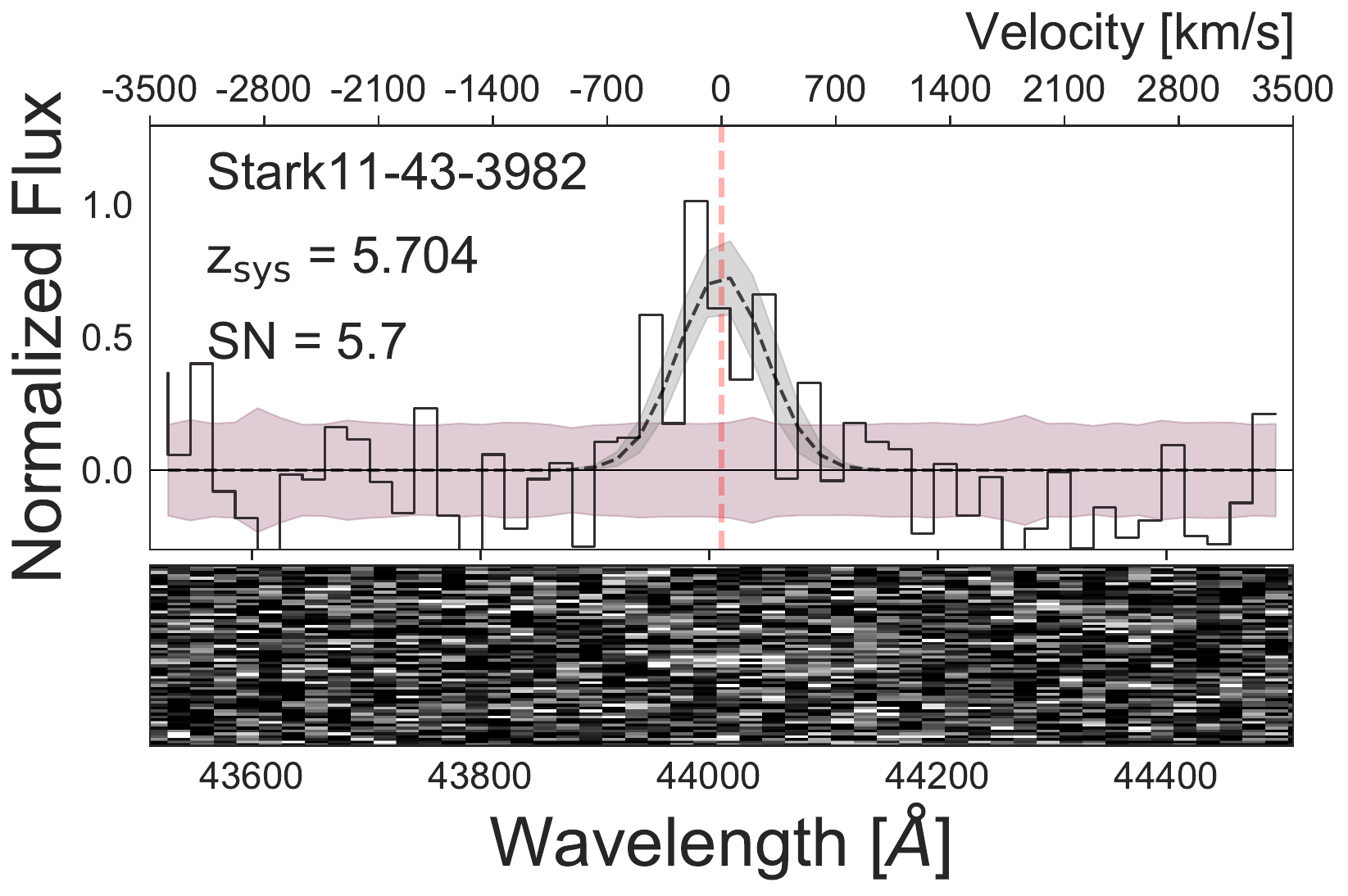}\hspace{0.2cm}\includegraphics[width=4.5cm, trim=0.1cm 1.5cm 0cm 0cm]{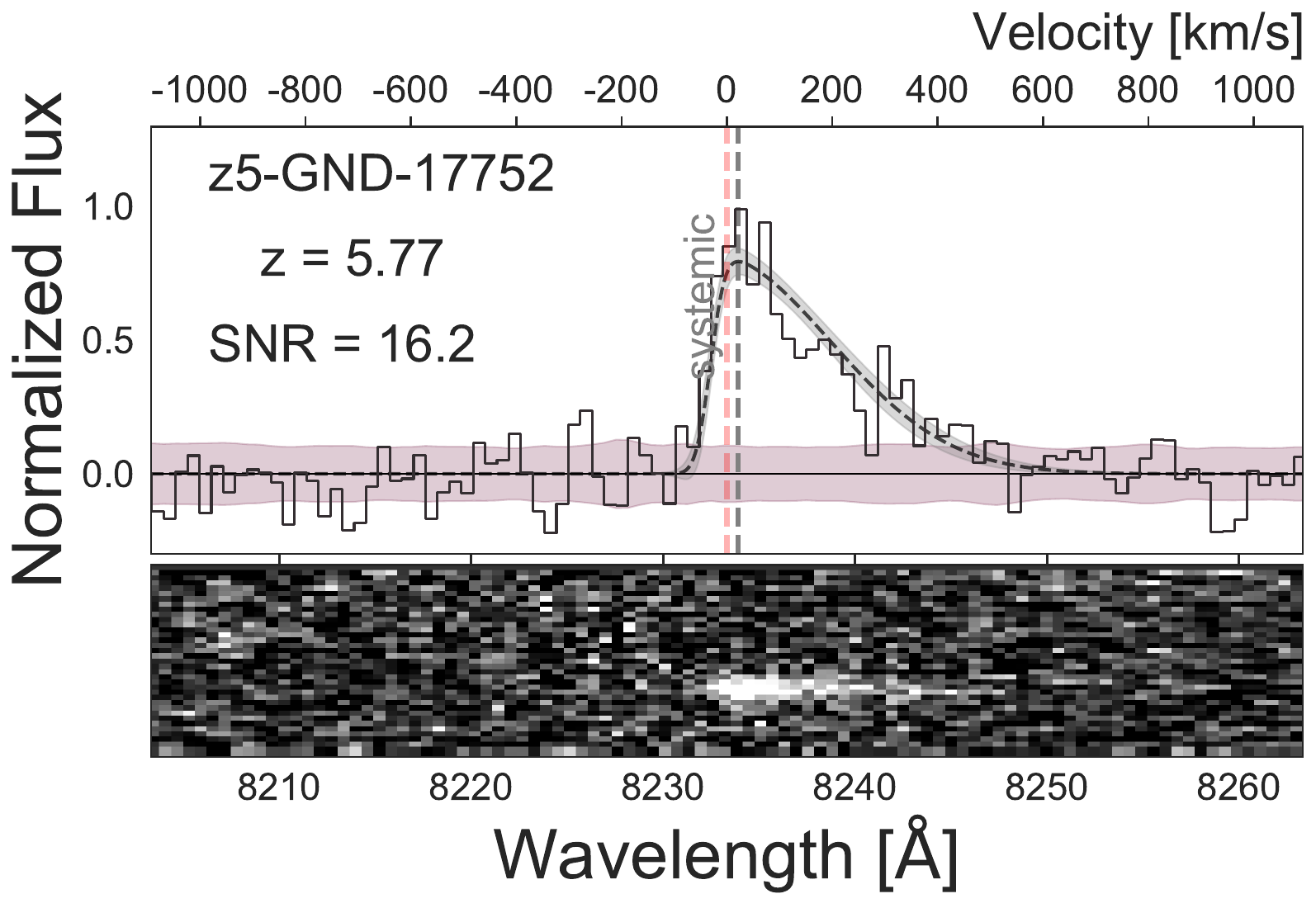}
\includegraphics[width=4.5cm, trim=0.1cm 1.5cm 1cm 0cm]{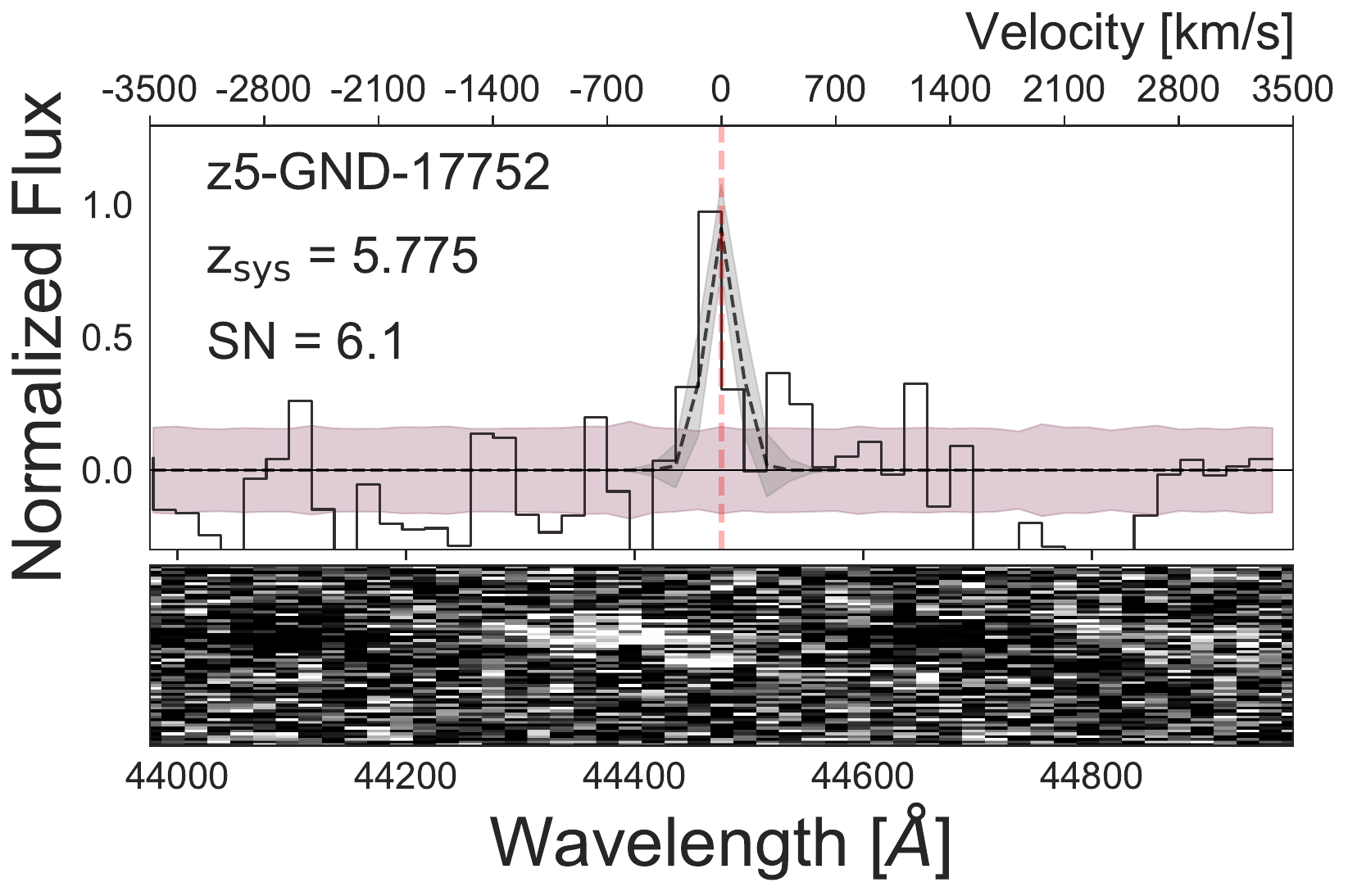}
\end{center}
\end{figure*}

\begin{figure*}[!h]
\begin{center}
\includegraphics[width=4.5cm, trim=0.1cm 1.5cm 0cm 0cm]{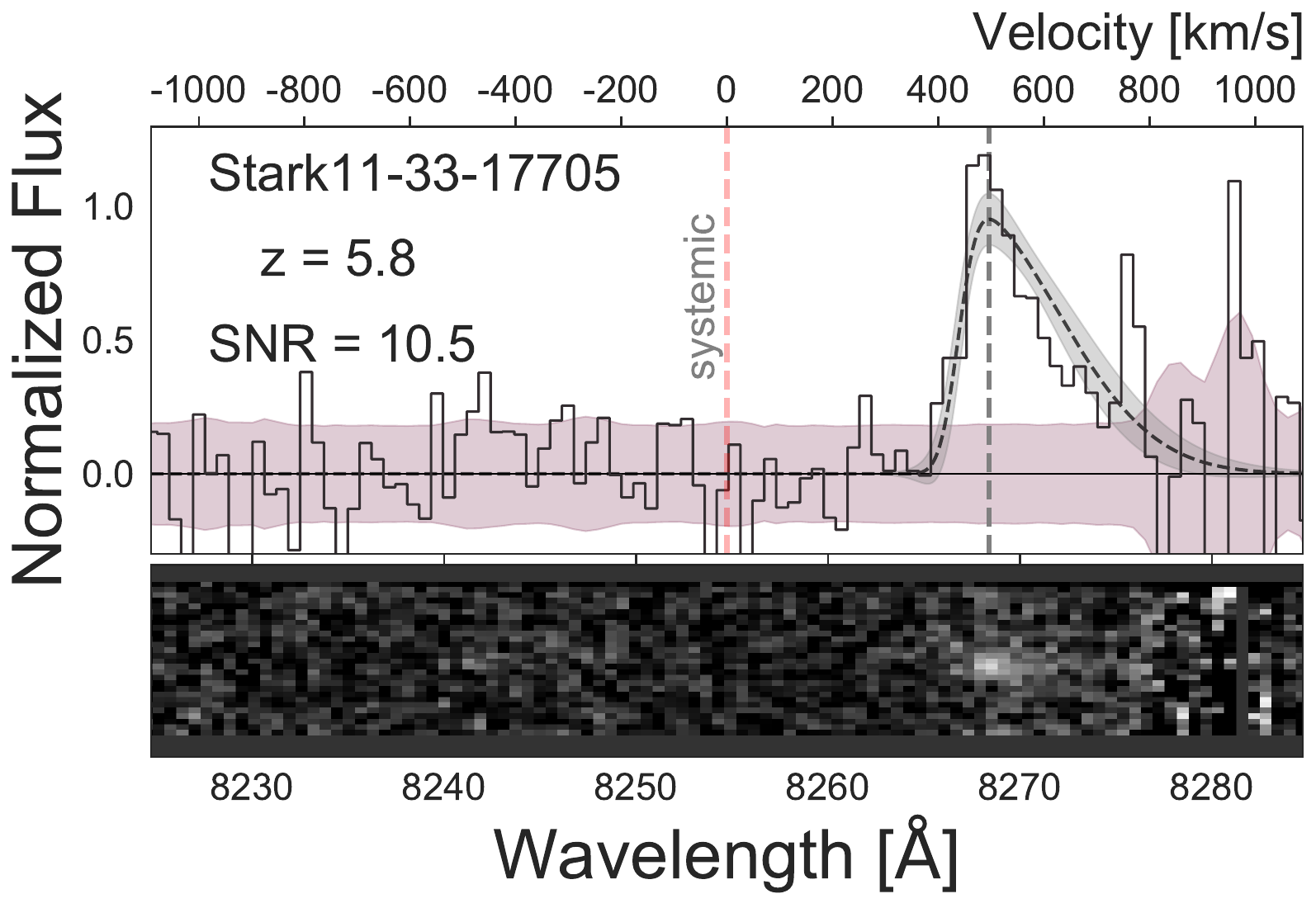}\includegraphics[width=4.5cm, trim=0.1cm 1.5cm 1cm 0cm]{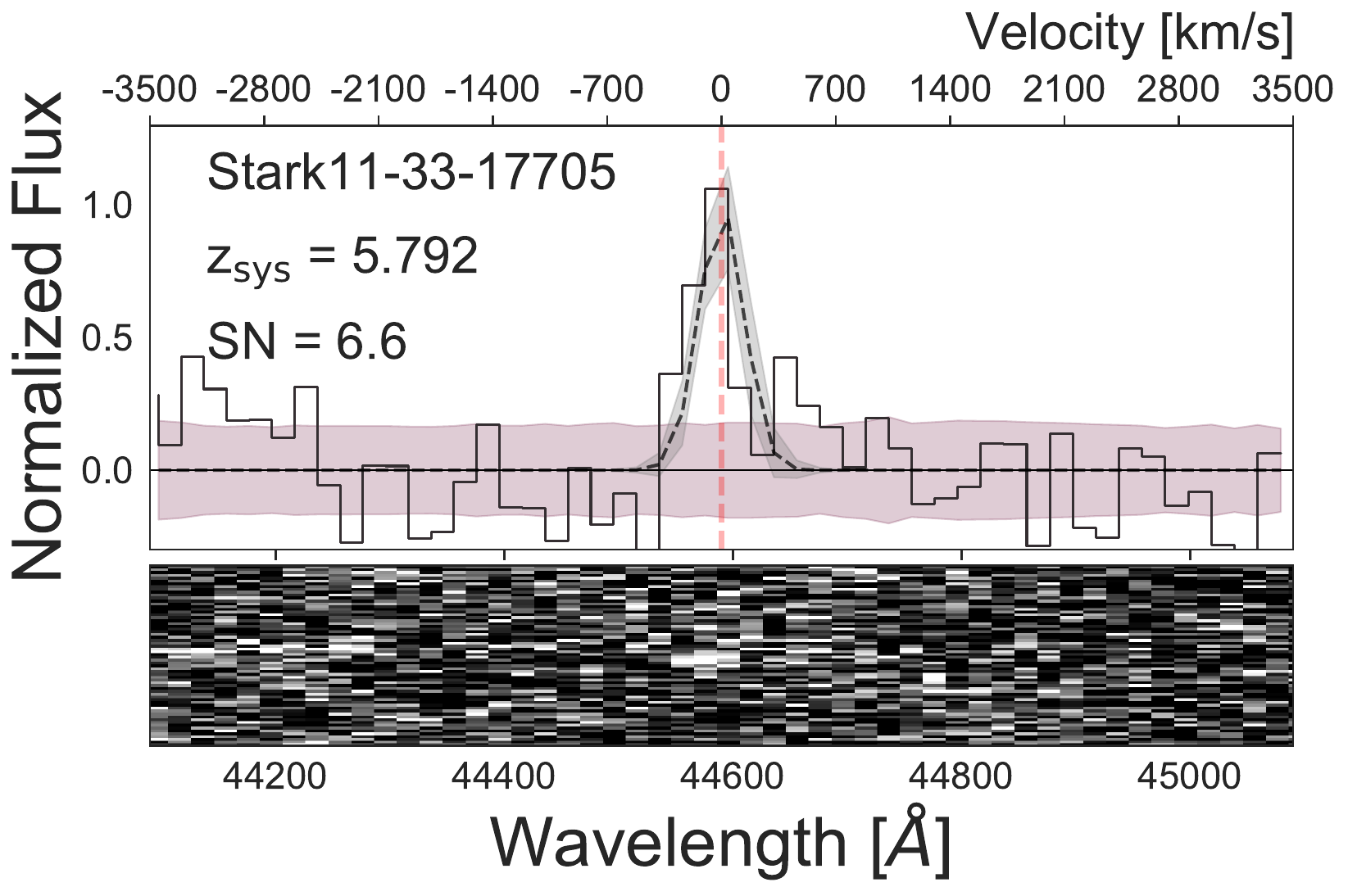}\hspace{0.2cm}\includegraphics[width=4.5cm, trim=0.1cm 1.5cm 0cm 0cm]{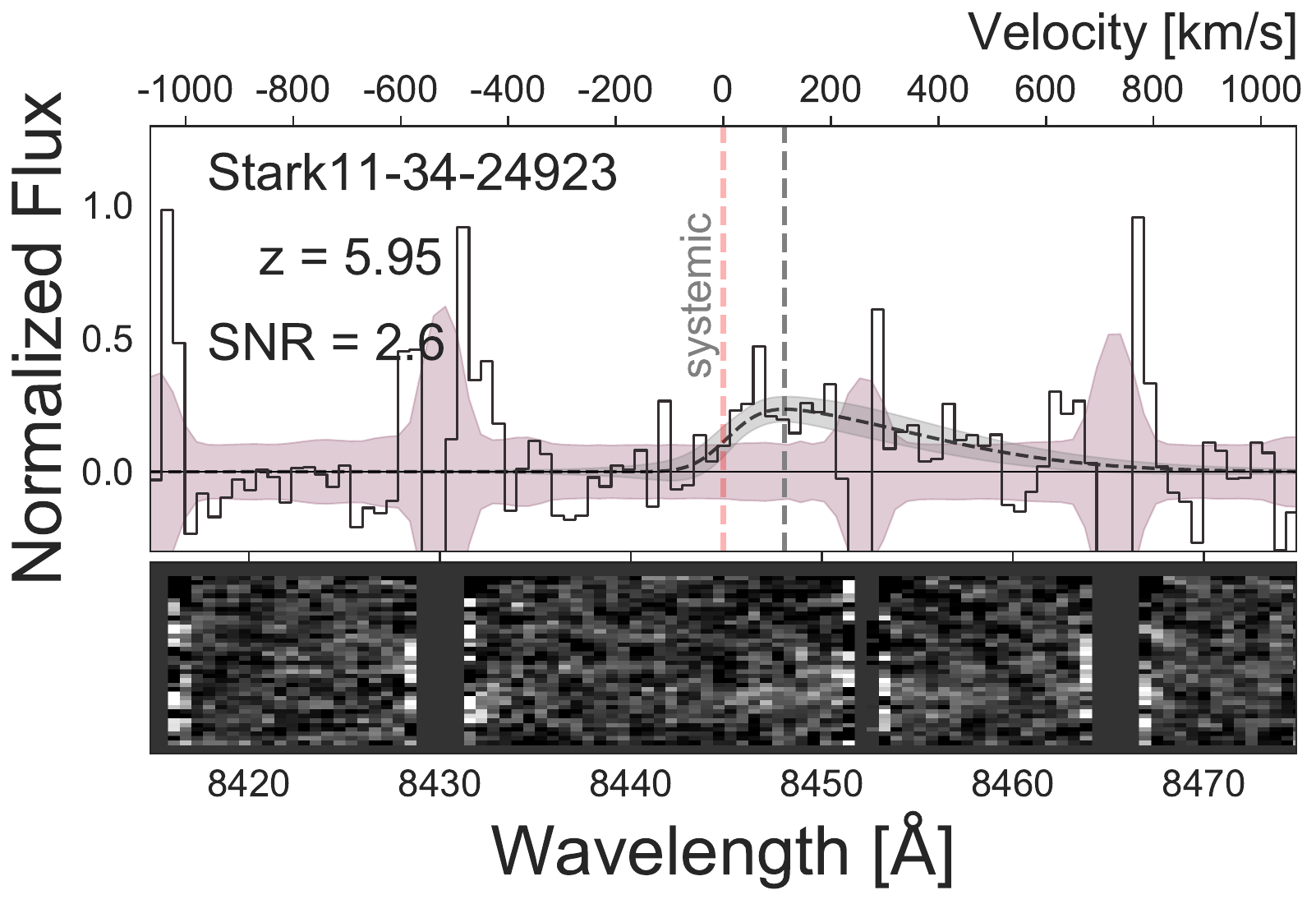}
\includegraphics[width=4.5cm, trim=0.1cm 1.5cm 1cm 0cm]{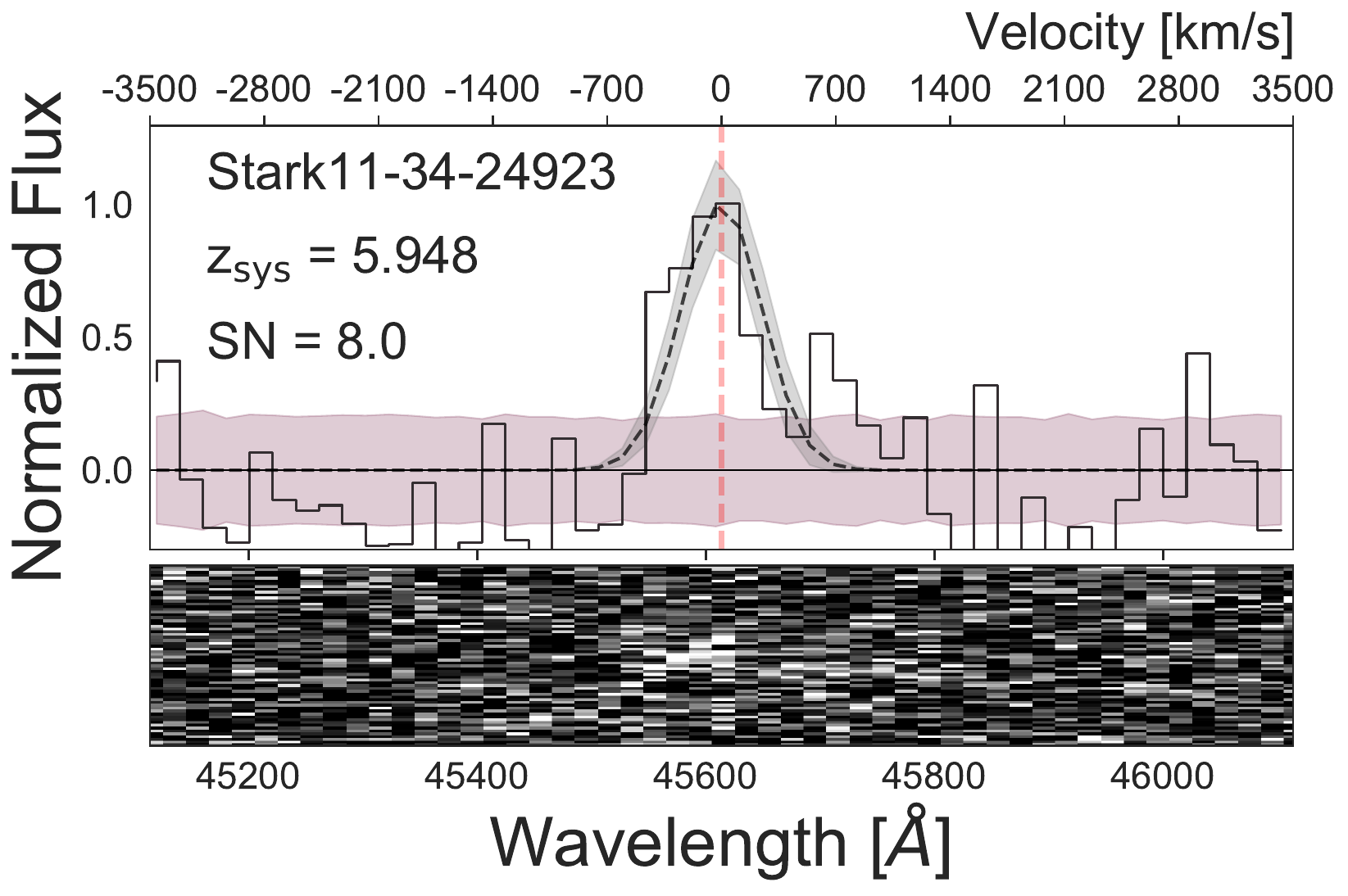}
\end{center}
\end{figure*}

\begin{figure*}[!h]
\label{fig:lae_spectra}
\begin{center}
\includegraphics[width=4.5cm, trim=0.1cm 1.5cm 0cm 0cm]{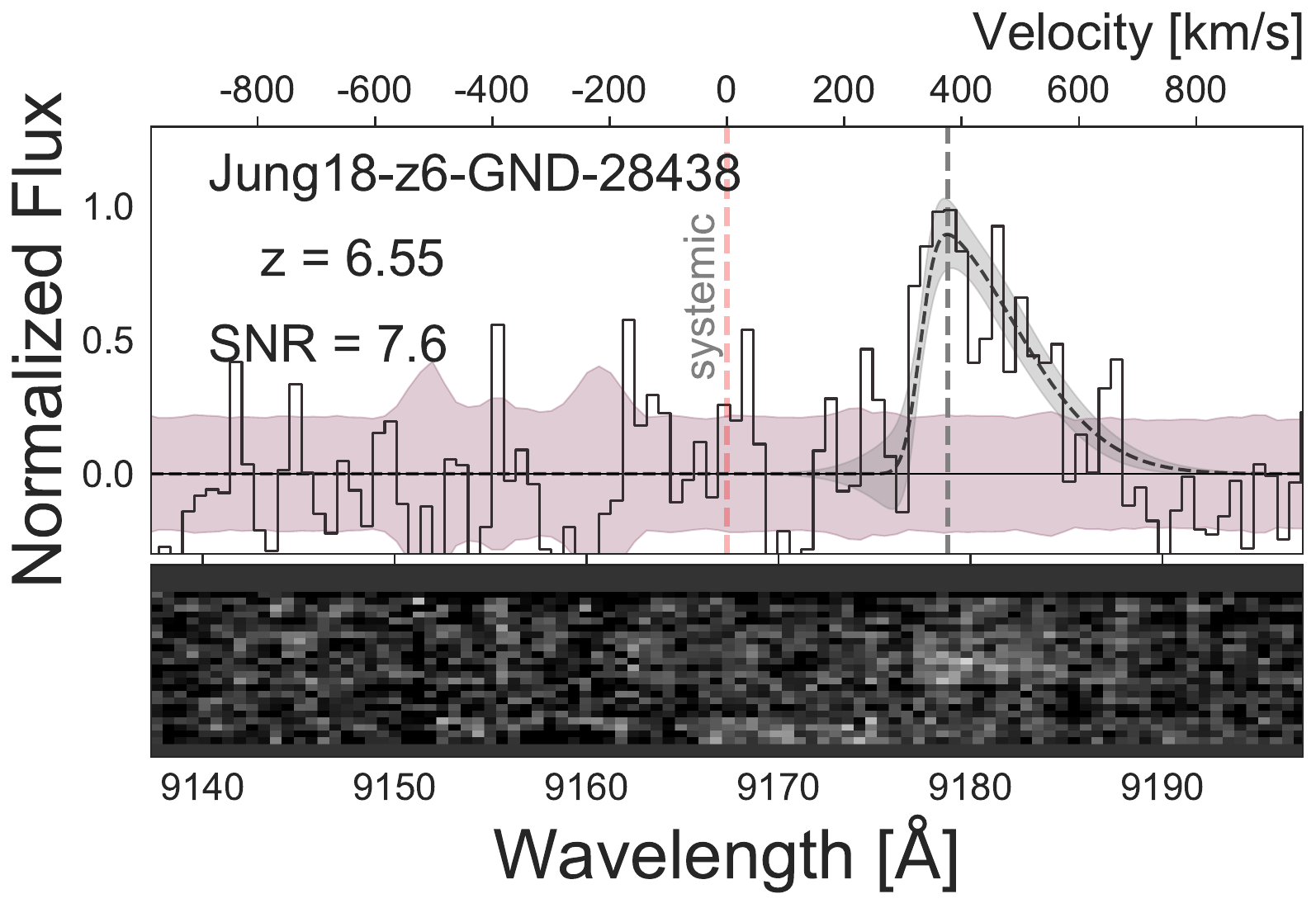}\includegraphics[width=4.5cm, trim=0.1cm 1.5cm 1cm 0cm]{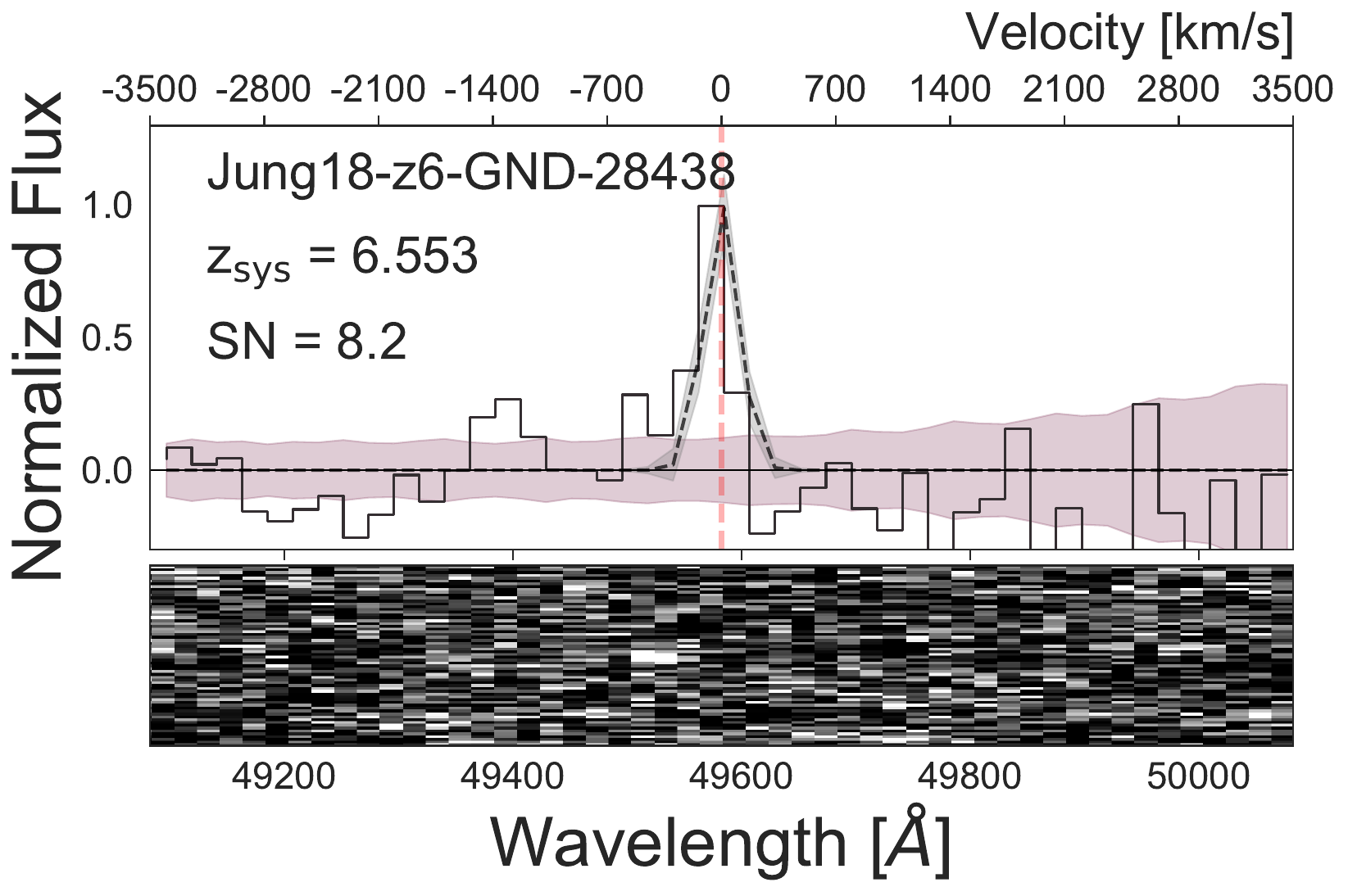}\hspace{0.2cm}\includegraphics[width=4.5cm, trim=0.1cm 1.5cm 0cm 0cm]{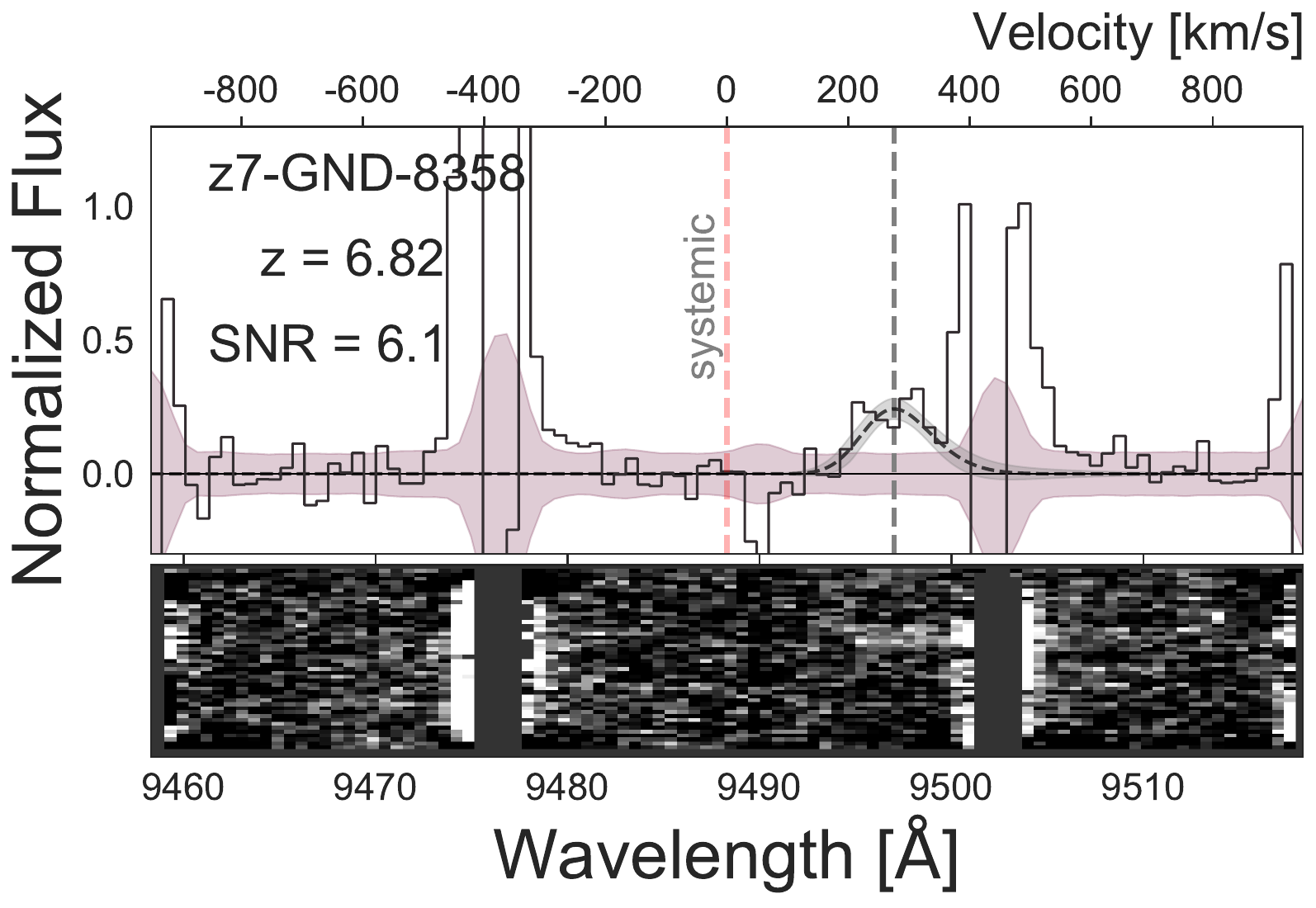}
\includegraphics[width=4.5cm, trim=0.1cm 1.5cm 1cm 0cm]{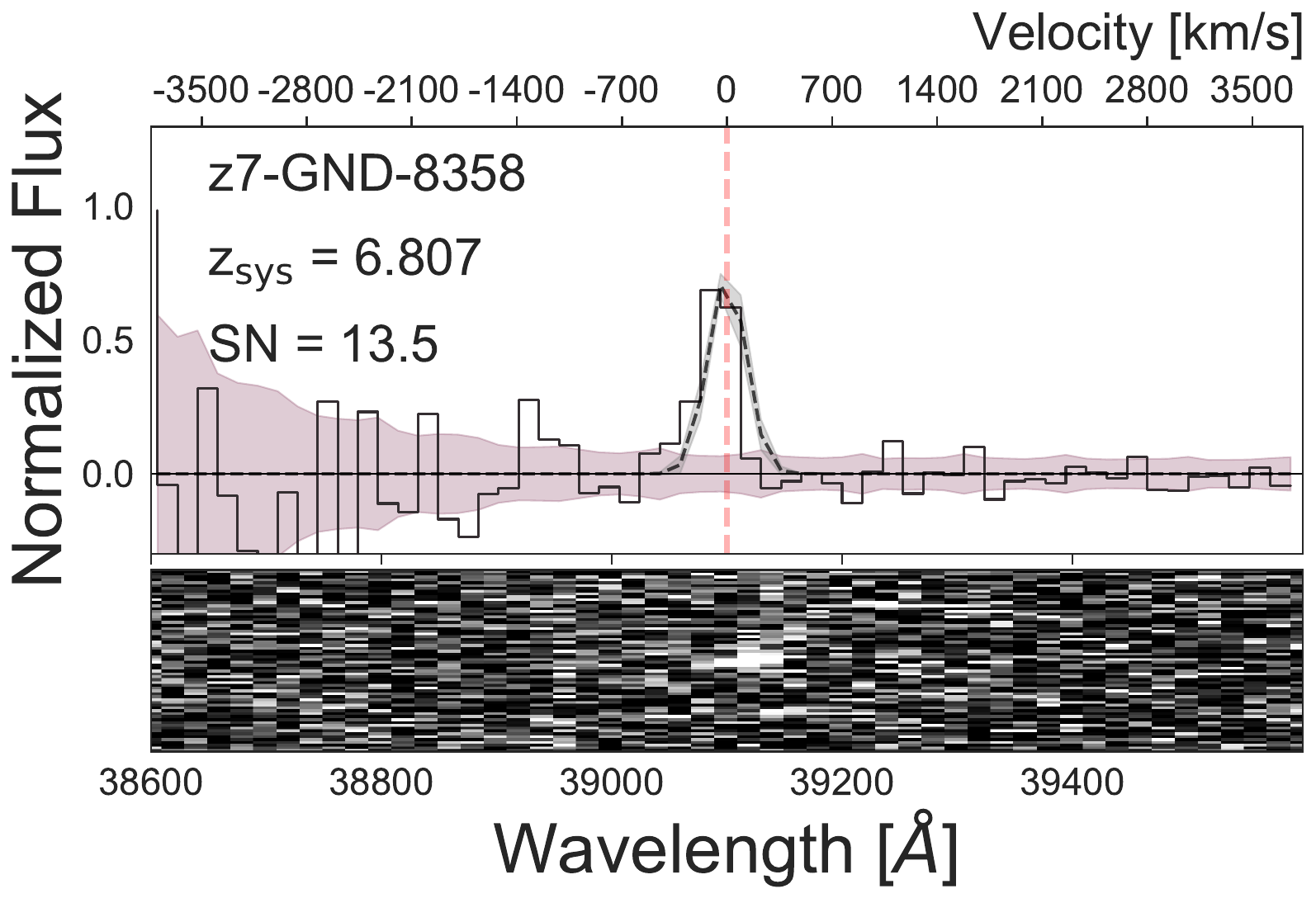}
\end{center}
\end{figure*}
\vspace{5cm}
\section{\lya asymmetry classification}\label{appendix:asymmetry}
\begin{figure*}[!h]
\begin{center}\includegraphics[width=7cm, trim=0.1cm 0cm 0cm 0cm]{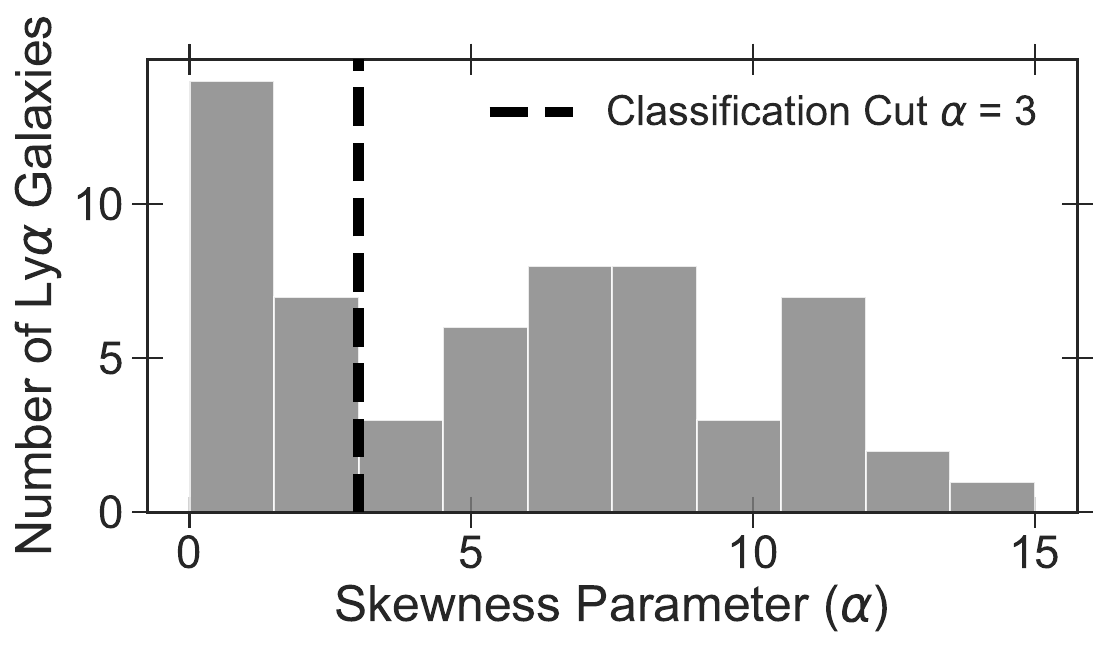}\includegraphics[width=7cm, trim=0.1cm 0cm 0cm 0cm]{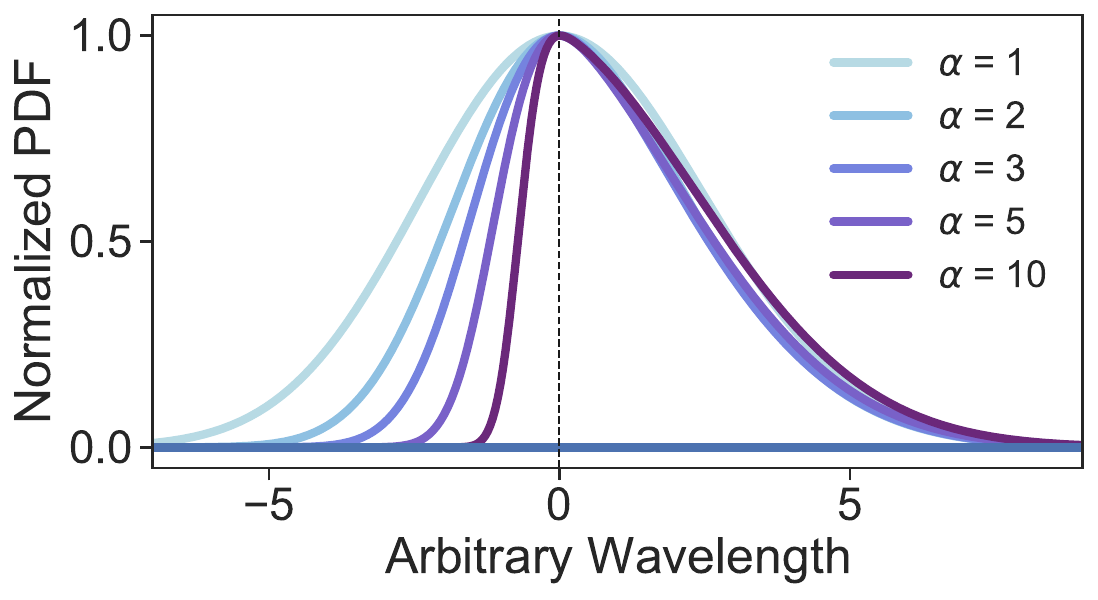}
\caption{\label{fig:classification_cut} Left: Histogram of the skewness ($\alpha$) of each observed  \lya profile. Right: Skewed Gaussian profiles as a function of $\alpha$.}

\end{center}

\end{figure*}

\normalsize
As discussed in Section~\ref{sec:results_profiles} $\sim70\%$ of our \lya profiles showed an asymmetric profile. We adopted a skewness parameter threshold of $\alpha > 3$ (See \ref{eq:SkewedGaussian}) to classify a \lya line as asymmetric. 
In Figure~\ref{fig:classification_cut}, we show the distribution of skewness parameters, $\alpha$, in our sample (left panel), along with example Skewed Gaussian profiles for different skewness parameters (right panel). 
The model line profiles demonstrate that as $\alpha > 3$ the profiles become increasingly asymmetric, with the profiles appearing nearly fully asymmetric by $\alpha = 5$.

\end{document}